%% file: STC_LHCILC.tex
\documentclass{desyprocA4}
\usepackage{color}
\usepackage{parskip}               
\usepackage{graphicx}				
\usepackage{xspace}
\usepackage{multirow}    
\usepackage{amssymb}
\usepackage{cite}
\usepackage{amsmath}
\usepackage{subfigure}
\usepackage{amsfonts}
\usepackage{booktabs}
\usepackage{abbreviations}
\usepackage[ , ,bf,it]{caption}    
\usepackage{url}

\def\to{\rightarrow}

\def\bi{\begin{itemize}}
 \def\ei{\end{itemize}}
\def\te{\widetilde e}

\def\c1p{C1^\prime}
\def\msq3{\overline{m}_{\widetilde{q}}(3)}

\def\tst{\widetilde t}
\def\ttau{\widetilde \tau}
\def\tmu{\widetilde \mu}

\def\be{\begin{equation}}  
\def\ee{\end{equation}}  
\def\bea{\begin{eqnarray}}  
\def\eea{\end{eqnarray}}

\def\tz{\widetilde\chi^0}
\def\XIPM#1{\ensuremath{ \widetilde{\chi}^{\pm}_#1}}
\def\XI0#1{\ensuremath{ \widetilde{\chi}^0_#1}}
\def\stau#1{\ensuremath{ \widetilde{\tau}_#1}}
\def\smu#1{\ensuremath{ \widetilde{\mu}_#1}}
\def\sel#1{\ensuremath{ \widetilde{e}_#1}}
\def\snu#1{\ensuremath{ \widetilde{\nu}_#1}}

\newcommand{\eeto}    {\ensuremath{ {\, e}^+ {e}^- \to}}
\input{newcom_mod.tex}

\begin{document}

\title{Non-Simplified SUSY: $\ttau$-Coannihilation at LHC and ILC}
\author{M.~Berggren$^1$, A.~Cakir$^{1,2}$, D.~Kr{\"u}cker$^1$, J.~List$^1$, I.-A.~Melzer-Pellmann$^1$, B.~Safarzadeh Samani$^{1,3}$, C.Seitz$^1$, S.~Wayand$^{4}$\\[1ex]
$^1$DESY, Notkestra{\ss}e 85, 22607 Hamburg, Germany\\
$^2$Istanbul Technical University, Department of Physics Engineering, 34469, Istanbul, Turkey\\
$^3$IPM, P.O. Box 19395-5531, Tehran, Iran \\
$^4$KIT IEKP, CS Geb. 30.23 Wolfgang-Gaede-Str. 1, 76131 Karlsruhe, Germany\\
}

\desyproc{DESY 15-145}
\doi
\maketitle


\begin{abstract}
If new phenomena beyond the Standard Model will be discovered at the 
LHC, the properties of the new particles could be determined with data from 
the High-Luminosity LHC and from a future linear collider like the ILC. 
We discuss the possible interplay 
between measurements at the two accelerators in a concrete example, namely 
a full SUSY model which features a small $\stau{1}$-LSP mass difference.
Various channels have been studied using the Snowmass 2013
combined LHC detector implementation in the Delphes simulation 
package, as well as simulations of the ILD detector concept from the 
Technical Design Report. We investigate 
both the LHC and ILC capabilities for discovery, separation and 
identification of various parts of the spectrum. While some parts
would be discovered at the LHC, there is substantial room 
for further discoveries at the ILC. We finally highlight examples where the precise 
knowledge about the lower part of the mass spectrum which could be acquired at the ILC 
would enable a more in-depth analysis of the LHC data with respect to the heavier states.

\end{abstract}


\input{Introduction}

\input{Model}

\input{LHC}

\input{ILC}

\input{Interpretation}  
\input{Summary}

\section*{Acknowledgement}
We would like to thank the {\sc Whizard} authors, the LC Generator Group as well as the ILD MC production team for their help in producing the large samples of events used in this work. 
The results presented could not be achieved without the National Analysis Facility and we thank the NAF team for their continuous support. 
We thankfully acknowledge the support by the DFG through the SFB 676 ``Particles, Strings and the Early Universe''.

\clearpage
\bibliographystyle{utphys}
\bibliography{STC_LHCILC}

\clearpage

\input{Appendix}

\end{document}

%% file: newcom_mod.tex
\def\leqsim{\mathbin{\;\raise1pt\hbox{$<$}\kern-8pt\lower3pt\hbox{$\sim$}\;}}
\def\geqsim{\mathbin{\;\raise1pt\hbox{$>$}\kern-8pt\lower3pt\hbox{$\sim$}\;}}

\def\MXN#1{\mbox{$ M_{\widetilde{\chi}^0_#1}                                $}}
\def\MXC#1{\mbox{$ M_{\widetilde{\chi}^{\pm}_#1}                            $}}

\def\XPM#1{\mbox{$ \widetilde{\chi}^{\pm}_#1                                $}}
\def\XMP#1{\mbox{$ \widetilde{\chi}^{\mp}_#1                                $}}
\def\XN#1{\mbox{$ \widetilde{\chi}^0_#1                                     $}}

\def\p#1{\mbox{$ \mbox{\bf p}_1                                         $}}

\newcommand{\Cnm}     {\mbox{$ \widetilde{\chi}^n_{m}                          $}}

\newcommand{\smul}    {\mbox{$ \widetilde{\mu}_{\mathrm L}                     $}}

\newcommand{\sell}    {\mbox{$ \widetilde{\mathrm e}_{\mathrm L}               $}}

\newcommand{\msnu}    {\mbox{$ m_{\widetilde\nu}                               $}}

\newcommand{\sle}     {\mbox{$ \widetilde{\ell}                                $}}
\newcommand{\sler}     {\mbox{$ \widetilde{\ell}_{\mathrm R}                     $}}
\newcommand{\slel}     {\mbox{$ \widetilde{\ell}_{\mathrm L}                     $}}

\newcommand{\ellell}  {\mbox{$ \, \ell^+ \ell^-                               $}}

\newcommand{\MZ}      {\mbox{$ m_{{\mathrm Z}}                             $}}

\newcommand{\ba}{\begin{array}}
\newcommand{\ea}{\end{array}}
\newcommand{\bc}{\begin{center}}
\newcommand{\ec}{\end{center}}
\newcommand{\eeq}{\end{eqnarray}}
\newcommand{\bes}{\begin{eqnarray*}}
\newcommand{\ees}{\end{eqnarray*}}
\newcommand{\Kz}{\ifmmode {\rm K^0_s} \else ${\rm K^0_s} $ \fi}
\newcommand{\xxbar}{\ifmmode {\rm x\bar{x}} \else ${\rm x\bar{x}} $ \fi}
\newcommand{\rphi}{\ifmmode {\rm R\phi} \else ${\rm R\phi} $ \fi}

\newcommand{\Lum}{${\cal L}\;$}

\newcommand{\Ecms}    {\mbox{$ E_{\mathrm{\small cms}}                      $}}

\def    \missEt      {\ifmmode{/\mkern-11mu E_t}\else{${/\mkern-11mu E_t}$}\fi}
\def    \missE       {\ifmmode{/\mkern-11mu E}\else{${/\mkern-11mu E}$}\fi}
\def    \missp       {\ifmmode{/\mkern-11mu p}\else{${/\mkern-11mu p}$}\fi}
\def    \misspt      {\ifmmode{/\mkern-11mu p_t}\else{${/\mkern-11mu p_t}$}\fi}


%% file: Introduction.tex
\section{Introduction} \label{sec:introduction}

Supersymmetry (SUSY)~\cite{Martin:1997ns,ref:SUSY0,ref:SUSY1,ref:SUSY2,ref:SUSY3,ref:SUSY4}
is one of the best motivated theories beyond the Standard Model (SM). In the past, at LEP, Tevatron and the 
early LHC, results were mainly interpreted in the constrained minimal
supersymmetric extension of the Standard Model (CMSSM)~\cite{CMSSM1,CMSSM2}, while later 
LHC search results are mostly presented in a simplified model approach~\cite{SMS1,SMS2,SMS3}. 
Simplified models offer more flexibility for comparing to predictions of different theoretical models, though they might not describe reality well, as 
they usually contain only one decay with $100\,\%$ branching ratio. In contrast, a real SUSY signal 
might comprise a large spectrum of SUSY particles and the
higher states of the spectrum may have many decay modes leading to
potentially long decay chains. This means that the simplified approach in general does not 
apply beyond the direct production of the next-to-lightest SUSY particle (NLSP) and the interpretation of exclusion limits formulated in
the simplified approach is non-trivial. Furthermore, also several production channels may be open, 
making SUSY the most serious background to itself. This becomes especially relevant for 
interpreting a future discovery of a non-SM signal.

In this paper, we investigate how a well-motivated full-spectrum $R$-parity~\cite{FarrarEtAl} conserving SUSY model could lead to 
signals beyond the SM at the LHC and at a future linear collider like the ILC.
A key feature of the investigated models is a $\ttau$-NLSP with a
small mass difference of about $10$~GeV to the $\tz_1$, the lightest SUSY particle (LSP) of the model. 
Such scenarios were favoured by fits to 
all pre-LHC experimental data within the CMSSM~\cite{Buchmueller:2009fn},
since the small mass difference allows to match the observed Dark Matter relic density
via a sizeable $\ttau$-coannihilation contribution. The $\ttau$-coannihilation remains among
the favoured regions of parameter space in the CMSSM and in the Non-Universal Higgs Mass Model (NUHM2) when including LHC
data~\cite{Buchmueller:2014check}. Furthermore, light $\ttau$'s will be an interesting candidate to enhance
the rate for $H\to \gamma\gamma$ with respect to the SM expectation~\cite{Endo:2014pja, Carena:2012gp}.

Within the context of constrained models, the masses of the fermion-partner particles (sfermions) in kinematic reach of the ILC are 
excluded with high confidence. However, this exclusion is in most cases based on the strongly
 interacting sector, which in constrained models is coupled to the electroweak sector by 
 GUT-scale mass unification. Without the restriction of {\it mass} unification, the part 
 of the spectrum which is of interest to electroweak and flavor precision observables as well
 as 
 dark matter, i.e.\ which is decisive for the fit outcome, is not at all in conflict with 
 LHC results. This applies in particular
to the $\ttau_1$ with a small mass difference to the LSP:
Although first limits on direct $\ttau$ pair production from the LHC have been 
presented~\cite{Aad:2014yka},
they rapidly loose sensitivity if the $\ttau$ is not degenerate with the $\te$ and $\tmu$, 
and has a small mass difference to the LSP. 
In fact, the current limit on the $\ttau$ without other assumptions on the mass difference to the LSP
than that it is larger than $m_{\tau}$
nor with any assumptions on the $\ttau$ mixing-angle comes from the DELPHI experiment at LEP
and is $M_{\ttau} > 26.3$\GeV~\cite{Abdallah:2003xe,Agashe:2014kda}.
Due the key feature of a small $\ttau$-LSP mas difference and the resulting sizeable $\ttau$-coannihilation contribution the series of $CP$-conserving model points considered here is called STC~\cite{Baer:2013ula}.

Motivated by solving the naturalness problem within the general MSSM~\cite{Naturalness_Ref}, 
the SUSY partner particles of third-generation quarks in STC have been chosen to be lighter 
than those of the first- and second-generation squarks.  When the first- and second-generation 
squarks and the gluino are rather heavy, 
$\gtrsim 2\TeV$, the size of the total SUSY
cross-section at the LHC strongly depends on the mass of the lightest top squark. 
We therefore consider in particular two model points, called STC8 and STC10, 
whose physical spectra differ only by the mass parameter of the partners of the 
right-handed third-generation squarks at a scale of 1\TeV. The mass parameter of 
the right-handed third-generation squarks is set to 800 and 1000\GeV, 
resulting in physical masses of the top squark of $m_{\tst_1} \approx 740\GeV$ and 
$m_{\tst_1} \approx 940\GeV$, and of the bottom squark of $m_{\sBot_1} \approx 800\GeV$ 
and $m_{\sBot_1} \approx 1000\GeV$, 
respectively.

\begin{figure}[htb]
  \begin{center}
\subfigure[]{  \includegraphics[width=0.45\linewidth]{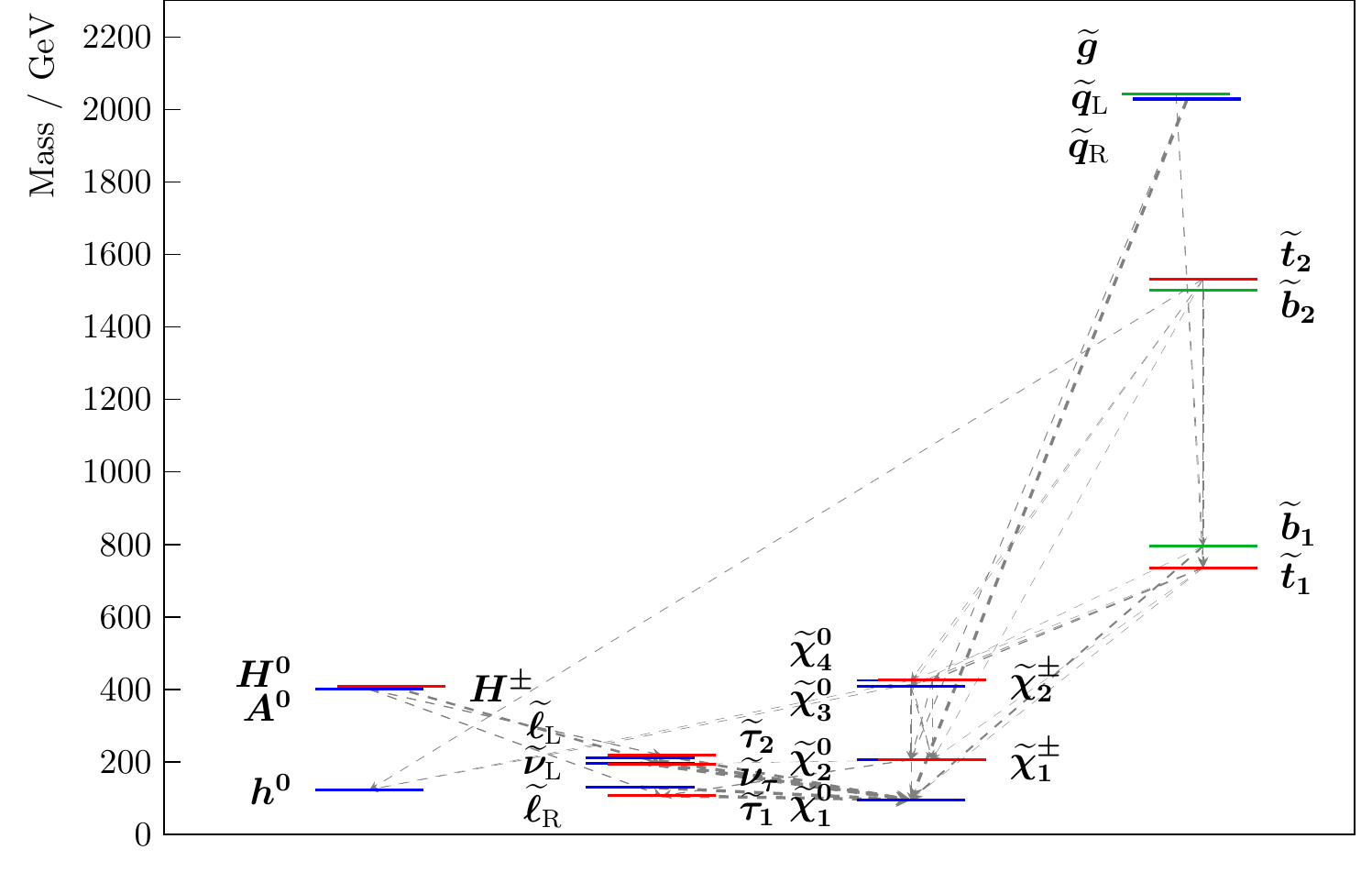}}
  \vspace{0.1cm}
\subfigure[]{  \includegraphics[width=0.45\linewidth]{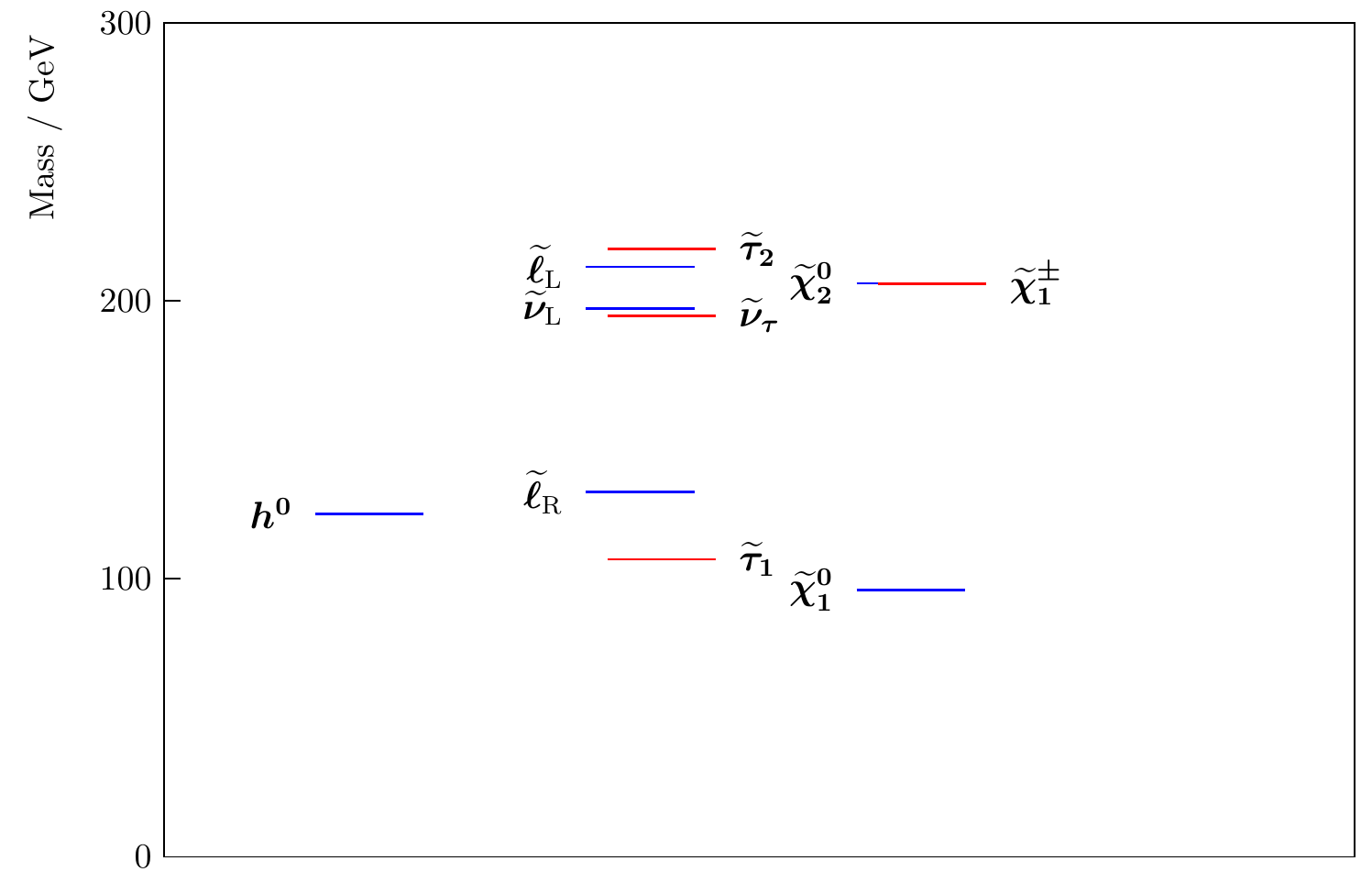}}
 \end{center}
  \caption{\label{fig:stc8} Full spectrum of STC8 and decay modes with a
   branching fraction of at least $10\,\%$ (a). The lower part of the spectrum of the STC
    scenarios, which features $M_{\stau{1}} - M_{\tz_1} \approx 10\GeV$ (b).}
\end{figure}

 
Figure~\ref{fig:stc8}(a) introduces the full mass spectrum of the benchmark scenario STC8, while Fig.~\ref{fig:stc8}(b) zooms into the part of the spectrum accessible at the ILC with $\Ecms=500\GeV$. The lightest higgs boson features SM-like branching ratios and has a mass in agreement with the LHC discovery within
the typical theoretical uncertainty of $\pm 3\GeV$ on MSSM higgs boson mass calculations. 

The dashed lines in Fig.~\ref{fig:stc8}(a) indicate those decay chains of 
the various sparticles which have branching fractions of at least $10\,\%$. The grey-scale 
of the lines indicates the size of the branching ratio. Only very few particles, 
namely the first and second generation squarks, the sneutrinos and the lighter set of 
charged sleptons have decay modes with $100\,\%$ branching ratio. 

In particular the top and bottom squarks, but also the the superpartners of the uncoloured bosons, 
called electroweakinos in the following, have various decay modes, none of them with a branching ratio 
larger than $50\,\%$, but many with less than $10\,\%$. This plethora of decay modes makes 
it challenging to separate the various production modes and identify each sparticle.
  

The final goals of this study comprise the following questions for both LHC and the ILC in the example of the
STC scenarios:

\begin{itemize}
    \item Which signature will lead to the first discovery of a discrepancy from the SM? How much integrated
    luminosity and operation time will be needed?
    \item Which other signatures will be observable? 
    \item Which production modes of which sparticles contribute to this signal? 
    \item Can we tell how many sparticles are involved? 
    \item Which observables (masses, BRs, cross sections) can be measured and with which precision?
    \item Can we show that it is SUSY?
    \item Can the $\tz_1$ be identified as Dark Matter particle?
\end{itemize}

In the next section, we will discuss the benchmark model and its phenomenology in more detail. In Section~\ref{sec:LHC} 
we will review its possible discovery in different analyses at LHC and summarise the
obtained simulation results, while we will do the same in Section~\ref{sec:ILC} for the ILC case. 
In Section~\ref{sec:interpretation}, we employ the combined results from LHC and ILC in order to investigate the 
questions raised above, before concluding in Section~\ref{sec:conclusion}.

%% file: Model.tex
\section{Collider phenomenology of the STC scenarios}
\label{sec:model}

In this section we present the parameters of the benchmark models in more detail, 
summarising the masses of the most important SUSY particles, their production cross sections 
and branching ratios. Based on this information, the phenomenology at the LHC and ILC will be discussed.

A corner-stone of SUSY is that the couplings of the standard-model particles
and their supersymmetric partners are the same, so that on tree-level the production cross sections
only depend on the masses and mixing angles of the produced and exchanged sparticles.
 
In the case of the LHC studies, we assumed $\Ecms = $ 14\TeV
and an integrated luminosity up to 300 fb$^{-1}$. We also
consider the high-luminosity upgrade of the LHC, the HL-LHC,
assumed also to be running at $\Ecms = $ 14\TeV, but
delivering a total integrated luminosity of 3~ab$^{-1}$.

For the ILC studies, the conditions presented in the Technical
Design Report (TDR)\cite{Adolphsen:2013kya} were used. This means that
the bulk of the data would be recorded at  $\Ecms = $ 500\GeV,
with an integrated luminosity of 250~fb$^{-1}$ per year. We extrapolate
our results to the running scenarios recently published by the Joint Working Group
on ILC Beam Parameters~\cite{paramgroup}. 
Since the ILC beam energy is tunable, we also consider the
option of running at different, lower energies, and performing energy scans
around thresholds. Very important are the
opportunities offered by the electron and positron beam polarisation, for which
the baseline design of the ILC foresees absolute values of 80\%
and 30\%, respectively.
As SUSY is a chiral theory, the possibility to have
polarised initial conditions is a very powerful tool to
disentangle different states, to enhance signal while reducing
SM background, and to study helicity-dependent predictions
of the theory.

\subsection{Mass spectrum and decay modes}
\label{subsec:massdecay}

The masses of the most important sparticles in the
benchmark points STC8 and STC10 are listed in Table~\ref{tab:masses}.
The two models differ mainly in the masses of the lighter bottom squark ($\sBot_1$) and 
the lighter top squark ($\sTop_1$). These sparticles are important for analyses at the LHC, 
where both would be accessible at the chosen masses and will be searched for with dedicated analyses. A bottom squark with
a mass as in the STC8 model is on the edge of already being detected at the LHC with a centre-of-mass energy 
of $8\TeV$ and an integrated luminosity of $20\fbinv$~\cite{Aad:2013ija,CMS-PAS-SUS-13-018}, 
provided that the branching ratio of the direct decay $\sBot \rightarrow$ b\ninoone is $100$\,\%. 
However, this is usually not the case for a full-spectrum model like the one we investigate here, unless the 
$\sBot$ is the NLSP. In particular, in our scenarios, the branching fraction of this decay is
about $60$\,\% in STC8 and about $50$\,\% in STC10, as listed in Table~\ref{tab:brcol}. The analogous decay of the $\sTop$ has a 
branching fraction of only about $10$\,\%. This will make it very hard to select a $\sTop$ sample in a specific decay mode as a prerequisite to identify a kinematic edge.

\begin{table*}[htb]
\caption{Sparticle masses for the models STC8 and STC10. The mass of
  the first two generation squarks (\sQua) varies by a few GeV, their average
mass is listed. We use \sle~when we refer to the first and second
generation sleptons, while the third generation is listed separately.}
\begin{center}
\begin{tabular}{l|r|r || l|r|r}
\hline \hline
{\bf Sparticle} & \multicolumn{2}{c||}{\bf Mass / GeV} & {\bf Sparticle} & \multicolumn{2}{c}{\bf Mass / GeV} \\ 
 & {\bf STC8} & {\bf STC10} & & {\bf STC8} & {\bf STC10} \\ \hline
$\sBot_1$ & 795  & 1008  &                 $\sTau_1$  & 107   & 107 \\
$\sBot_2$ & 1500  & 1500 &                $\sTau_2$  & 219   & 219 \\
$\sTop_1$ & 736  & 944 &                 $\snu{\tau}$ & 196 & 196 \\
$\sTop_2$ & 1532  & 1537 &       \ninoone   & 96   & 96  \\           
\gluino   & 2042  & 2042 &     \ninotwo   & 206  & 206  \\           
$\sQua$  &  2028  & 2028 &    \ninothree & 410  & 410  \\            
\slel & 212  & 212 & \ninofour  & 425  & 426  \\
\sler & 131 & 131  &    \chipmone  & 206  & 206  \\
$\snu{\mathrm{L}}$ & 198 & 198  &   \chipmtwo  & 426  & 427  \\
\hline \hline
\end{tabular}
\end{center}
\label{tab:masses}
\end{table*}

\begin{table*}[htb]
\caption{Branching ratios (BR) of the gluinos and the third generation squarks in  the models STC8 and STC10. 
Only branching ratios above 1\,\% are shown.\label{tab:brcol}}
\begin{center}
\begin{tabular}{l|r|r || l|r|r}
\hline \hline
{\bf Decay} &  \multicolumn{2}{c||}{\bf BR / \%} & {\bf Decay} &  \multicolumn{2}{c}{\bf BR / \%} \\
  & {\bf STC8} & {\bf STC10} &   & {\bf STC8} & {\bf STC10} \\ \hline
$\gluino \rightarrow \topq\sTop_1$    & 37.8 & 35.5 &           $\gluino \rightarrow \botq\sBot_1$ & 39.1 & 36.9 \\              
$\gluino \rightarrow \topq\sTop_2$    & 11.5 & 13.8 &           $\gluino \rightarrow \botq\sBot_2$ & 11.5 & 13.7 \\ \hline          
$\sTop_1 \rightarrow \topq\ninoone$   & 13.2 & 9.9 &            $\sBot_1 \rightarrow \botq\ninoone$   & 58.2 & 51.4 \\ 
$\sTop_1 \rightarrow \topq\ninotwo$   & 4.5 &  4.3 &            $\sBot_1 \rightarrow \botq\ninotwo$   & 3.1 & 3.0 \\    
$\sTop_1 \rightarrow \topq\ninothree$ & 22.4 & 24.3 &           $\sBot_1 \rightarrow \botq\ninothree$ & 10.7 & 12.1 \\   
$\sTop_1 \rightarrow \topq\ninofour$  & 12.0 & 15.1&            $\sBot_1 \rightarrow \botq\ninofour$  & 9.2 & 10.5 \\    
$\sTop_1 \rightarrow \botq\chipone$   & 10.8 & 9.8 &            $\sBot_1 \rightarrow \topq\chimone$   & 5.1 & 5.2  \\   
$\sTop_1 \rightarrow \botq\chiptwo$   & 37.1 & 36.5&            $\sBot_1 \rightarrow \topq\chimtwo$   & 13.7 & 17.9\\    
\hline \hline
\end{tabular}
\end{center}
\end{table*}

The actual masses of the $\sell$ and $\smul$ in STCx have been excluded by the ATLAS experiment for the case of $100$\,\% branching ratio for the direct decay
to the corresponding lepton and the LSP~\cite{Aad:2014vma}. As can be seen in Table~\ref{tab:brslep}, the STCx branching
ratios for these decays are quite close to $100\%$, thus this particular 
part of the spectrum is most likely excluded. It should be noted, however, that this large BR is a special 
case due to the small mass difference of only $6$\,GeV between the \slel\ and the \XPM{1} / \XN{2}, which leads to a strong phase space suppression for the cascade decays $\slel \to l \XN{2}$ and $\slel \to \nu_l \XPM{1}$.  With increasing mass difference, the branching ratio for the direct decay $\slel \to l \XN{1}$ drops rapidly in favour of these cascade decays. Therefore, the results of this study remain highly relevant in the broader picture. Cases where the specific mass and branching ratio combination of the \slel\ plays a role will be pointed out. 

The STCx masses of the right-handed sleptons are by far not excluded, although their BR to lepton and LSP is 100\%. This has to be attributed to their smaller mass difference to the LSP, which leads to softer leptons and thus a significantly smaller acceptance. Among the other slepton decays listed in Table~\ref{tab:brslep},
it is interesting to note that the $\sNu_{\tau}$ features $5$\,\% of visible decays to $\sTau_1$W.

\begin{table*}[htb]
\caption{Branching ratios (BR) of the sleptons in  the models STC8 and STC10. We use \sle~when we refer to the first and second
generation sleptons, while the third generation is listed separately.\label{tab:brslep}}
\begin{center}
\begin{tabular}{l|r|r || l|r|r}
\hline \hline
{\bf Decay} &  \multicolumn{2}{c||}{\bf BR / \%} & {\bf Decay} &  \multicolumn{2}{c}{\bf BR / \%} \\ 
  & {\bf STC8} & {\bf STC10}  & & {\bf STC8} & {\bf STC10} \\ \hline
$\sler \rightarrow l \ninoone$        & 100  & 100  &   $\sTau_1 \rightarrow \tau \ninoone$         & 100  & 100     \\  \hline         
$\slel \rightarrow l \ninoone$        & 95.3 & 95.4 &   $\sTau_2 \rightarrow \tau \ninoone$         & 81.3 & 81.5    \\        
$\slel \rightarrow l \ninotwo$        & 1.7  &  1.6 &   $\sTau_2 \rightarrow \tau \ninotwo$         &  3.4 &  3.4    \\ 
$\slel \rightarrow \nu_{l} \chipmone$ & 3.0  &  3.0 &   $\sTau_2 \rightarrow \nu_{\tau} \chipmone$  &  6.2 &  6.2    \\    
                                        &      &       &   $\sTau_2 \rightarrow \stau{1} Z$            &  9.1 &  9.0    \\  \hline
$\snu{\ell} \to \nu_{\ell} \ninoone$    & 100  & 100  &   $\snu{\tau} \rightarrow \nu_{\tau} \ninoone$ & 94.2 & 94.6    \\           
                                        &      &      &   $\sNu{\tau} \rightarrow \sTau_1 W$          &  5.8 &  5.4    \\  \hline
\hline \hline
\end{tabular}
\end{center}
\end{table*}

\begin{table*}[htb]
\caption{Branching ratios (BR) of the charginos and neutralinos in the models STC8 and STC10.}
\begin{center}
\begin{tabular}{l|r|r || l|r|r}
\hline \hline
{\bf Decay} &  \multicolumn{2}{c||}{\bf BR / \%} & {\bf Decay} &  \multicolumn{2}{c}{\bf BR / \%} \\
  & {\bf STC8} & {\bf STC10} &   & {\bf STC8} & {\bf STC10} \\ \hline
$\chipone \rightarrow \sMu^+_{\rm R}\nu_{\mu}$ & 0.2 & 0.2 &        $\ninotwo \rightarrow  \sEl_{\rm R}^{\pm}$e$^{\mp}$  & 2.1 & 2.1 \\                      
$\chipone \rightarrow \sTau_1 \nu_{\tau}$ & 67.9 & 67.5 &           $\ninotwo \rightarrow  \sMuR^{\pm} \mu^{\mp}$ & 2.1 & 2.2 \\                             
$\chipone \rightarrow \sNu_{\rm e}$e$^+$ & 6.6 & 6.8 &              $\ninotwo \rightarrow  \sTau_1^{\pm} \tau^{\mp}$ & 73.2 & 72.9 \\                        
$\chipone \rightarrow \sNu_{\mu} \mu^+$ & 6.6 & 6.8  &              $\ninotwo \rightarrow  \sNu_{\rm e} \nu_{\rm e} $  & 5.7 & 5.8 \\                        
$\chipone \rightarrow \sNu_{\tau} \tau^+$ & 11.3 & 11.5 &           $\ninotwo \rightarrow  \sNu_{\mu} \nu_{\mu} $  & 5.7 & 5.8 \\                             
$\chipone \rightarrow \ninoone \Wp$ & 7.2 & 7.0 &                   $\ninotwo \rightarrow  \sNu_{\tau} \nu_{\tau} $  & 9.5 & 9.7 \\                           
     & & &                                                          $\ninotwo \rightarrow  \ninoone \Z $  & 1.2 & 1.1 \\                       
$\chiptwo \rightarrow \sEl^+_{\rm L}\nu_{\rm e}$ & 4.6 & 4.6 &      $\ninothree \rightarrow  \chipmone \W^{\mp} $  & 58.3 & 58.4 \\                                                                                                                            
$\chiptwo \rightarrow \sMu^+_{\rm L}\nu_{\mu}$ & 4.6 & 4.6 &        $\ninothree \rightarrow  \ninoone \Z $  & 10.3 & 10.2 \\                                                      
$\chiptwo \rightarrow \sTau_2 \nu_{\tau}$ & 5.1 & 5.1 &             $\ninothree \rightarrow  \ninotwo \Z $  & 23.2 & 23.2 \\                                                      
$\chiptwo \rightarrow \sNu_{\rm e}$e$^+$ & 1.7 & 1.7 &              $\ninothree \rightarrow  \ninoone \HO $  & 2.2 & 2.2 \\                                                       
$\chiptwo \rightarrow \sNu_{\mu} \mu^+$ & 1.7 & 1.7 &               $\ninothree \rightarrow  \ninotwo \HO $  & 1.2 & 1.2 \\                               
$\chiptwo \rightarrow \sNu_{\tau} \tau^+$ & 2.5 & 2.5 &             $\ninofour \rightarrow  \sTau_2^{\pm} \tau^{\mp}$ & 3.2 & 3.2 \\         
$\chiptwo \rightarrow \ninoone \Wp$ & 7.8 & 7.7    &                $\ninofour \rightarrow  \sNu_{\rm e} \nu_{\rm e} $  & 4.3 & 4.3 \\        
$\chiptwo \rightarrow \ninotwo \Wp$ & 28.3 & 28.3  &                $\ninofour \rightarrow  \sNu_{\mu} \nu_{\mu} $  & 4.3 & 4.3 \\           
$\chiptwo \rightarrow \chipone \Z$ & 25.0 & 25.0   &                $\ninofour \rightarrow  \sNu_{\tau} \nu_{\tau} $  & 4.3 & 4.3 \\        
$\chiptwo \rightarrow \ninoone \HO$ & 18.8 & 18.8  &                $\ninofour \rightarrow  \chipmone \W^{\mp} $  & 51.9 & 52.0 \\               
     & &   &                                                        $\ninofour \rightarrow  \ninoone \Z $  & 2.3 & 2.2 \\                          
     & &   &                                                        $\ninofour \rightarrow  \ninotwo \Z $  & 2.0 & 2.0 \\                        
     & &   &                                                        $\ninofour \rightarrow  \ninoone \HO $  & 6.7 & 6.7 \\                                                                                      
     & &   &                                                        $\ninofour \rightarrow  \ninotwo \HO $  & 15.8 & 15.9 \\     

\hline \hline
\end{tabular}
\end{center}
\label{tab:brewk}
\end{table*}

Table~\ref{tab:brewk} lists the decay modes of the electroweakinos. The largest branching ratios
of about $70$\,\% occur for $\ninotwo \to \sTau_1 \nu_{\tau}$ and $\chipmone \to \sTau_1 \tau$, both always 
followed by $\sTau_1 \to \tau \ninoone$. Since the $\sTau_1$ NLSP is only about $10\GeV$ more massive than
the $\ninoone$ LSP, the $\tau$ leptons from the $\sTau_1$ are typically very soft, making these
decays challenging to detect. Although the mass splitting between $\ninotwo$ ($\chipmone$) and LSP is large
enough to allow decays to on-shell Z (\Wpm) bosons, these decay modes occur only at the level of a few 
percent due to the presence of the light sleptons. The heavier neutralinos $\ninothree$ and $\ninofour$, 
however, feature sizeable branching ratios to $\chipmone \Wpm$ of about $60$\,\% and $50$\,\%, respectively,
while the $\chipmtwo$ decays most frequently to $\ninotwo \Wpm$, $\chipmone$Z and $\chipmone$h$^0$ with
branching fractions of about $30$\,\%, $25$\,\% and $20$\,\%, respectively. 
Thus, although the $\sTau_1$ is the NLSP, in particular the heavier electroweakinos have 
sizeable branching fractions to other final states
than the notoriously difficult $\tau$-lepton. This also means that signatures with 
electrons or muons in the final state can originate either from slepton or electroweakino 
production.

\subsection{Production cross-sections at the LHC}
\label{subsec:lhcxsec}

The cross sections for the main production processes of STC8 at the LHC are given in 
Table~\ref{tab:masses_STC8} for $\Ecms=14\TeV$, while Table~\ref{tab:masses_STC10} lists
the cross sections which are considerably different in STC10. All cross sections have been calculated 
with {\sc Prospino}2~\cite{Beenakker:1996ch,Beenakker:1999xh}. The particles produced with the 
largest production cross sections are the neutralinos and charginos.
Nevertheless, their observation at the LHC is challenging due to their various decay chains via the $\sTau_1$,
which lead to very soft $\tau$ leptons due to the small mass difference between the $\sTau_1$ and the LSP. 
The cross sections for direct selectron and smuon pair production is sizeable as well, such that these could be discovered
in final states with two leptons and missing transverse energy. 

\begin{table*}[htb]
\caption{Production cross sections for the benchmark model STC8 at LHC with $\Ecms=14\TeV$. The leading order (LO), 
next-to-leading order (NLO) and the $K$-factor between LO and NLO are shown.}
\begin{center}
\begin{tabular}{ l||r|r|r  }
 \hline \hline
\textbf{Process}    & \textbf{LO} \textbf{(fb)} & \textbf{NLO} \textbf{(fb)} & \textbf{$K_{\rm NLO/LO}$} \\
 \hline
pp $\rightarrow$ \gluino \gluino    & 0.18    & 0.67 &   3.58 \\
pp $\rightarrow$ \sQua \sQua  &   5.1 & 6   & 1.16 \\
pp $\rightarrow$ \sQua \gluino  & 3 & 5.4 &  1.80 \\
pp $\rightarrow$ \sQua$\bar{\sQua}$   &16.4 & 25.4&  1.54 \\\hline 

pp $\rightarrow$ $\sBot_1 \sBot_1$  &   22.8  & 38.3 & 1.67\\
pp $\rightarrow$ $\sBot_2 \sBot_2$  & 0.2  & 0.37   & 1.90 \\
pp $\rightarrow$ $\sTop_1 \sTop_1$  &  37.7  &  62.7 & 1.66 \\
pp $\rightarrow$ $\sTop_2 \sTop_2$  & 0.16  & 0.31   & 1.94\\\hline

pp $\rightarrow$ \slel \slel  & 15.5  &  19.4 & 1.25 \\
pp $\rightarrow$ \sler \sler  &  32.4 &  41.8 &  1.29 \\
pp $\rightarrow$ $\widetilde{\nu_{\ell}}$$\widetilde{\nu_{\ell}}$  &  19.4  &  24.6 & 1.26 \\
pp $\rightarrow$ \sle$\widetilde{\nu_{\ell}}$ &  63.3  &  79.6 & 1.26 \\
pp $\rightarrow$ $\sTau_1$$\sTau_1$ &  59.1  &  77.4 & 1.30 \\
pp $\rightarrow$ $\sTau_2$$\sTau_2$ &  11.7  &  14.6 & 1.24 \\
pp $\rightarrow$ $\sTau_1$$\sTau_2$ &  34.7  &  44.7 & 1.28 \\
pp $\rightarrow$ $\widetilde{\nu_{\tau}}$$\widetilde{\nu_{\tau}}$ &  20.4  &  25.9 & 1.26 \\
pp $\rightarrow$ $\widetilde{\tau}$$\widetilde{\nu_{\tau}}$ &  73.9  &  93.6 & 1.26 \\ \hline

pp $\rightarrow$ $\ninoone \ninoone$  & 0.34  & 0.42   & 1.25 \\
pp $\rightarrow$ $\widetilde{\chi}^{0}_{i}$$\widetilde{\chi}^{0}_{j}$ (except $\ninoone \ninoone$)  & 15.8  & 19.6   & 1.24 \\
pp $\rightarrow$ $\chipmone \chipmone$  &  585  &  747 & 1.28 \\
pp $\rightarrow$ $\widetilde{\chi}^{\pm}_{k}$$\widetilde{\chi}^{\pm}_{m}$ (except $\chipmone \chipmone$) &  17.3  & 20.2  & 1.17 \\
pp $\rightarrow$ $\widetilde{\chi}^{0}_{1}$$\widetilde{\chi}^{\pm}_{1}$  & 10.0  & 12.9 & 1.29 \\ 
pp $\rightarrow$ $\widetilde{\chi}^{0}_{1}$$\widetilde{\chi}^{\pm}_{2}$  & 1.07  & 1.36 & 1.27 \\ 
pp $\rightarrow$ $\widetilde{\chi}^{0}_{2}$$\widetilde{\chi}^{\pm}_{1}$  & 1170  & 1492 & 1.28 \\ 
pp $\rightarrow$ $\widetilde{\chi}^{0}_{2}$$\widetilde{\chi}^{\pm}_{2}$  & 3.51  & 4.33 & 1.23 \\ 
pp $\rightarrow$ $\widetilde{\chi}^{0}_{3}$$\widetilde{\chi}^{\pm}_{1}$  & 6.10  & 7.63 & 1.25 \\ 
pp $\rightarrow$ $\widetilde{\chi}^{0}_{3}$$\widetilde{\chi}^{\pm}_{2}$  & 22.0  & 27.0 & 1.25 \\ 
pp $\rightarrow$ $\widetilde{\chi}^{0}_{4}$$\widetilde{\chi}^{\pm}_{1}$  & 3.45  & 4.25 & 1.25 \\ 
pp $\rightarrow$ $\widetilde{\chi}^{0}_{4}$$\widetilde{\chi}^{\pm}_{2}$  & 24.4  & 29.8 & 1.25 \\ 

 \hline \hline
\end{tabular}
\end{center}
\label{tab:masses_STC8}
\end{table*}

\begin{table*}[htb]
\caption{Cross sections of the processes in the model STC10 at LHC with $\Ecms=14\TeV$.  Only 
cross sections significantly different from those of STC8 given in Table~\ref{tab:masses_STC8} are listed. The leading order (LO), 
next-to-leading order (NLO) and the $K$-factor between LO and NLO are shown.}
\begin{center}
\begin{tabular}{ l||r|r|r  }
 \hline \hline
\textbf{Process}    & \textbf{LO} \textbf{(fb)} & \textbf{NLO} \textbf{(fb)} & \textbf{$K_{\rm NLO/LO}$} \\
 \hline
pp $\rightarrow$ $\sBot_1 \sBot_1$  &   4.4  & 7.7 & 1.74\\
pp $\rightarrow$ $\sTop_1 \sTop_1$  &  7.08  &  12.0 & 1.72 \\
 \hline \hline
\end{tabular}
\end{center}
\label{tab:masses_STC10}
\end{table*}

In addition, in the direct decay $\sBot_1 \rightarrow \botq\ninoone$ leads to 
a characteristic edge in the contransverse mass distribution~\cite{mCT1,mCT2},  which depends on the masses of \sBot and \ninoone. This edge is of high interest for determining the parameters of the model and could be observed if the direct decay mode can be selected with sufficiently high purity.
In STC8, the cross section for $\sBot_1$ pair production is sizeable, which gives hope to be able to observe
the edge. Due to the larger $\sBot_1$ mass in STC10, the edge position is in a region with lower SM backgrounds,
but on the other hand the cross section is lower. Thus, observing the edge will be challenging.
Due to the rapidly decreasing cross section, we expect that bottom-squarks with masses well beyond 1\TeV cannot be seen at the LHC.

The top squark with masses as chosen here will also be produced at the LHC, but it will be difficult to distinguish
direct top-squark production from its production in the decay of the gluino (\gluino). 
The masses of the heavier coloured sparticles, the gluino and the squarks of the first and second 
generation (\sQua), are chosen such that they will also be produced at reasonably rates at the LHC, and dedicated
analyses will be able to detect them with the full luminosity delivered by the LHC.

The by far largest cross section  in electroweakino production, above  $1\;$pb, is obtained for $\ninotwo \chipmone$ production.
This channel will therefore certainly be discovered at the LHC, by a multilepton search. The cross sections for other electroweak processes is lower, the largest
cross section for neutralino-neutralino production appears for $\ninothree \ninofour$ production with almost $14\;$fb.
While the lighter electroweakinos could be well discovered at the ILC, the LHC searches would profit from the exact mass and cross section
information of these in order to specifically search for the heavier electroweakinos that would not be accessible at the ILC at a centre-of-mass
energy of 500\GeV. One example is the production of \chipmtwo, which will be discussed later.

\subsection{Production cross-sections at the ILC}
\label{subsec:ilcxsec}
\begin{figure}[htb]
  \begin{center}
\subfigure[]{\includegraphics[width=0.49\linewidth]{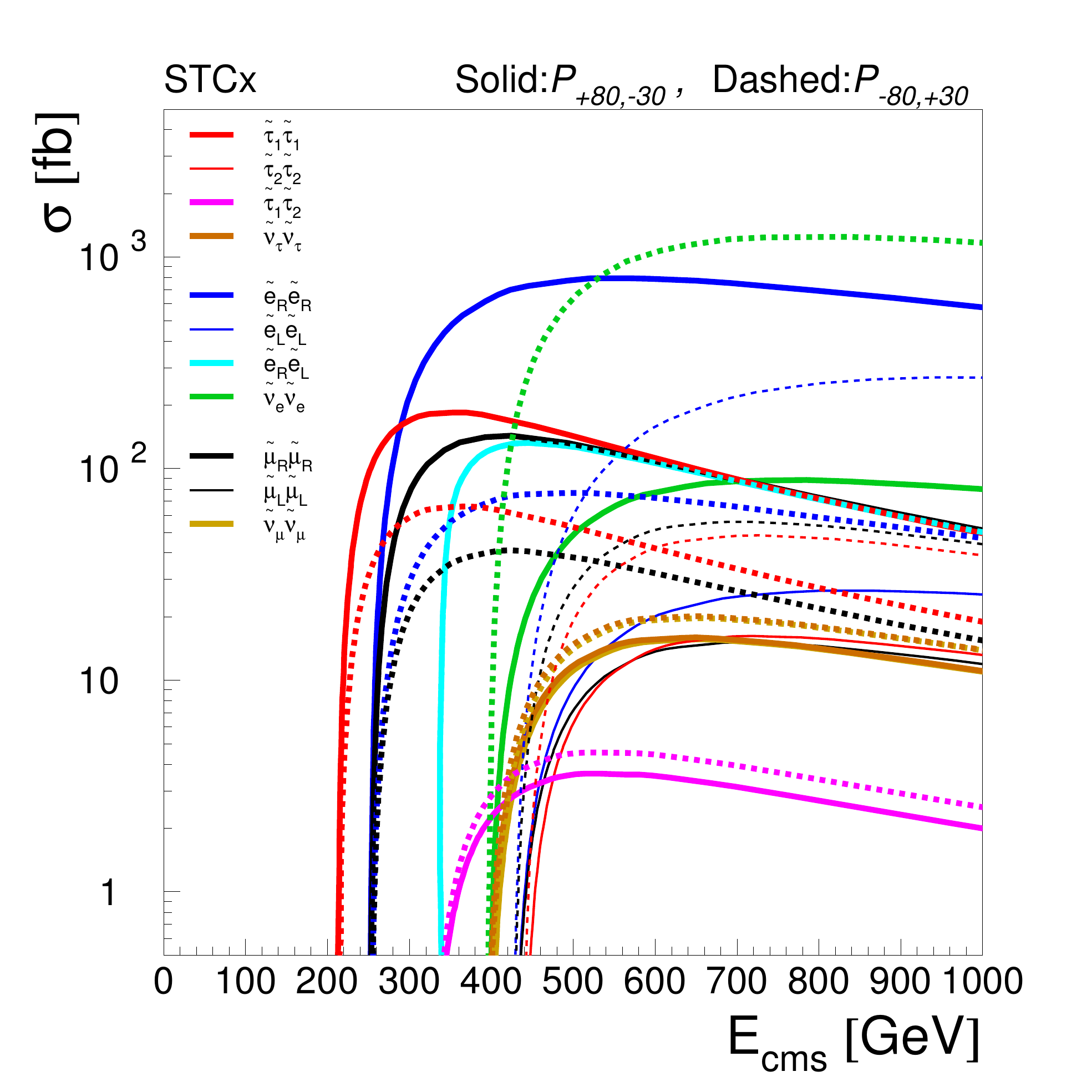}}
\subfigure[]{\includegraphics[width=0.49\linewidth]{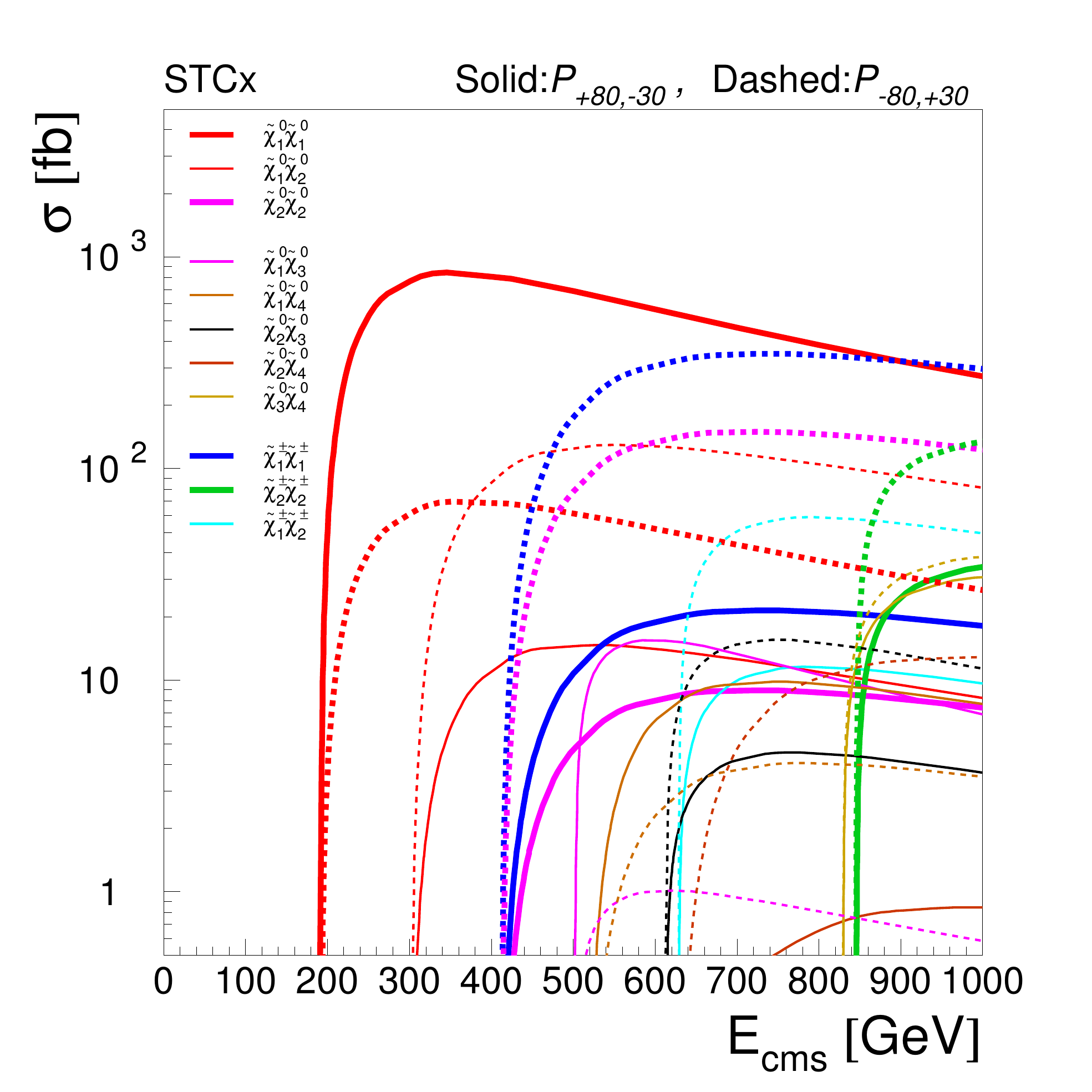}}
 \end{center}
  \caption{\label{fig:xsect-ilc} STC8 cross sections for sparticle production as a 
  function of $\Ecms$ at the ILC, separated in (a) sfermion production, and 
  (b) electroweakino production.
  }
\end{figure}

The key feature of the STCx models for the ILC is the mass spectrum of the sleptons and the lighter electroweakinos,
and thus at tree-level STC8 and STC10 are identical from the ILC point-of-view. 
Figure~\ref{fig:xsect-ilc} shows the polarised cross sections\footnote{
Here, and in the following, $\mathcal{P}_{p^-,p^+}$ denotes the beam-polarisation configuration
$\mathcal{P}(e^-,e^+)=(p^-,p^+)$, with $p^-$ and $p^+$ given in percent (for brevity
fully left(right) handed beams are denoted by L(R)).} for various STC processes in $e^+e^-$ collisions
as a function of the  centre-of-mass energy. In the part
kinematically accessible at the ILC, they do not differ among the two models. While a few processes open up already below $\Ecms=250\GeV$ and thus would be accessible even when
running near the higgs-strahlung threshold, a plethora of thresholds of slepton, sneutrino and electroweakino production appears between $\Ecms=250\GeV$ and 500\GeV. In most cases, these
can be observed and even distinguished from each other in the clean ILC environment. The ability to operate at any desired centre-of-mass
energy between 200 and 500\GeV (or even 1\TeV) and to switch the sign of the 
beam polarisations are unique tools to identify each of these processes.
The low SM background levels allow in many cases a full and unique kinematic reconstruction of cascade decays.

In particular at the ILC running at $\Ecms = 500\GeV$, 
all sleptons and the lighter set of electroweakinos of the STCx scenarios can be produced. 
\XI0{3} and \XI0{4} become accessible in associated production around
$\Ecms = 600\GeV$. Pair production of  \XIPM{2} appears  
at around $\Ecms = 850\GeV$.
At this energy, also  pair-production of \XI0{3} and \XI0{4} is possible;
however because these two states are mainly higgsino, the rate is very low.
At $\Ecms = 500\GeV$,
the cross sections are sizeable, as can be seen in Table~\ref{tab:xsect-ilc}. Only 
one of the kinematically allowed processes,
$\eeto  \stau{1}\stau{2} $, would have
a production cross section below $10\,$fb for both beam polarisation configurations.
The total SUSY cross section is well over $1\,$pb in both cases.

\begin{table*}[htb]
\caption{Production cross sections for the benchmark model STC8 at the ILC,
for different degrees of beam-polarisation.
The ILC TDR beam-spectrum is used with a
nominal centre-of-mass energy is 500\GeV.
All channels accessible at this energy are shown. Channels with no detectable final states are marked with~$(*)$. 
In addition, the cross section for $\eeto  \seR^+\seL^- (\seL^+\seR^-)$ is 335.85 fb for $\mathcal{P}_{R,R}$($\mathcal{P}_{L,L}$);
for all other processes the cross-section vanishes for both $\mathcal{P}_{L,L}$ and 
$\mathcal{P}_{R,R}$.}
\begin{center}
\begin{tabular}{ l||r|r|r|r  }
 \hline \hline
\textbf{Process}        & $\mathcal{P}_{R,L}$ \textbf{(fb)} 
                                           &  $\mathcal{P}_{L,R}$ \textbf{(fb)} 
                                                       &  $\mathcal{P}_{+80,-30}$ \textbf{(fb)} 
                                                                     &  $\mathcal{P}_{-80,+30}$ \textbf{(fb)} \\ 
\hline
$\eeto  \XN{1}\XN{1}(*)$      &    1203.57 &      34.47 &     705.29 &      62.29 \\ 
$\eeto  \XPM{1}\XMP{1}$       &       0.38 &     259.37 &       9.30 &     151.74 \\ 
$\eeto  \XN{1}\XN{2}$         &      11.97 &     206.67 &      14.24 &     121.32 \\ 
$\eeto  \XN{2}\XN{2}$         &       0.03 &     115.71 &       4.07 &      67.69 \\ 
$\eeto  \seL\seL$             &       8.01 &      84.07 &       7.63 &      49.46 \\ 
$\eeto  \seR\seR$             &    1313.73 &      52.32 &     770.36 &      76.59 \\ 
$\eeto  \sMuL\sMuL$           &       8.05 &      39.11 &       6.08 &      23.16 \\ 
$\eeto  \sMuR\sMuR $          &     222.47 &      52.23 &     131.97 &      38.34 \\ 
$\eeto  \snu{\tau}\snu{\tau}$ &      16.93 &      22.02 &      10.67 &      13.47 \\ 
$\eeto  \snu{e}\snu{e}(*)$    &      15.86 &     973.97 &      43.37 &     570.33 \\ 
$\eeto  \snu{\mu}\snu{\mu}(*)$&      15.93 &      20.71 &      10.04 &      12.67 \\ 
$\eeto  \stau{1}\stau{1} $    &     244.33 &      77.61 &     145.65 &      53.95 \\ 
$\eeto  \stau{1}\stau{2} $    &       5.46 &       7.10 &       3.44 &       4.34 \\ 
$\eeto  \stau{2}\stau{2} $    &       7.14 &      26.10 &       5.09 &      15.52 \\ 
$\eeto  \seL\seR$             &       0.00 &       0.00 &     127.62 &     127.62 \\
\hline
\hline
\end{tabular}
\end{center}
\label{tab:xsect-ilc}
\end{table*}




%% file: LHC.tex
\section{LHC projections}
\label{sec:LHC}
The searches for new physics beyond the Standard Model at the LHC are either kept as inclusive as possible, or tailored
to search for a specific scenario of new physics. We present here a representative selection of typical studies, starting
with search for and excess of large hadronic activity caused by the heavy new particles, in connection with large missing
transverse energy due to the escaping lightest SUSY particles (LSPs). Such searches have been performed by both the ATLAS
and CMS Collaborations based on data taken at 7 and 8\TeV with and without b-tagging
requirements~\cite{Aad:2011qa,Aad:2011ib,Aad:2012fqa,Aad:2012hm,Aad:2012pq,Aad:2014wea,Aad:2013wta,Aad:2014lra,RA2_2010,RA2_2011,RA2_2012,RA2b_2010,RA2b_2011,RA2b_2012} in the full-hadronic final state.
Exclusive searches for top and bottom squarks rely on the use of special variables, as the 
production cross section for third-generation sparticles is about an order of magnitude lower than the one for first- 
or second-generation squarks of similar mass and the signal is hidden below a large background by SM top-quark production.
We discuss here a full-hadronic search for direct bottom squark production, similar to searches performed by both the ATLAS
and CMS Collaborations based on data taken at 7 and 8\TeV~\cite{Aad:2011cw,Aad:2013ija,CMS-PAS-SUS-13-018}, 
where it is assumed that the pair-produced bottom squark decays directly
to a bottom quark and the lightest neutralino which is the LSP. Furthermore, we present the results of search for direct 
top squark production in the single-lepton channel, also similar to previously published analyses by CMS and ATLAS~\cite{Aad:2012xqa,Aad:2014kra,Chatrchyan:2013xna}.
Multilepton searches are sensitive to decays of the electroweak-produced sparticles, e.g.\ in case of 
neutralino-chargino production, as discussed for $7$ and $8$\,\TeV by the ATLAS and CMS collaborations.
The recent results are interpreted in the SMS approach which reflects a best case 
scenario~\cite{Khachatryan:2014qwa,Aad:2014iza,Aad:2014nua}, assuming a branching ratio of 100\,\% 
for one specific decay. In nature the 100\,\% are not realised, as also shown for our benchmark models
in Table~\ref{tab:brcol}. For such cases the bounds for the chargino and neutralinos are weaker.  We also present the results
of a multilepton search and the possible interpretation in case of a signal by the discussed model.

The detector response is simulated using the {\sc Delphes}~3.0.9 fast simulation program~\cite{deFavereau:2013fsa},
both for signal and background events. This {\sc Delphes} version has been used for the Snowmass studies and is able 
to include pileup from minimum bias collisions that are randomly selected 
from a file containing inelastic proton-proton interactions produced with 
{\sc Pythia}6~\cite{Sjostrand:2006za}. These events are randomly distributed along the beam axis (also called $z$-axis)
according to a Gaussian distribution with a width of 5\cm.
If the $z$-position of a pileup vertex is less than the 1\mm from the primary vertex 
(corresponding to the resolution), the pileup interaction is not separated 
from the primary vertex, and all particles from both the pileup and primary interactions 
are included in the object reconstruction. For pileup interactions with a larger $z$-vertex difference 
to the primary vertex, the subtraction of charged pileup particles within the tracker 
volume is applied with an efficiency of unity. The FastJet area method~\cite{Cacciari:2007fd} is applied 
to correct measurements of jets and energy in the calorimeters for the
contribution from neutral pileup particles and charged pileup particles outside the tracker acceptance.

About 10 to 100 million events per background process that were produced for the Snowmass effort~\cite{Snowmass_BG_samples} with 
{\sc Madgraph}5~\cite{Alwall:2011uj}, including up to four extra partons from initial 
and final state radiation, matched to {\sc Pythia}6 for fragmentation and hadronisation, are used in this paper. 
The background cross section is normalised 
to next-to-leading order (NLO) in the background production process, which is based on the work in preparation for the Snowmass summer
study 2013 and discussed in more detail in Ref.~\cite{Anderson:2013kxz,Avetisyan:2013onh,Avetisyan:2013dta}. 
While we studied all the major sources of background events,
background processes with low cross sections that might become
relevant at 3000\fbinv are not included. 
The signal samples are generated with {\sc Pythia}6 and passed through the {\sc Delphes} simulation.
For {\sc Pythia}6 the tune $\mathrm{Z2}^{*}$~\cite{ref:TuneZ2} is used.
The signal cross sections are calculated at NLO with {\sc Prospino}2~\cite{Beenakker:1996ch,Beenakker:1999xh}.

Assuming systematic uncertainties of the same order as in the existing 8\TeV analyses, we determine
for each search the discovery sensitivity, using the Binomial significance $Z_{\mathrm {Bi}}$~\cite{stat_1,stat_2,stat_3} in 
Roostats~\cite{Moneta:2010pm}. Here, the sensitivity is calculated in a frequentist way in one-sided 
Gaussian standard deviations, performing a hypothesis test between background-only 
and signal-plus-background, where the uncertainty on the background estimate is taken as Poisson distributed.   

\subsection{Full-hadronic search}
\input{Hadronic}

\input{Sbot}

\subsection{Search for direct top squark production in the
  single-lepton channel}

\input{Stop}

\subsection{Search in the multilepton channel}

\input{Multilepton}

%% file: Hadronic.tex
\label{sec:LHChadronic}
Heavy squark and gluino production in R-parity conserving SUSY scenarios can lead to long 
decay chains with multiple jets and therefore a large amount of hadronic energy, and 
large missing transverse momentum. A typical search for such a scenario is based on the 
variable \HT, the scalar sum of the momenta of all jets with $\pt> 50\GeV$ and $|\eta| <  2.5$, and missing hadronic
transverse energy (MHT), which is defined as absolute value of the negative
vectorial sum of all jets with $\pt> 30\GeV$ and $|\eta| <5$. The SM background to this 
SUSY search arises mainly from the following processes: Z($\nu\nu$) + jets events, and W($l\nu$) + jets 
events from W, or \ttbar + jets, where at least one W boson decays leptonically. 
The W($l\nu$) + jets events pass the search selection when the e/$\mu$ escapes detection or 
when a top decays hadronically. QCD multijet events also contribute to the background when 
jet-energy mismeasurements or leptonic decays of heavy-flavour hadrons inside jets produce large MHT. 
However, the QCD background generally becomes negligible at very high MHT as required here.

The analysis follows the baseline selections motivated by the 8\TeV analysis~\cite{RA2_2012}, 
requiring at least three jets with $\pt >50\GeV$ and $|\eta|< 2.5$, $\HT>1000\GeV$ and MHT $> 500\GeV$. 
In order to remove the QCD background, events are required to satisfy the following cuts for the azimuthal angle 
difference between the leading jets and MHT direction: $|\Delta \phi (j_n$, MHT$)| > 0.5$ for $n = 1, 2$ and   
$|\Delta \phi (j_3$, MHT$)| > 0.3$. Events are vetoed when they contain isolated muons satisfying 
$\pt > 10\GeV$ and $|\eta| < 2.4$ or electrons with $\pt > 10\GeV$ and $|\eta| <2.5$, which suppresses mainly the 
\ttbar and W(l$\nu$) + jets background. The baseline cut flow is given in the Appendix, in Table~\ref{tab:had_app}.

\begin{table} [!ht]
\caption{LHC all-hadronic inclusive search: Background and signal
  event yields corresponding to 300\fbinv. The notation "V" refers to W,
  Z and $\gamma$. Four signal regions (SR) are shown here.
} 
\centering 
\begin{tabular}{l | r r r r | r | r r} 
\hline \hline

\textbf{Signal regions} & \textbf{\ttbar+jets} & \textbf{V+jets} & \textbf{VV+jets} & \textbf{Top+jets} & \textbf{Total SM} &\textbf{STC8} & \textbf{STC10} \\ 
\hline
\textbf{SR A:} & & & & & & & \\
\HT $>$ 2000 GeV;& 0.29 & 43 & 2.1 & 0.01 & 45 & 45 & 35 \\
MHT $>$ 1500 GeV & & & & & & & \\ \hline
\textbf{SR B:}  & & & & & & & \\
\HT $>$ 2000 GeV;  & \multirow{2}{*}{4.4} & \multirow{2}{*}{14} & \multirow{2}{*}{0.92} & \multirow{2}{*}{0.15} & \multirow{2}{*}{20} & \multirow{2}{*}{51} & \multirow{2}{*}{35} \\
MHT $>$ 1000 GeV; & & & & & & &  \\ 
n(Bjets) $\ge$ 2  & & & & & & & \\ \hline
\textbf{SR C:}  & & & & & & & \\
\HT $>$ 3500 GeV; & 0.49 & 9.5 & 0.54 & 0.01 &  10 & 8.4 & 7.6 \\
MHT $>$ 1000 GeV  & & & & & & & \\ \hline
\textbf{SR D:}  & & & & & & & \\
\HT $>$ 3000 GeV; & 0.1 & 6.3 & 0.41 & 0.002 & 6.8 & 8.2 & 7.2 \\ 
MHT $>$ 1500 GeV & & & & & & & \\
\hline \hline
\end{tabular}
\label{tab:had} 
\end{table}

\begin{figure}[htb!]
\centering
\subfigure[]{\includegraphics[width=0.44\linewidth]{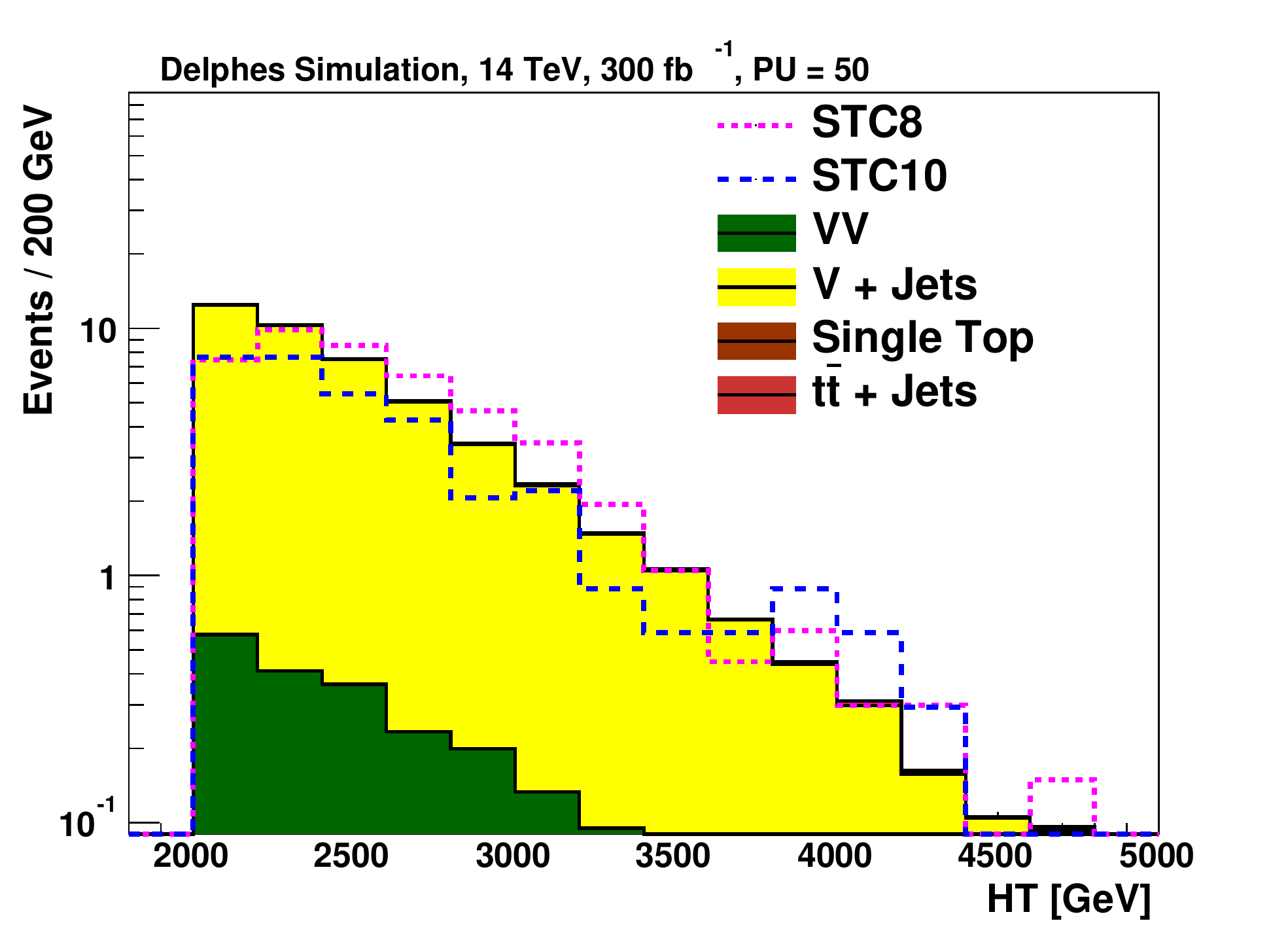} }
\subfigure[]{\includegraphics[width=0.44\linewidth]{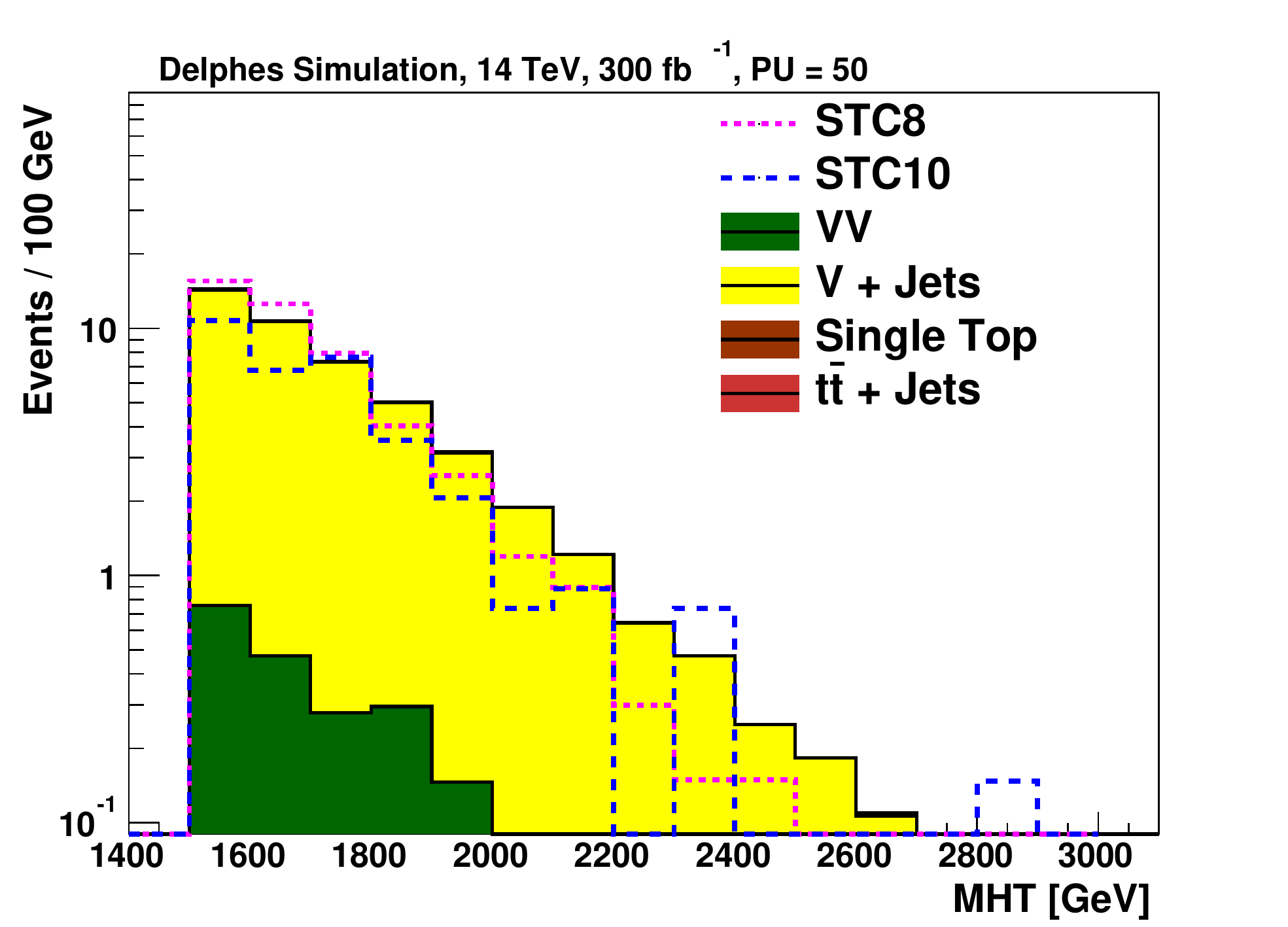} }\\
\subfigure[]{\includegraphics[width=0.44\linewidth]{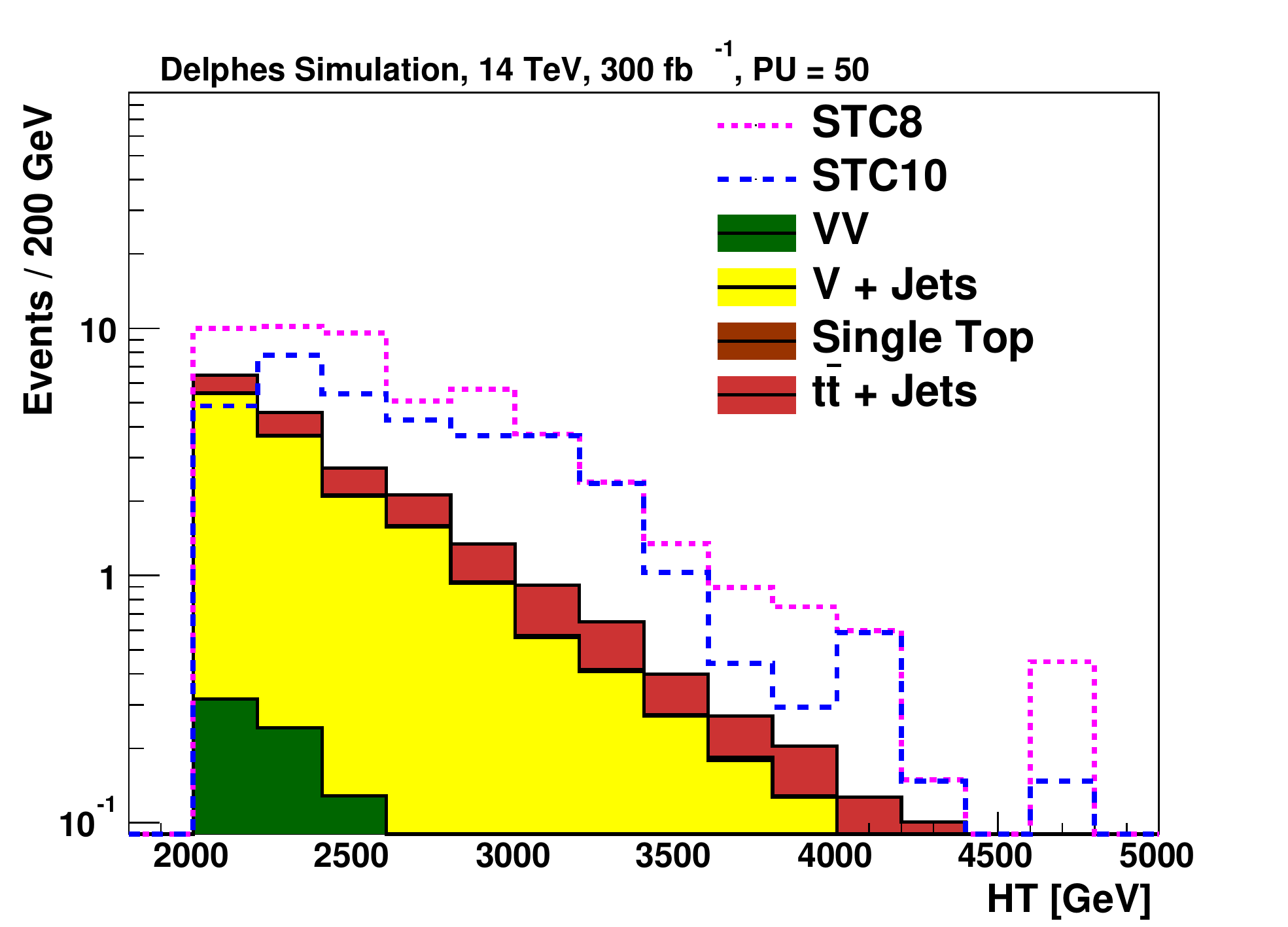} }
\subfigure[]{\includegraphics[width=0.44\linewidth]{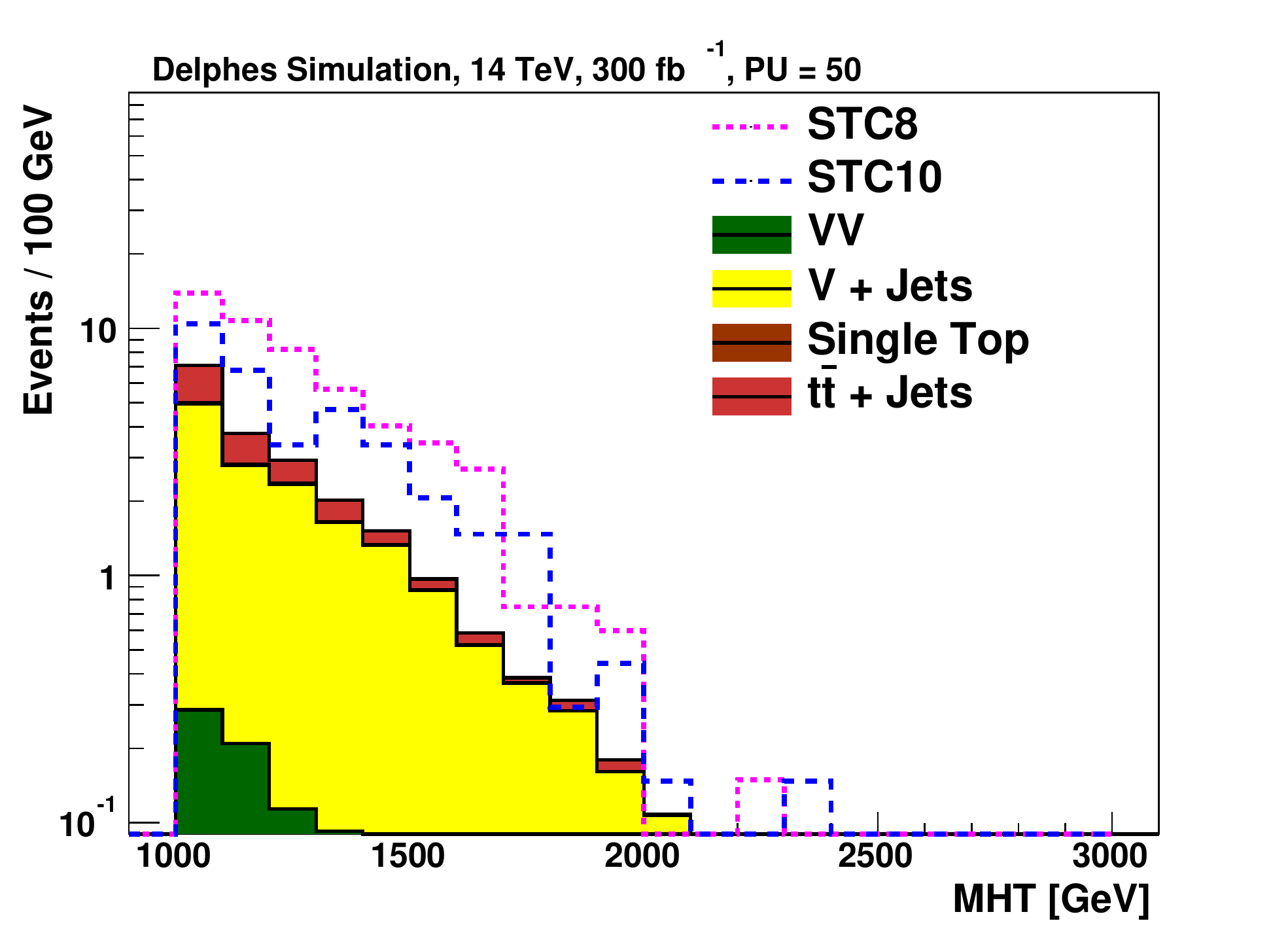} }
\caption{LHC all-hadronic inclusive search: \HT and MHT for signals and SM backgrounds after all
  selection requirements are applied. (a),(b): inclusive search region SR~A; 
  (c), (d) signal region SR~B including the requirement of at least two b-tagged jets.}. 
\label{fig:had_plots}
\end{figure}

The recent searches by the CMS Collaboration split the events from the baseline selection into several 
exclusive search regions according to their \HT, MHT and b-tag multiplicity, while for this study we keep
only four well-motivated signal regions listed in Table~\ref{tab:had}. The most promising signal region targeting the inclusive production 
of heavy gluinos and squarks of all generations is SR~A, where we require $\HT > 2000\GeV$ and MHT $>1500\GeV$. 
Figure~\ref{fig:had_plots}(a) and (b) shows the \HT and MHT distribution for this selection.
Two other signal regions, SR~C and SR~D, are characterised by 
higher \HT requirements and reject too much signal, which could be taken as a hint on the (not too high) squark or gluino mass.
In SR~A we find 45 (35) signal events for STC8 (STC10) over 45 background events for an integrated luminosity of 300\fbinv.
About 80\,\% of these signal events are first- and second-generation squarks. Assuming a systematic uncertainty of 20\,\%, we determine
a discovery sensitivity of about 3$\,\sigma$.   
 
Another signal region, SR~B, defined by $\HT > 2000\GeV$, 
MHT $>$ 1000\GeV and $N_{\rm b-tags} \geq 2$, is tailored to SUSY signals with a light third generation, being 
sensitive to either gluinos decaying through (virtual) top or bottom squarks or to directly produced third-generation squarks.
The results for this signal region are shown in Fig.~\ref{fig:had_plots}(c) and (d).
We find 51 (35) signal events in STC8 (STC10), expecting 20 background events for 300\fbinv. A discovery sensitivity of 5$\,\sigma$ will
be reached with 200\fbinv, assuming a systematic uncertainty of 20\,\%. The signal consists mainly of 
gluino-associated (\gluino\gluino and \gluino\squark) and direct heavy-squark production (about~$65\,\%$ of the final events). 

Further studies of the kinematic variables may shed more light 
on the nature of the new physics seen in this scenario and are discussed in the following sections.

%% file: Sbot.tex
\subsection{Search for direct bottom squark production in the
  final states with two b-quark and missing energy}
\label{subsec:LHCsbot}

In this section, we investigate the discovery potential for third-generation squarks in the
final state with two b-jets and missing energy in the LHC.
Of particular interest for this search is the decay $\sBot_1 \rightarrow$ b\ninoone, which is the dominant
decay mode of bottom squark in the investigated models, with branching fractions of more than 50\,\%.
Assuming that the bottom squarks are pair-produced, a final state containing exactly
two b-quarks and two neutralinos is expected for about 25\,\% of the signal events.
As the masses of the \chipmone and the \ninoone are not degenerate, the contribution of the top squark to
this final state is small the STCx models.

\begin{figure}[!htb]
  \begin{center}
  \subfigure[]{\includegraphics[width=0.49\linewidth]{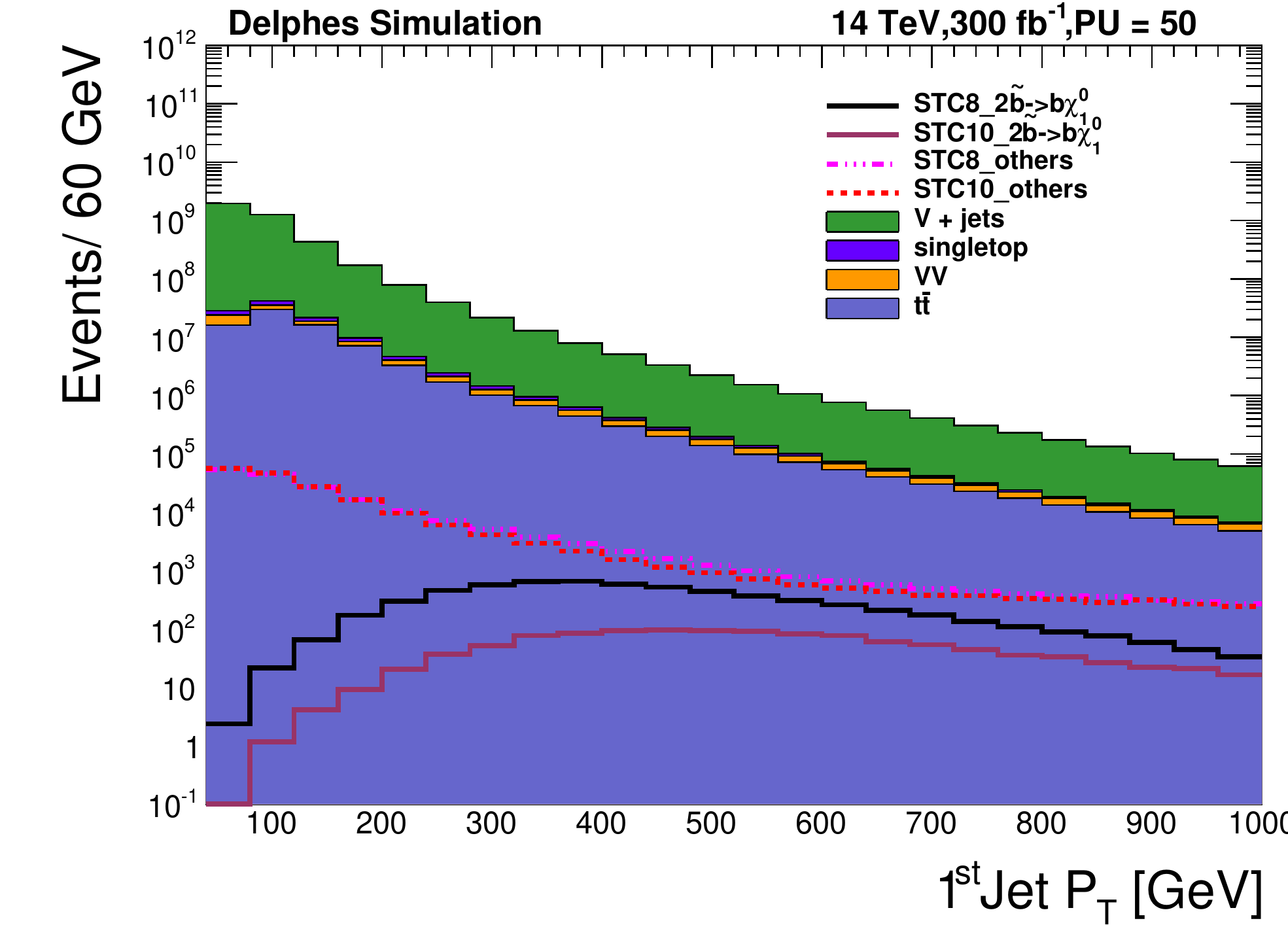}\label{fig:sbottom:a}}
  \subfigure[]{\includegraphics[width=0.49\linewidth]{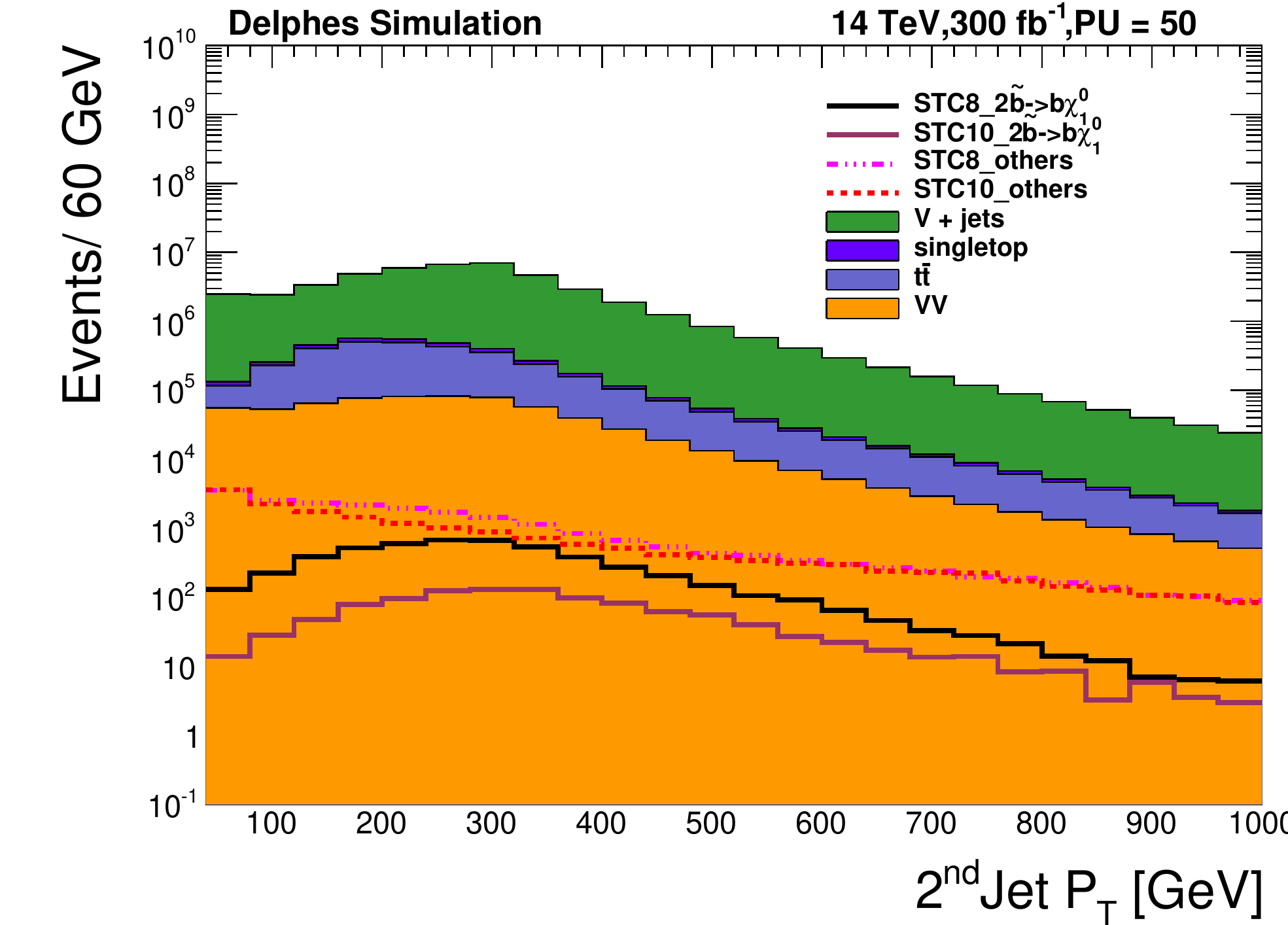}\label{fig:sbottom:b}}
  \subfigure[]{\includegraphics[width=0.49\linewidth]{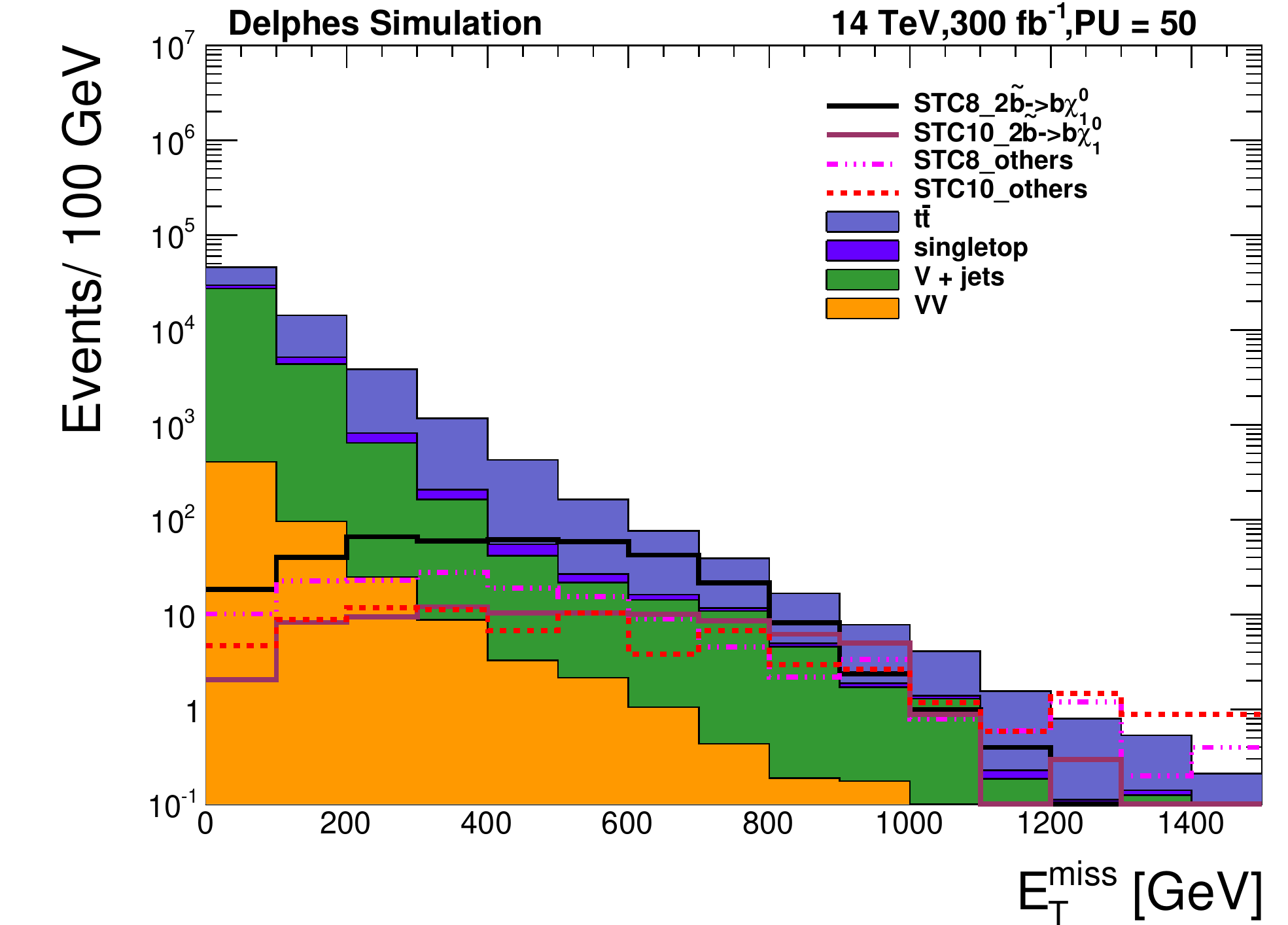}\label{fig:sbottom:c}}
  \subfigure[]{\includegraphics[width=0.49\linewidth]{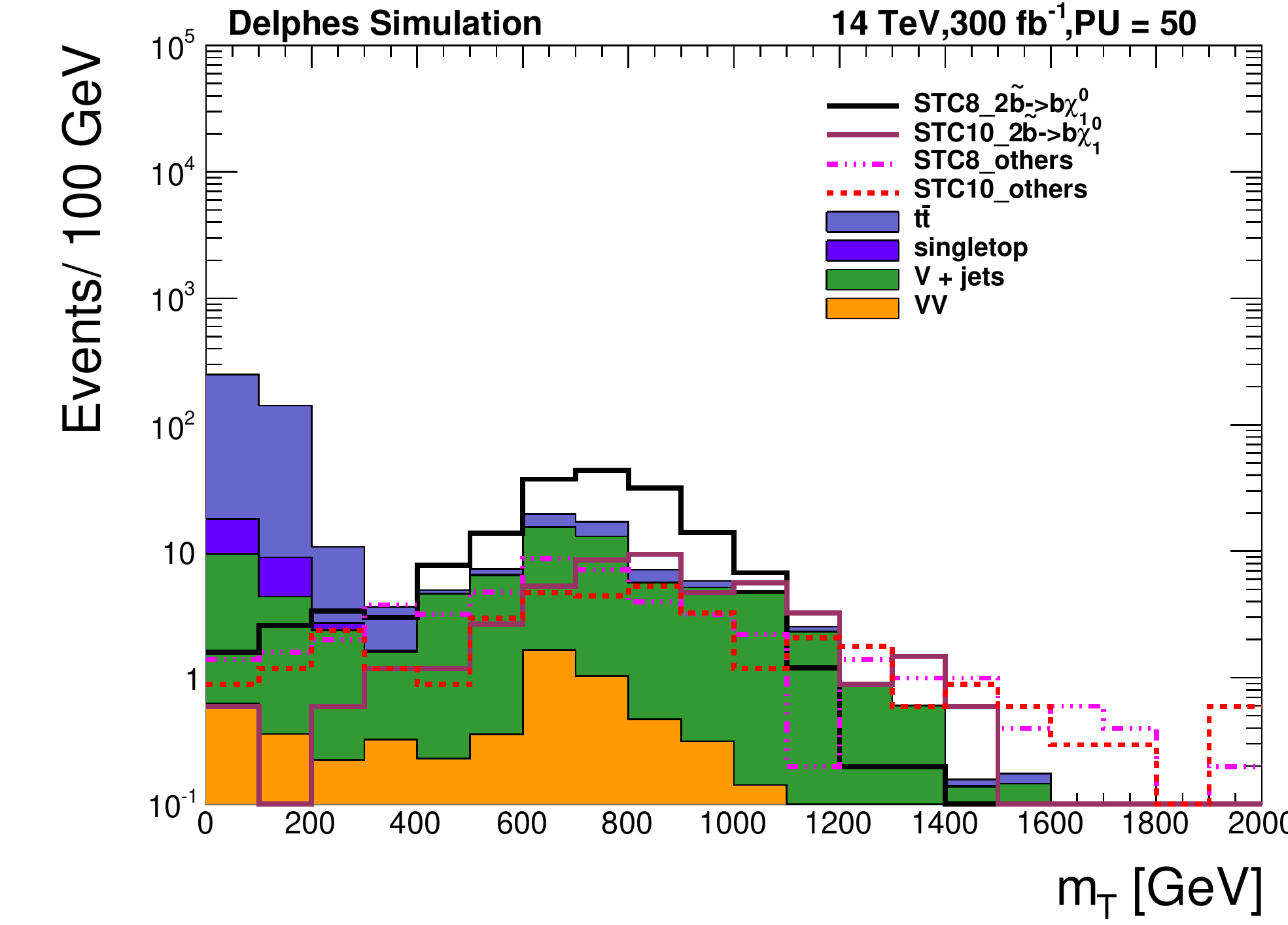}\label{fig:sbottom:d}}
  \end{center}
  \caption{\label{fig:sbottom} LHC bottom squark search: (a) \pt of the leading jet for events with exactly two central jets;
  (b) \pt of the next-to-leading jet after applying the leading jet \pt requirement; (c) \ETm after requiring for exactly two b-tagged jets; (d) \MT after requiring for exactly two b-tagged jets and \ETm$>$ 450\GeV. All signal events where at least one bottom squark do not decay to $\sBot_1 \rightarrow$ b\ninoone, plus all other SUSY processes are classified as ``STC other''.}
\end{figure}

Two jets which originate directly from bottom squark production should
exhibit sizeable transverse momentum. Figure~\ref{fig:sbottom:a} and (b) display the \pt of the two leading
jets for all the standard model processes and the signal. When the bottom squark decays directly to a b quark and a neutralino,
the resulting b-jet \pt is on average harder than for SM and other SUSY processes.
Therefore, we require the events to contain exactly two central jets with
$|\eta| < 2.4$ and $\pt > 300$ and 200\GeV, respectively. Events including any additional jet 
with $\pt > 70\GeV$ are rejected. In addition, both leading jets are required to be identified
as originating from a bottom quark. 

By vetoing events with at least one lepton with $\pt > 10\GeV$, we
suppress the main standard model background processes such as \ttbar and \Wjets. 
The distribution of the missing energy for the events passing the above requirement is shown in Fig.~\ref{fig:sbottom:c}. 
We require missing transverse energy to be $\MET>450\GeV$.

The \ttbar and \Wjets backgrounds are suppress further by requiring the minimum invariant transverse mass \mT of one of the two leading jets (jet$_i$) and the missing transverse momentum to be greater than 500\GeV. 
The invariant transverse mass is 
a quantity that is often used in LHC analyses, and it is defined as function of the transverse momenta of two objects $a$ and $b$ as follows:
\begin{eqnarray}
\displaystyle \MT(\pt^a,\pt^b) = \sqrt{2 \pt^a \pt^b \left(1-\cos(\Delta\phi(\pt^a,\pt^b))\right)}\;\; . \label{eqn:mt}
\end{eqnarray}
The distribution of $\min \left ( \MT(\pt^{\mathrm{jet}_1}, \ETm), \MT(\pt^{\mathrm{jet}_2},\ETm) \right )$ as shown in Fig.~\ref{fig:sbottom:d} is expected to have a kinematic edge at the mass
of the top quark when the jet and \ETm originate from semileptonic decay of a top quark. A cutflow is shown in the Appendix, in Table~\ref{tab:baselineEventYields}.

With an integrated luminosity of 300\fbinv, we find 74 (44) signal events in STC8 (STC10) for a background yield of 28 events.
73\,\% of the selected signal events in the STC8 scenario are from direct bottom squark production with the desired decay, $\sBot_1\rightarrow\botq\ninoone$.
In STC10, this number reduces to 58\,\% due to higher contamination from decays to \botq quarks and \ninotwo, \ninothree, or \ninofour.
In both scenarios, we observe about 5\,\% SUSY background due to $\sTop_1$ decays.
The other bottom squark decay modes and all events originating from top squark decays are 
treated as SUSY background in this analysis. Assuming a systematic uncertainty of 15\,\%, this channel reaches a discovery sensitivity of 5$\,\sigma$ with about 90 and 300\fbinv of data for STC8 and STC10, respectively.

%

%% file: Stop.tex
\label{sec:LHCstop}

In this section, we discuss the search for direct top squark pair production. 
Previously conducted searches by the CMS collaboration~\cite{Chatrchyan:2013xna} focus on simplified SUSY models, where
only the process of interest is considered while all other sparticles masses are assumed to be out of reach.
In these simplified models two top squarks are produced, which decay into either t\ninoone or b\chipmone, 
with varying branching ratios.
The two STC models considered here, however, only a very small fraction of events where two top squarks
are produced exhibit the sought after decay process. As an example, in the STC8
model only 1.6\% of $\sTop_1\sTop_1^*$ events are expected to have a decay mode
$\sTop_1\sTop_1^*\to(\topq\ninoone)(\bar{\topq}\ninoone)$.

Additionally, bottom squark pair production
enters as a sizeable intrinsic SUSY background when one of the bottom squarks decays to b\ninoone and the other one decays to t\chipmone. Gluino and squark production also enter as an intrinsic background with top quarks in the decay chain resulting in similar signatures as expected for process of interest.
Thus, we face two major challenges in this analysis, one being large SM background and the other one being background from
other SUSY processes.

The analysis method follows the aforementioned search performed by the CMS collaboration at
8\TeV~\cite{Chatrchyan:2013xna}, but with tighter selection requirements.
We require a single isolated electron or muon with $\pt > 30\GeV$ and $|\eta| < 2.4$. 
Events are vetoed if there are additional 
isolated leptons with $\pt > 20\GeV$. In addition, we
require at least five jets with $\pt > 40\GeV$ and $|\eta| < 2.4$,
which enhances the fraction of $\sTop_1\sTop_1^*$ events with respect to $\sBot_1\sBot_1^*$ events. 
One or two of these jets must satisfy at least a medium b-tag requirement.
To further reduce the SM background, we require $\MET > 400$\GeV.

Additionally, we introduce an angular variable $\min\Delta \phi$, the minimum azimuthal angle between the leading or sub-leading jet and 
the \MET. For this variable we require events to have a value greater than 0.8 in order to reduce backgrounds from SM 
processes. Another variable that aids in reducing backgrounds is centrality, defined as the sum of the \pt of the lepton and 
jets divided by their total momentum $\frac{\sum_i{{\rm jet}_i(\pt)}+{\rm lepton}(\pt)}{\sum_i{{\rm jet}_i(p)}+{\rm lepton}(p)}$. For SUSY events we 
expect this variable to be shifted towards higher values, while SM backgrounds are less central. Events are selected that satisfy centrality $>$ 0.6.

\begin{figure}[htb!]
\centering
\subfigure[]{\includegraphics[width=0.44\linewidth]{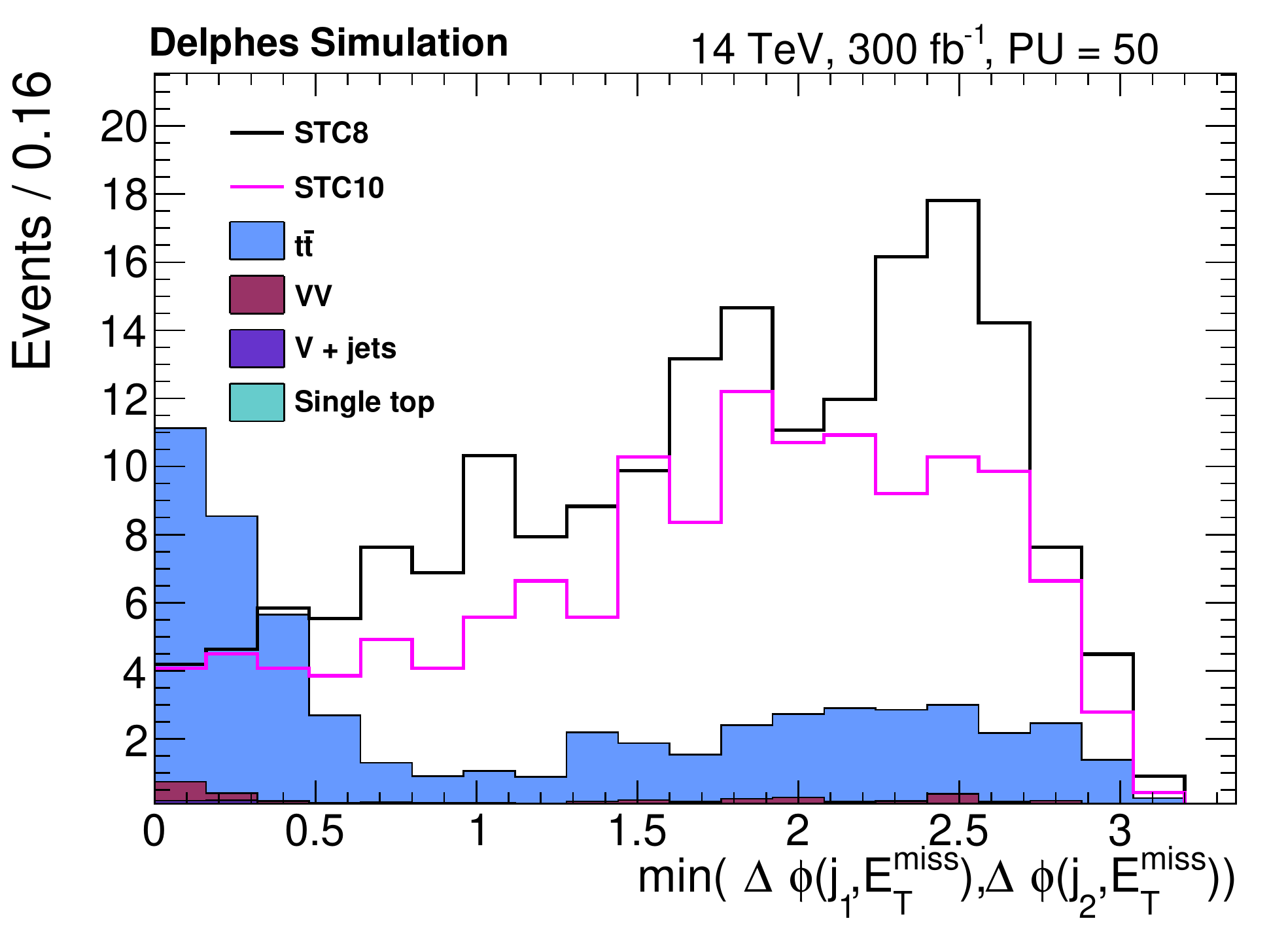} }
\subfigure[]{\includegraphics[width=0.44\linewidth]{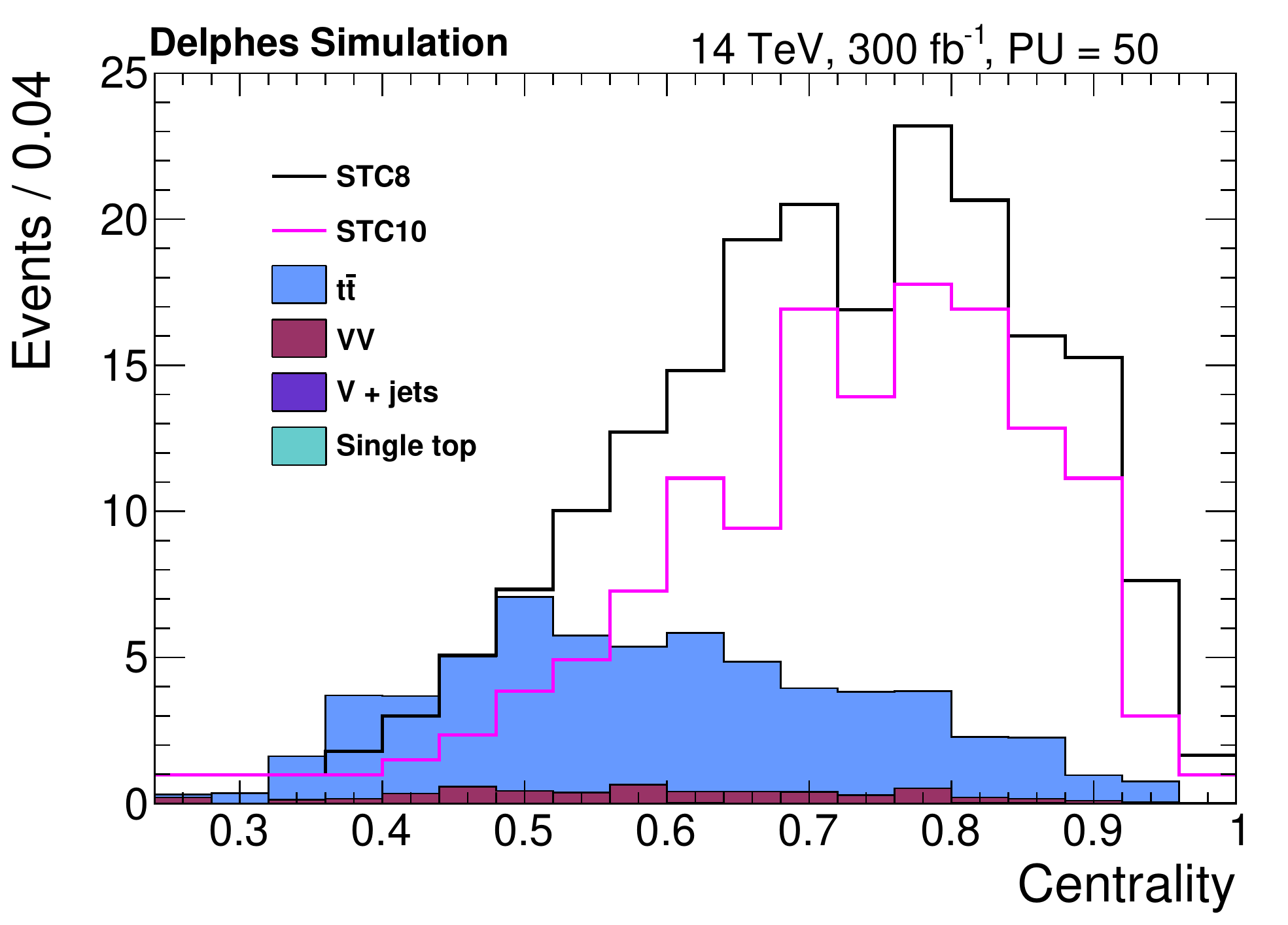} }\\
\subfigure[]{\includegraphics[width=0.44\linewidth]{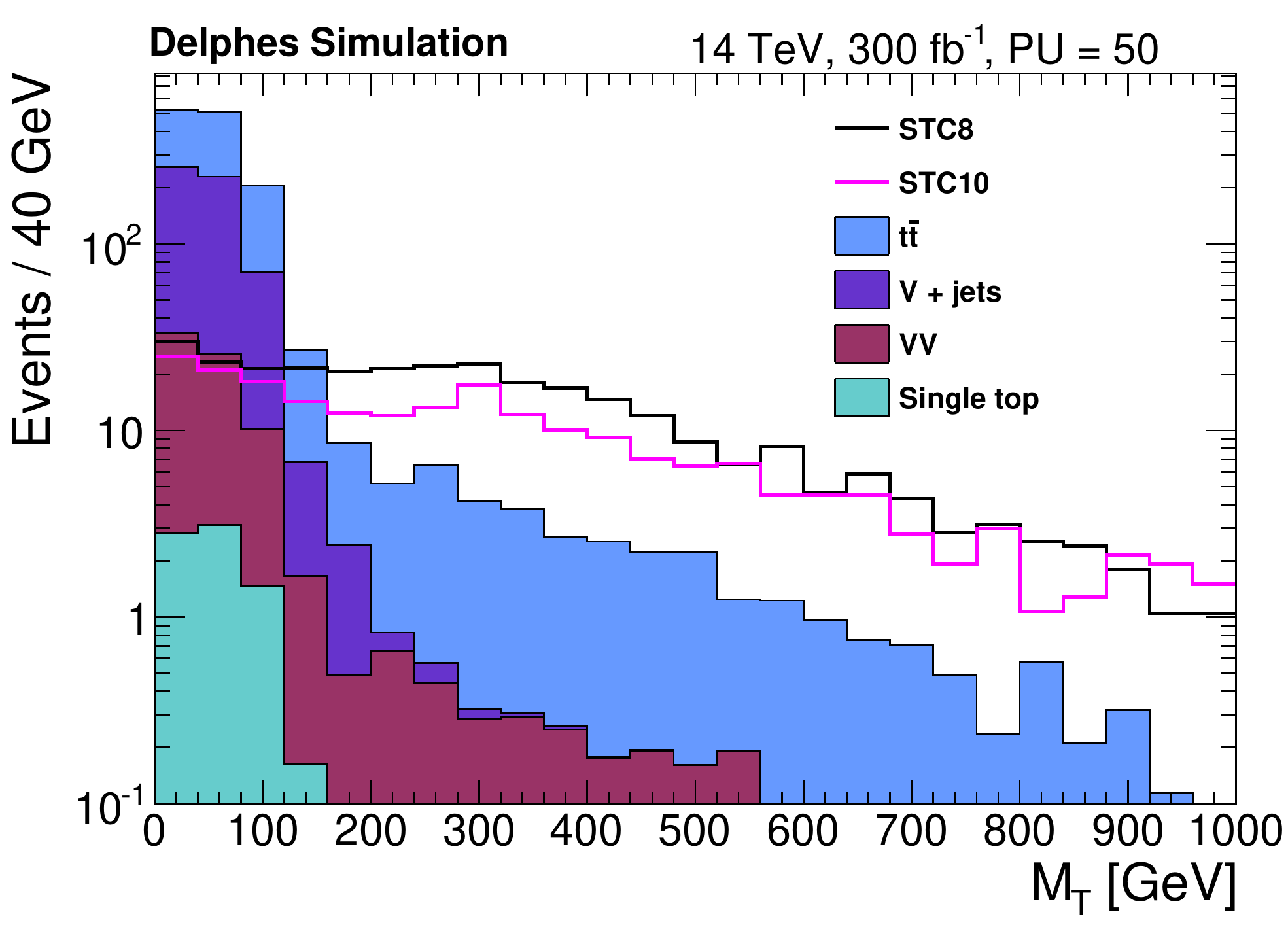} }
\subfigure[]{\includegraphics[width=0.44\linewidth]{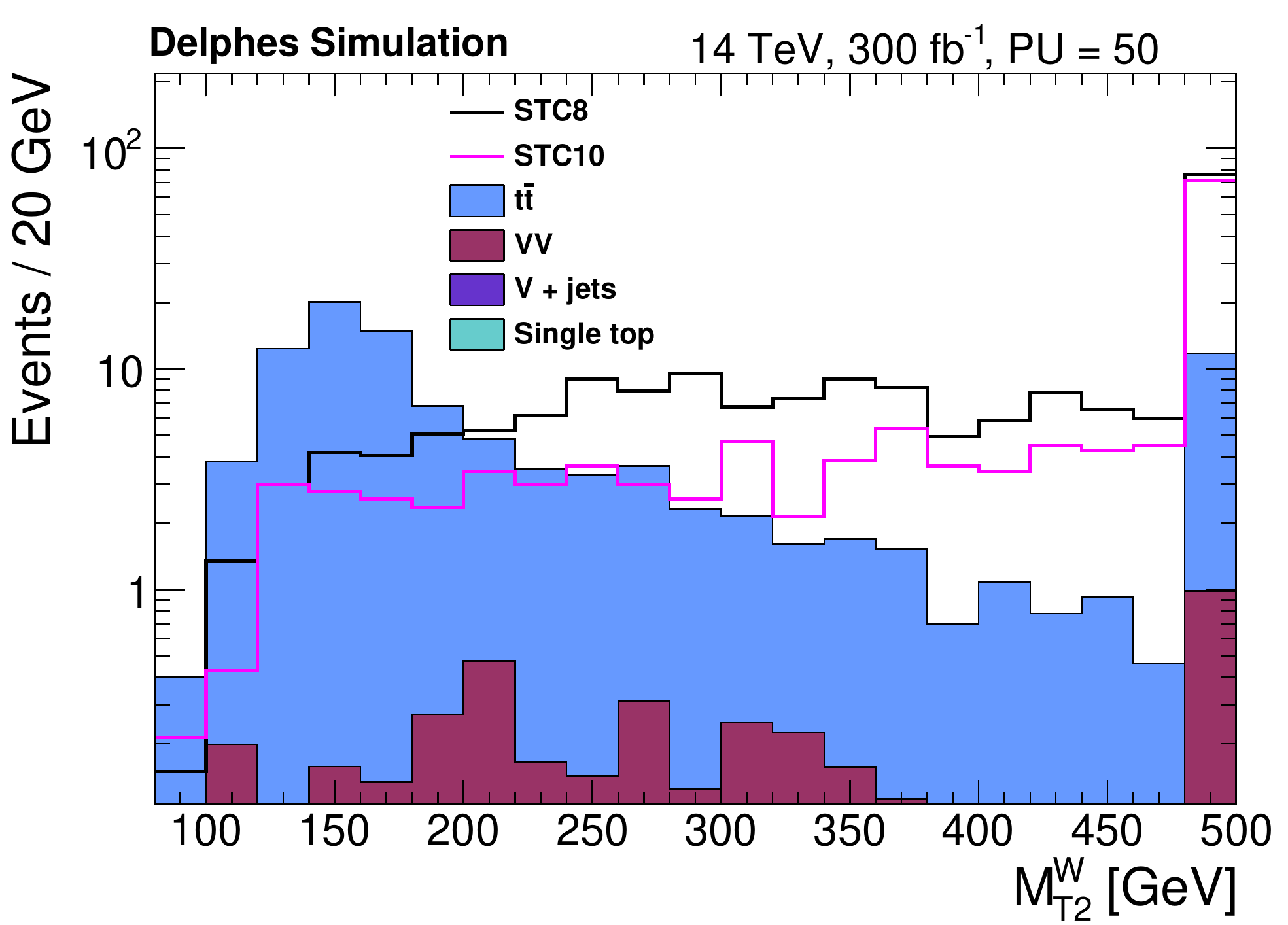} }
\caption{LHC top squark search: Comparison of (a) $\Delta \phi$, (b) centrality, (c) \mT, 
and (d) \MTtW for signals and SM backgrounds after all
  selection requirements are applied except on the variables themselves. The last bin in 
in the variable \MTtW holds all events where the minimisation, which is performed during the 
calculation of the variable, does not converge within a reasonable amount of steps. 
}
\label{fig:stop_plots}
\end{figure}

After requiring the transverse mass, $\mT$, calculated with Equation~\ref{eqn:mt} for the system consisting of the
lepton \pt and the missing transverse momentum vector, to satisfy $\mT > 260\GeV$, the 
background arises predominantly from two sources: \ttbar events in which both W bosons decay
leptonically but one lepton is lost, and diboson events. 
In order to suppress the \ttbar background, we require $\MTtW$, defined as the
minimum ``mother'' particle mass compatible with all the transverse momenta and mass-shell
constraints~\cite{MT2W}, to be above 260\GeV. By construction, for the dilepton \ttbar background without mismeasurement effects,
$\MTtW$ has an endpoint at the top quark mass, while for semi-leptonic \ttbar events and signal it has a large tail. 
Figure~\ref{fig:stop_plots} shows the $\Delta \phi$, centrality, $\mT$, and the $\MTtW$ distributions after all previously 
mentioned selection criteria are applied, except on the variables themselves. A cutflow can be found in the Appendix, in Table~\ref{tab:cutflow_stop}.

After all selection requirements, we find for 300\fbinv 155 (STC8) and 113 (STC10) signal events with a very low background
of 28 events. Assuming a systematic uncertainty of 15\,\%, the STC8
model could be discovered with 5\,$\sigma$ with an integrated
luminosity of 25\fbinv, and STC10 with 40\fbinv. While the search targets direct top squark pair production,
the final composition of the selected events contains only about 42\% of this type for STC8 and 23\% for STC10.
The additional events originate from squark-gluino production and bottom-squark pair production.

%% file: Multilepton.tex
In the STCx models, the cross section for direct \ninotwo\chipmone production in proton-proton collisions amounts to more than 1~pb. The golden channel for the discovery
of this process is the multilepton channel, with final states containing three 
or more prompt leptons. The leptons for the signal are produced by slepton mediated decays or by 
leptonic decays of W or Z bosons which are produced in the decay chain. 
However, in the STC8 model the $\sTau_1$ is the NLSP and almost mass degenerated with \ninoone, leading to
soft leptons in the dominant decay modes of \XN2 and \chipmone decays via $\sTau_1$ (c.f.\ Table~\ref{tab:brewk}). Most of these leptons will not pass the selection criteria, and thus we expect only a small sensitivity to these decay modes.

In the following we define lepton to be either an isolated muon or an isolated electron, which 
includes the leptonic decay modes of the $\tau$ leptons. As hadronic decays of $\tau$ leptons are not optimally modelled
in the used {\sc Delphes} version, we do not consider these here. All leptons must have a transverse momentum $\pt> 10\GeV$. 
In order to comply with the trigger, the leading (sub-leading) lepton must satisfy $\pt> 25\GeV$ (15\GeV). 
We present here the analysis of the three-lepton final state requiring exactly three leptons. The analysis of events with four or more leptons does not increase the sensitivity and is not discussed further.

The WZ and ZZ production are summarised as VV background. 
Non-prompt backgrounds cover all events which include leptons from misidentified objects 
(also known as 'fake' sources), or leptons which are not produced by the hard scattering. 
For example, dileptonic decays of \ttbar, where one of the b-jets produces an isolated lepton, 
leads to final states with three leptons. All other processes which contribute to the three- or four-lepton 
final states, e.g.\ SM higgs-boson production or triple-boson production, are summarised as rare backgrounds. 
The signal is subdivided into four production mechanisms, the direct chargino-neutralino production of  
the second-lowest mass electroweakinos \XN2\chipmone, and the production of at least one higher-mass chargino, \chipmtwo,
summarised as \XPM2\Cnm. Other direct production modes of charginos and/or neutralinos are comprised in 'other EWK',
while we label all events from other sources than electroweakino production, such as 
slepton-pair production or production of coloured SUSY particles, as 'noEWK'.

\begin{figure}[htb]
\centering
\subfigure[]{\includegraphics[width=0.44\linewidth]{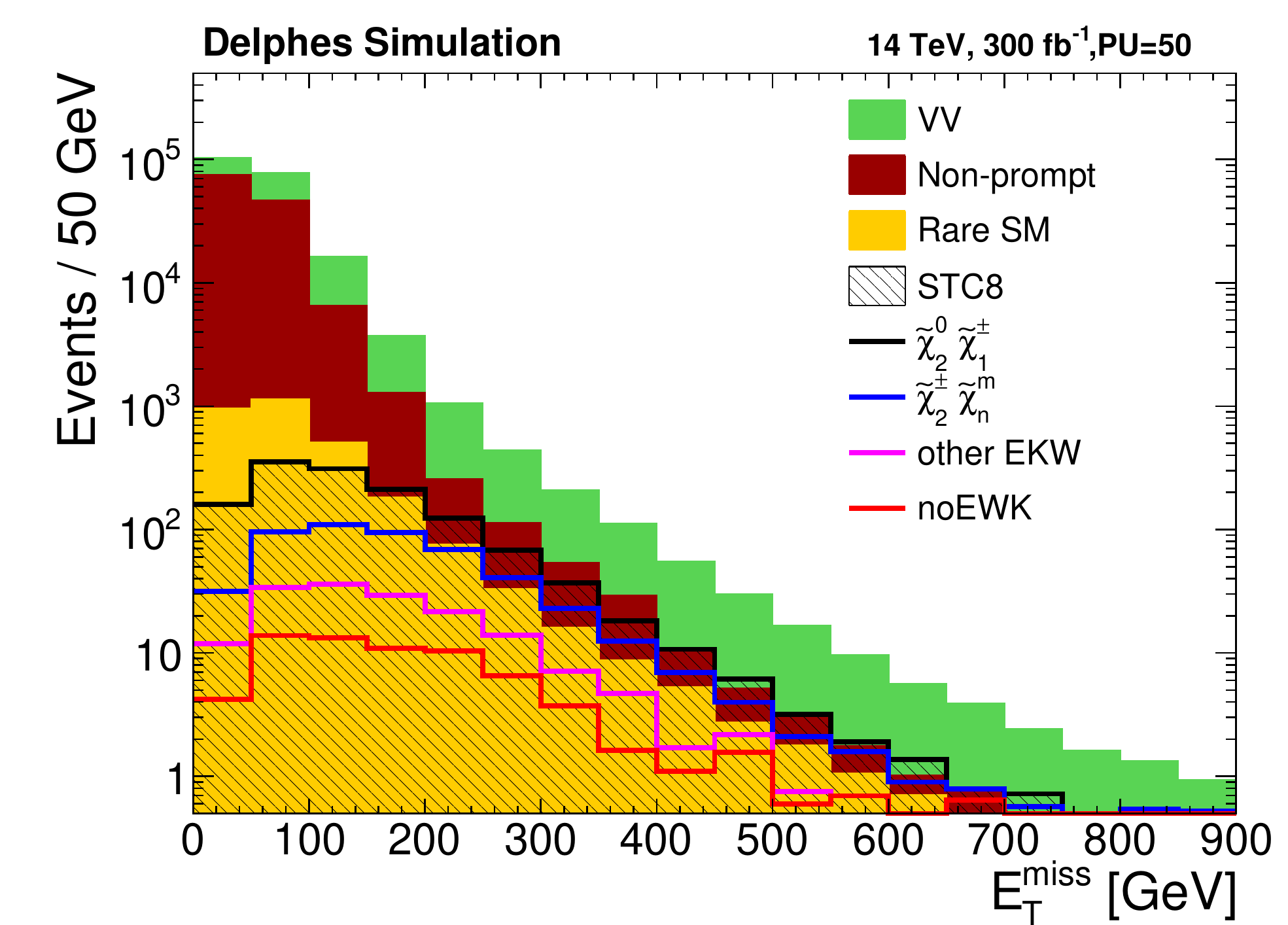} }
\subfigure[]{\includegraphics[width=0.44\linewidth]{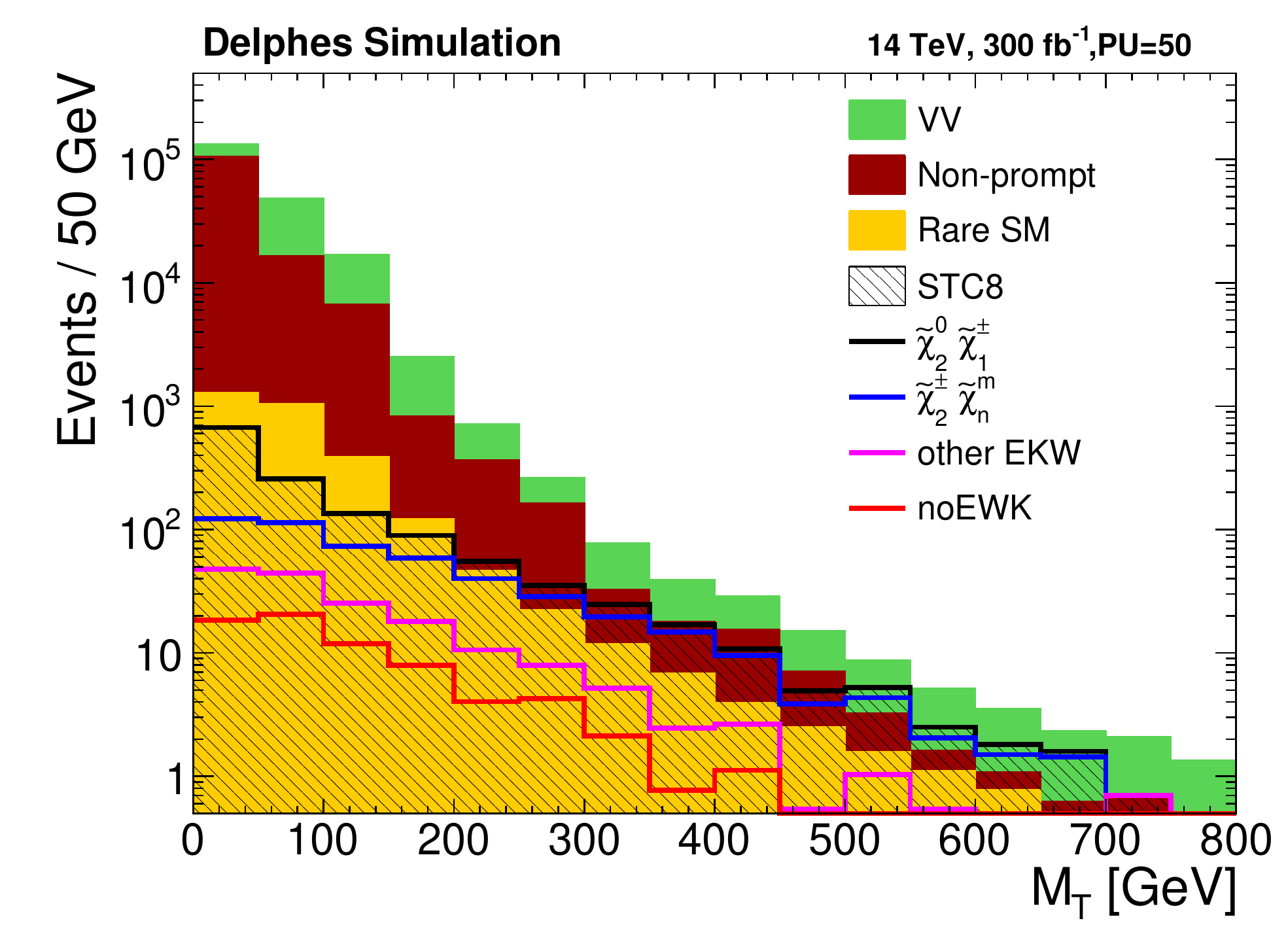} }\\
\subfigure[]{\includegraphics[width=0.44\linewidth]{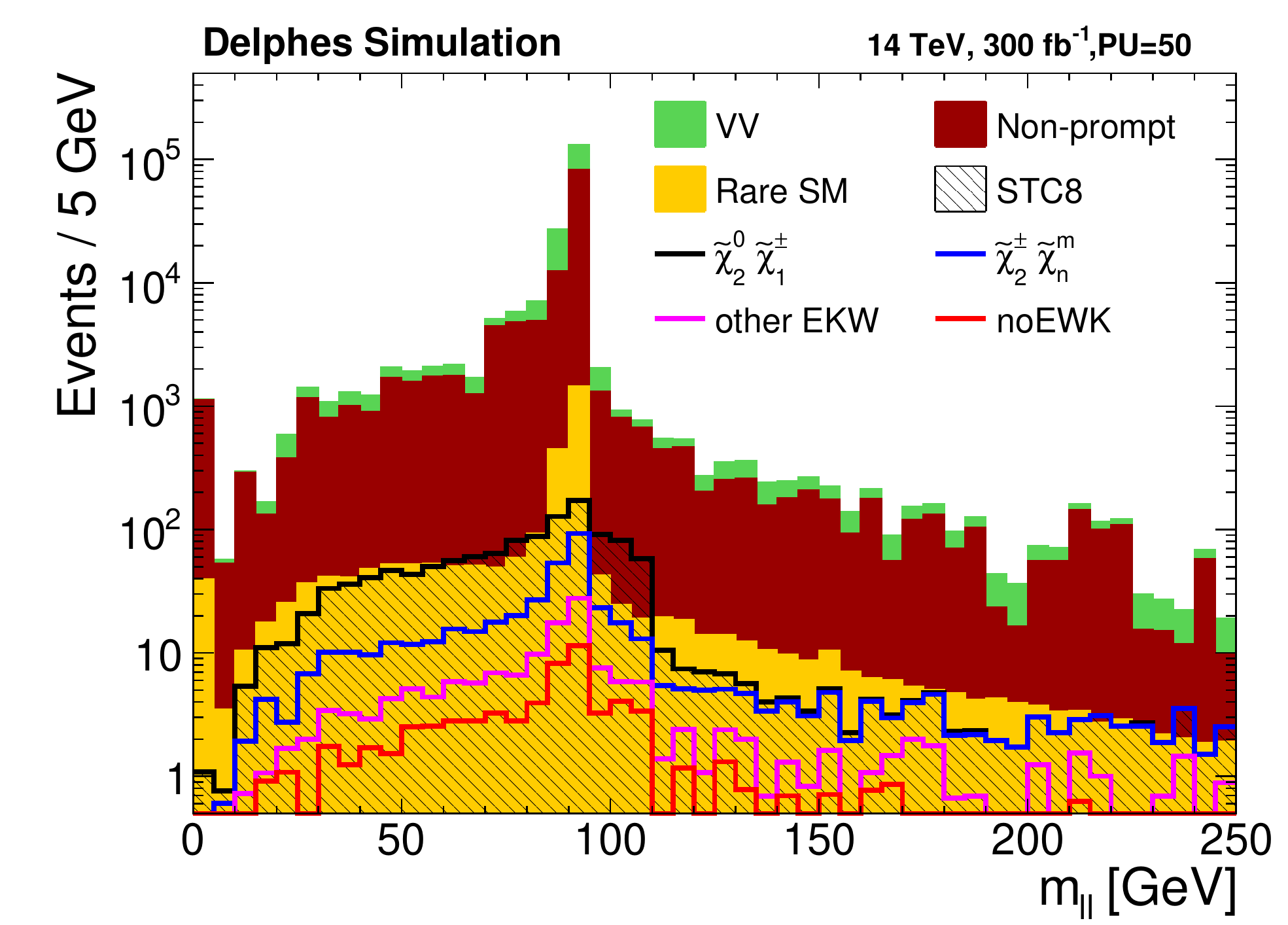} }
\subfigure[]{\includegraphics[width=0.44\linewidth]{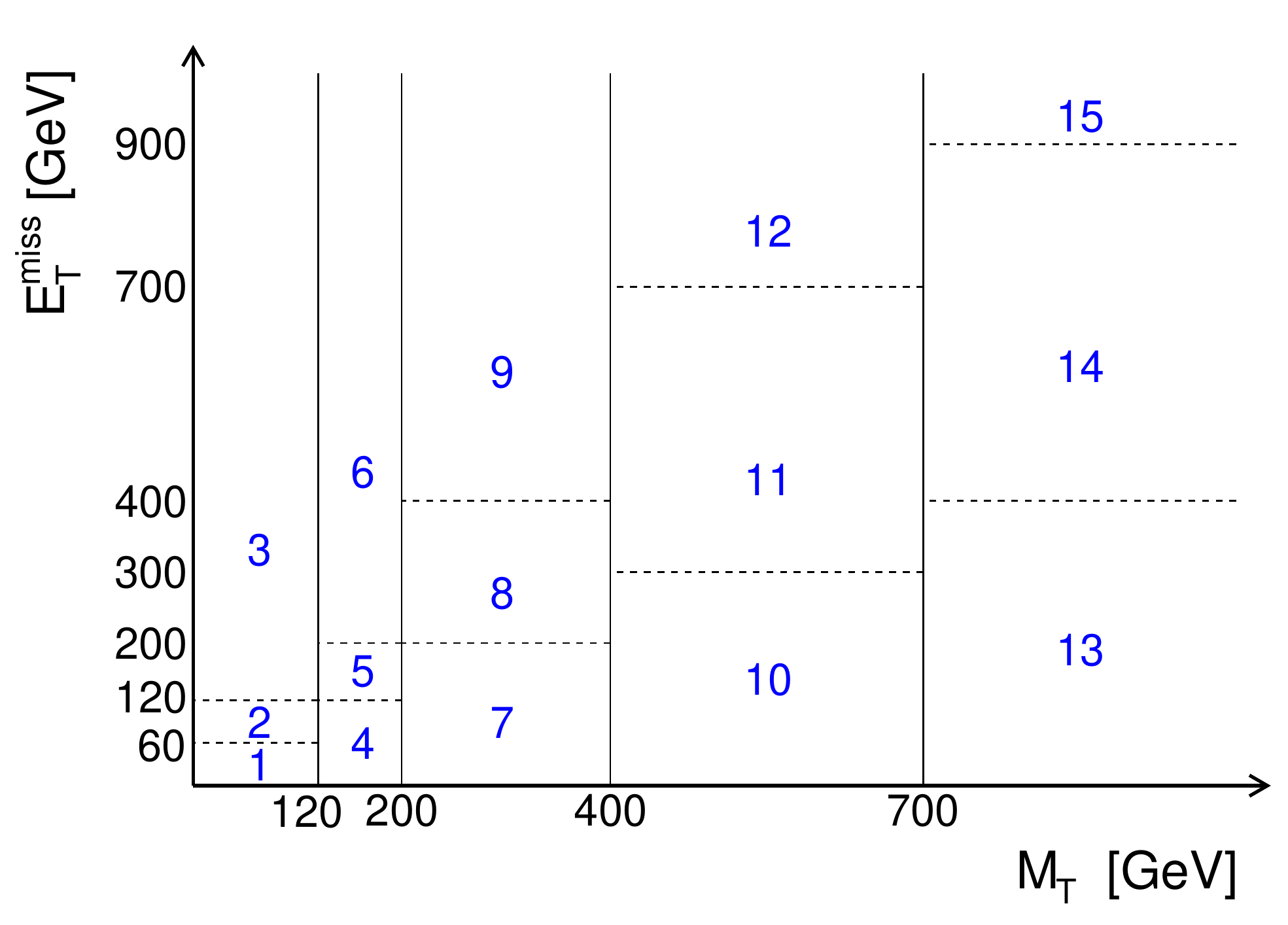} }
\caption{LHC Multilepton search: \MET (a), \MT (b) and m$_{\ellell}$ (c) distribution for three lepton events. 
Figure (d) illustrates the numbering scheme of the search regions. Detailed results for these are given in the Appendix, in Table~\ref{tab:3l}. 
}
\label{fig:multilepton_plots}
\end{figure}

The main background in the three-lepton final state originates from WZ production, where both bosons 
decay into leptons. The transverse mass, calculated with Equation~\ref{eqn:mt} for the missing transverse energy and 
the \pt of the lepton from the W boson, 
is used to suppress this kind of background. The second most important background are three-lepton events 
from non-prompt sources. Most of those events stem either from \ttbar or Drell-Yan events. The
latter are suppressed by the requirement of large \MET, whereas the \ttbar events are reduced by a 
b-jet veto. In order to maximise the sensitivity we use a three-dimensional binning in the variables 
m$_{\ellell}$ (invariant mass closest to \MZ), \MET and \MT. In Fig.~\ref{fig:multilepton_plots} the 
expected distributions for SM and STC8 are shown as well as definition of our search regions. The events 
are further separated by the value of the invariant mass of the opposite sign same flavour lepton pair: 
m$_{\ellell}< 75\GeV$, $75\GeV <$  m$_{\ellell}< 105\GeV$ and m$_{\ellell}> 105\GeV$. 

Overall we define 45 independent search regions, which are listed in the Appendix in Table~\ref{tab:3l} and
which all contribute to the final sensitivity. 
The applied uncertainties include lepton uncertainties (3\% per lepton), uncertainties on the \MET shape 
(15\%-25\%) and uncertainties of the MC statistics (0\%-30\%).
When combining all search regions, we can discover STC8 with less than 200\fbinv. 
The most sensitive individual regions are at medium \mT (200-400\GeV) and medium \MET (200-400\GeV), either requiring
m$_{\ellell}$ to be around the Z boson mass, which is mainly driven by direct \XN2\XPM1 production followed by $\XN2 \to Z \XN1$,
or for m$_{\ellell}$ higher than the Z boson mass. In the latter signal region long-chain decays of
the higher-mass neutralinos and charginos give important contributions, whereas the role of decays of the \XN2 via smuon or selectron is reduced due to their kinematic edge at m$_{\ellell}< 110\GeV$.


%% file: ILC.tex
\section{ILC projections} \label{sec:ILC}

Due to the clean conditions at an $e^+e^-$ collider, the searches for
SUSY at the ILC are exclusive, with separate analyses
adapted to each specific channel searched for.

Out of the plethora of exclusive processes, in particular pair production of
sleptons has been analysed in context of this paper, assuming an integrated 
luminosity of 1000\fbinv at $\Ecms= 500$\GeV,
which corresponds to four years of data taking at the design
luminosity of the ILC.
The data is assumed to be collected in
two equal-size samples, with beam polarisations of
$\mathcal{P}_{+80,-30}$ or $\mathcal{P}_{-80,+30}$, respectively.
The beam-conditions 
described in the ILC TDR~\cite{Adolphsen:2013kya} are applied for the simulation. 
We use the same SM background samples as for 
the detector benchmarking, presented in the ILC TDR\cite{Behnke:2013lya}.
These samples are produced with v.1.95 of the
{\sc Whizard}~\cite{Kilian:2007gr} event-generator.
For one background category, $\gamma\gamma$ processes, the statistics in the TDR sample was insufficient,
and a dedicated simulation, using the same version of {\sc Whizard} is done
to extend also this sample to correspond to 
the above-mentioned integrated luminosity of 500\fbinv per polarisation.
Under these conditions the total expected SM background is 1270 million events.
The SUSY signal is also generated with {\sc Whizard 1.95}, using the same
settings as for the SM samples, with the
exception that the study of the threshold scans of $\sMuR$ and $\seR$,
where the signal is generated using {\sc Pythia6.422}~\cite{Sjostrand:2006za}.
The detector simulation used is the fast
simulation program {\sc SGV}~\cite{Berggren:2012ar},
adapted to the ILD detector concept at ILC~\cite{Behnke:2013lya}.
In this program, 
the detector geometry is described in a complete but simplified model,
and is used to calculate the helix parameters and covariances
for all charged particles individually. The calculation includes
description of the point-resolutions at each layer, the magnetic
field, and the effect of multiple scattering.
In addition,
brems-strahlung and $\gamma$ conversions in the tracker material is simulated at
each layer.
The measurements by the calorimeter-system are in a first pass simulated parametrically
for each particle reaching it. 
In a second pass, the individual showers are split or merged by
a procedure adjusted to as well as possible emulate the performance
of the particle-flow algorithm as implemented in the the full simulation and
reconstruction of the ILD detector. 

Beyond slepton pair production, we summarise the results from previous studies performed on similar benchmark models and discuss them in the context of STCx. Most of these studies have been performed in full {\sc Geant4}-based simulation of the ILD detector concept, using samples simulated for the ILD LoI~\cite{Abe:2010aa}.

\subsection{Analysis of direct electroweakino production}

At the ILC, direct electroweakino production occurs at rates in the order of 100~pb,
c.f.\ Fig.~\ref{fig:xsect-ilc}. At a centre-of-mass energy of $500\GeV$, \XI0{1}\XI0{1}, \XI0{1}\XI0{2}, \XI0{2}\XI0{2} and \XIPM{1}\XIPM{1} are kinematically accessible in the STCx scenarios.
By measuring their polarised production cross sections and the masses of the involved sparticles, the parameters of the electroweakino sector  
($M_1$, $M_2$, $\mu$, $\tan{\beta}$) can be determined~\cite{Choi:2002mc}, given sufficient precision of the experimental observations.

As a particular challenge of the STCx models, the \XI0{2} and  \XIPM{1} decay dominantly via the 
\stau{1} NLSP. Nevertheless, almost all electroweakinos have branching fractions at the few percent level to other final 
states than the notoriously difficult $\tau$ lepton.
The use of beam polarisation and tunable $\Ecms$ will further enhance the
power of the observations and allow to disentangle production modes with very similar final states.

\subsubsection{\XI0{1}\XI0{2} production}
The dominating decay $\XI0{2}\to \stau{1}\tau$, with a branching ratio of about $70$\,\%, leads to the same final 
state content ($\tau^+\tau^-\XI0{1}\XI0{1}$) as for \stau{1} pair production. However, they can be disentangled by their 
different beam polarisation dependency and kinematic properties. The di-$\tau$ invariant mass can be employed to measure 
the mass of the \XI0{2}. This channel has 
been studied based on four-vector smearing, indicating that precisions of 1-2\GeV could be achievable~\cite{bib:thesis_ball}. In our scenario, the di-$\tau$ final state also receives background from
$\XIPM{1}\XIPM{1} \to \stau{1}\nu_{\tau}\stau{1}\nu_{\tau}$, which features a similar cross section $\times$ BR 
at $\Ecms=500\GeV$. The two processes can in principle be disentangled by the different kinematic features, but
the achievable resolutions would need to be studied. However, thanks to the tunable centre-of-mass energy
of the ILC, this problem can be avoided altogether by collecting data below the threshold for chargino pair production, in our case e.g.\ between $\Ecms=350$ and $400$\,GeV, or by scanning the \XI0{1}\XI0{2} 
production threshold.

\begin{figure}[htb]
  \begin{center}
\subfigure[]{\includegraphics[width=0.45\linewidth]{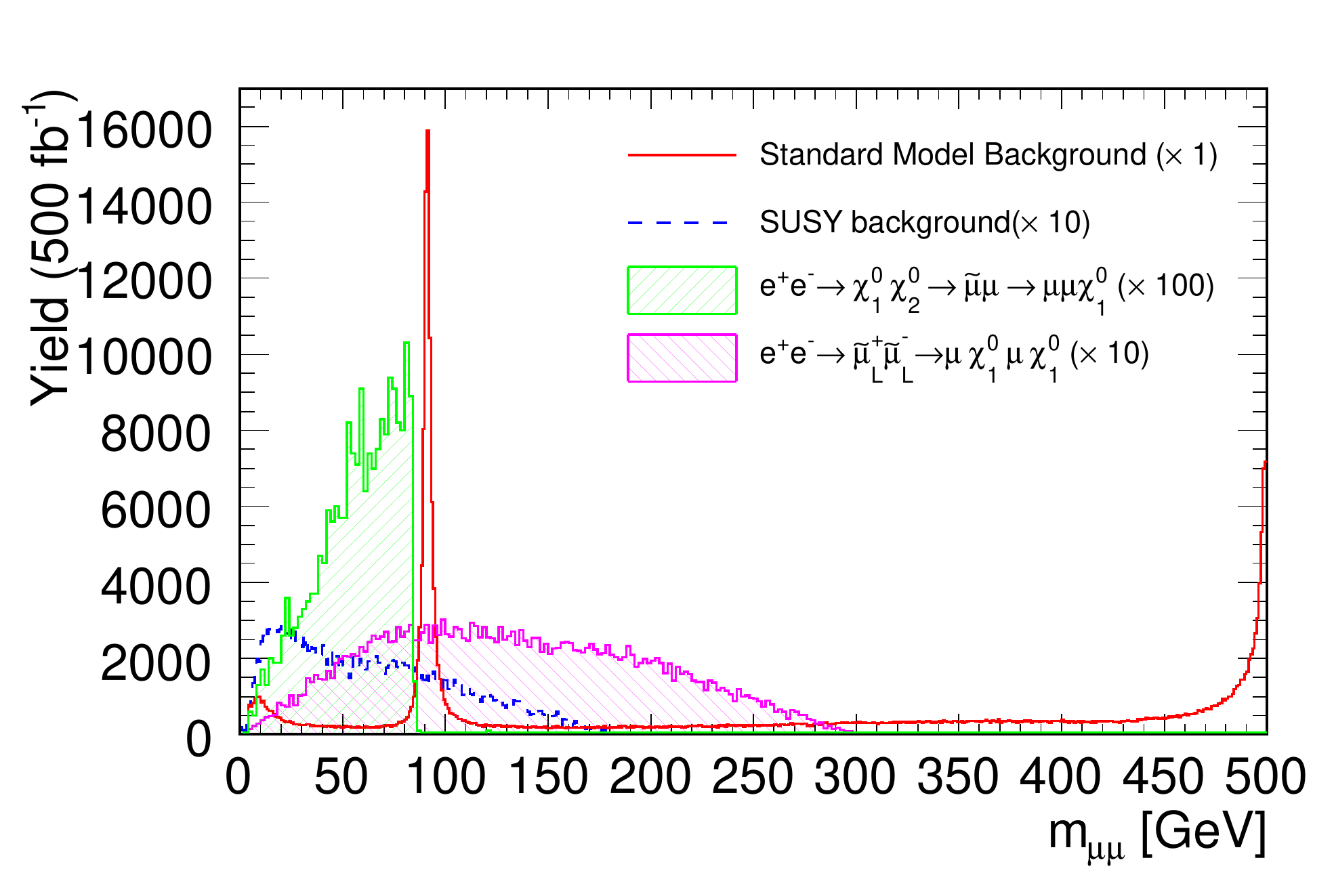}}
\subfigure[]{\includegraphics[width=0.45\linewidth]{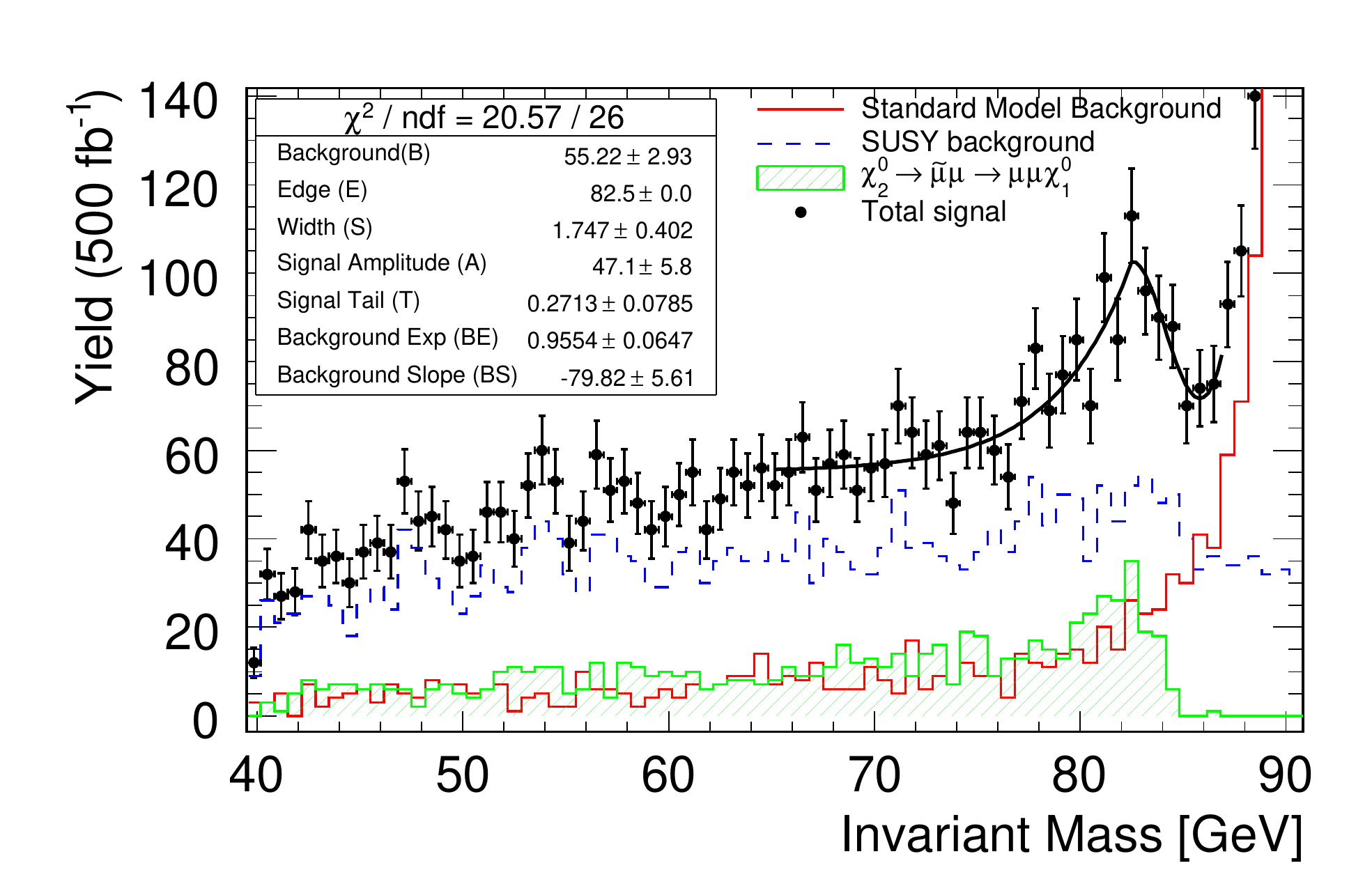}}
 \end{center}
  \caption{\label{fig:chi02mu} Determination of the $\XI0{2}$ mass from the di-muon invariant mass spectrum, 
   with the full spectrum for inclusive di-muon sample (a) and a zoom into signal region after 
  dedicated selection (b). From~\cite{bib:thesis_dascenzo}. }
\end{figure}

More recently, it has been shown in full {\sc Geant4}-based simulation of the ILD detector that the contribution from $\XI0{1}\XI0{2}\to \mu \mu \XI0{1}\XI0{1}$, 
which is about a factor of $10$ smaller, leads to competitive mass and cross section measurements. Figure~\ref{fig:chi02mu} shows for instance 
the invariant mass spectrum of the two muons before ((a), signal is scaled by 
factor $100$) and after event selection (b). From this channel alone, the mass of the
$\XI0{2}$ can be determined to a precision of about $1\GeV$ for an integrated luminosity of $500\fbinv$ collected 
with $\mathcal{P}_{-80,+60}$, depending on the assumed precision
for the mass of $\sMuR$ and $\XI0{1}$~\cite{bib:thesis_dascenzo}.

Within the same study, the corresponding uncertainty on $\sigma(\eeto\XI0{1}\XI0{2})\times BR(\XI0{2}\to \mu \mu \XI0{1})$ has been determined to about $20\,\%$, while the precision on $\sigma(\eeto\XI0{1}\XI0{2})\times BR(\XI0{2}\to \tau \tau \XI0{1})$ has been estimated
to $2\,\%$~\cite{bib:thesis_ball}. We thus conclude that the corresponding studies of the di-$\tau$
channel should be repeated with up-to-date simulation of the expected detector and accelerator performance.


\subsubsection{\XIPM{1}\XIPM{1} production}
\label{subsubsec:chipm1}
As discussed above, the final states of \XIPM{1} pair production will be dominated by di-$\tau$ plus missing four-momentum final states from $\XIPM{1} \to \stau{1} \nu_{\tau}$ or $\XIPM{1} \to \sNu_{\tau} \tau$ followed by an invisible decay of $\sNu_{\tau}$, thus be similar to those from $\XI0{1}\XI0{2}$. While at $\Ecms=500$\GeV 
the cross section $\times$ BR should be clearly measurable above the $\XI0{1}\XI0{2}$ background, especially
once the latter is known from running below the \XIPM{1}\XIPM{1} threshold, kinematic mass reconstruction in this
channel needs further study, like in the case of $\XI0{1}\XI0{2}$. However, there are several alternatives
to measure the mass: 

On one hand, the \XIPM{1} features a branching ratio of about $7\%$ to $\XI0{1}W^{\pm}$. Events where this decay, followed by $W \to q \bar{q}'$, occurs for one of the \XIPM{1}, while the other could e.g.\ decay to $\stau{1} \nu_{\tau}$, give a very unique signature. The edges in the energy spectrum of the $W$ bosons
can then be used to determine the mass of the \XIPM{1} if the \XI0{1} mass is known (see Section~\ref{subsec:sleptons}). This reconstruction method
has been studied in full detector simulation by ILD~\cite{Abe:2010aa, Suehara:2009bj} and SiD~\cite{Aihara:2009ad} for a SUSY scenario where the sleptons are heavier than the electroweakinos and thus both $\XIPM{1} \to \XI0{1} W$ and
$\XI0{2} \to \XI0{1} Z$ have branching ratios close to unity. These studies achieved mass resolutions of about $1.5\%$ for the \XIPM{1} based on $125 \cdot 10^3$ $\XIPM{1} \to \XI0{1} W$ decays, assuming the \XI0{1} mass is known from another source. In STCx, with an integrated luminosity of $500$\fbinv, only $11 \cdot 10^3 $ decays would be available and the \XI0{1} mass would be known to permille level precision from slepton pair production, c.f.\ Section~\ref{subsubsec:selsmu}. However, backgrounds
would also be lower, since a) the branching ratio for $\XI0{2} \to \XI0{1} Z$ is $100$ times lower than in the 
original scenario of these studies and b) the dominating SM background does not originate from fully hadronic 
$W$ pairs anymore, but from semi-leptonic ones, where the charge of the lepton in conjunction with the forward-backward asymmetry provides an additional, very effective suppression mechanism~\cite{Bechtle:2009em}. But
even neglecting these expected benefits, the pure scaling according to the number of decays yields a projected 
uncertainty of $5\%$ on the $\XIPM{1}$ mass. With the full running program of the ILC~\cite{paramgroup}, this would
shrink to about $2.5\%$.

In addition, the decays $\XIPM{1} \to \sNu_e e$ and $\XIPM{1} \to \sNu_{\mu} \mu$ give important information not only on the \XIPM{1} mass, but also on the mass of the mostly invisible $\sNu$. Since the
mass difference between \XIPM{1} and $\sNu$ is only $10$\GeV, lower as well as the upper edge
of the lepton energy spectrum are significantly below the lower edge from pair production of the left-handed sleptons (c.f.~\ref{subsubsec:selsmu}), at $8$ and $16$\GeV, respectively. Pair production of the right-handed
selectrons is heavily suppressed by the appropriate choice of beam polarisation. Selectron and smuon backgrounds can be
reduced further by selecting chargino decays to different lepton flavours in the two decay chains of an
event. While $\stau{1}$ pair production leads to different flavour leptons in the final state, the decay leptons
of the $\tau$ decays will be distributed over a wide range of energies (c.f.\ Fig~\ref{fig:stc-ilc}), whereas the signal
leptons lead to a sharp-edged, narrow box with the above edge positions. With the branching ratios of STCx, about $20 \cdot 10^3 $ decays to $\sNu_{\ell} \ell$ will be available from $500$\fbinv of data. Since this is of similar size than the available statistics for e.g.\ the $\sMuL$ (c.f.~sec~\ref{subsubsec:selsmu}), with steeper edges due to the smaller range of lepton energies, a precision of $1\%$ or better should be achievable for both the \XIPM{1} and the $\sNu$ mass.

\subsubsection{Radiative \XI0{1}\XI0{1} production}
\label{subsubsec:WIMP}

Additional information can be gained from the analysis of radiative \XI0{1}\XI0{1} production, employing a similar technique as in WIMP dark matter searches in mono-photon signatures~\cite{Bartels:2012ex}. This has been studied in full {\sc Geant4}-based simulation
of the ILD detector concept in~\cite{bib:thesis_bartels}. With an integrated luminosity of $500\fbinv$, equally split between all four beam helicity combinations for $|P(e^-,e^+)|=(80\%,30\%)$,
the unpolarised cross section as well as the cross section for purely right-handed electrons and left-handed positrons could be determined with a precision of about $5\,\%$ each. The mass of the
\XI0{1} could be determined to about $2\GeV$, completely dominated by a rather conservative estimate
of the systematic uncertainty due to the limited knowledge of the shape of the beam energy spectrum. 

\begin{figure}[htb]
  \begin{center}
\includegraphics[width=0.4\linewidth]{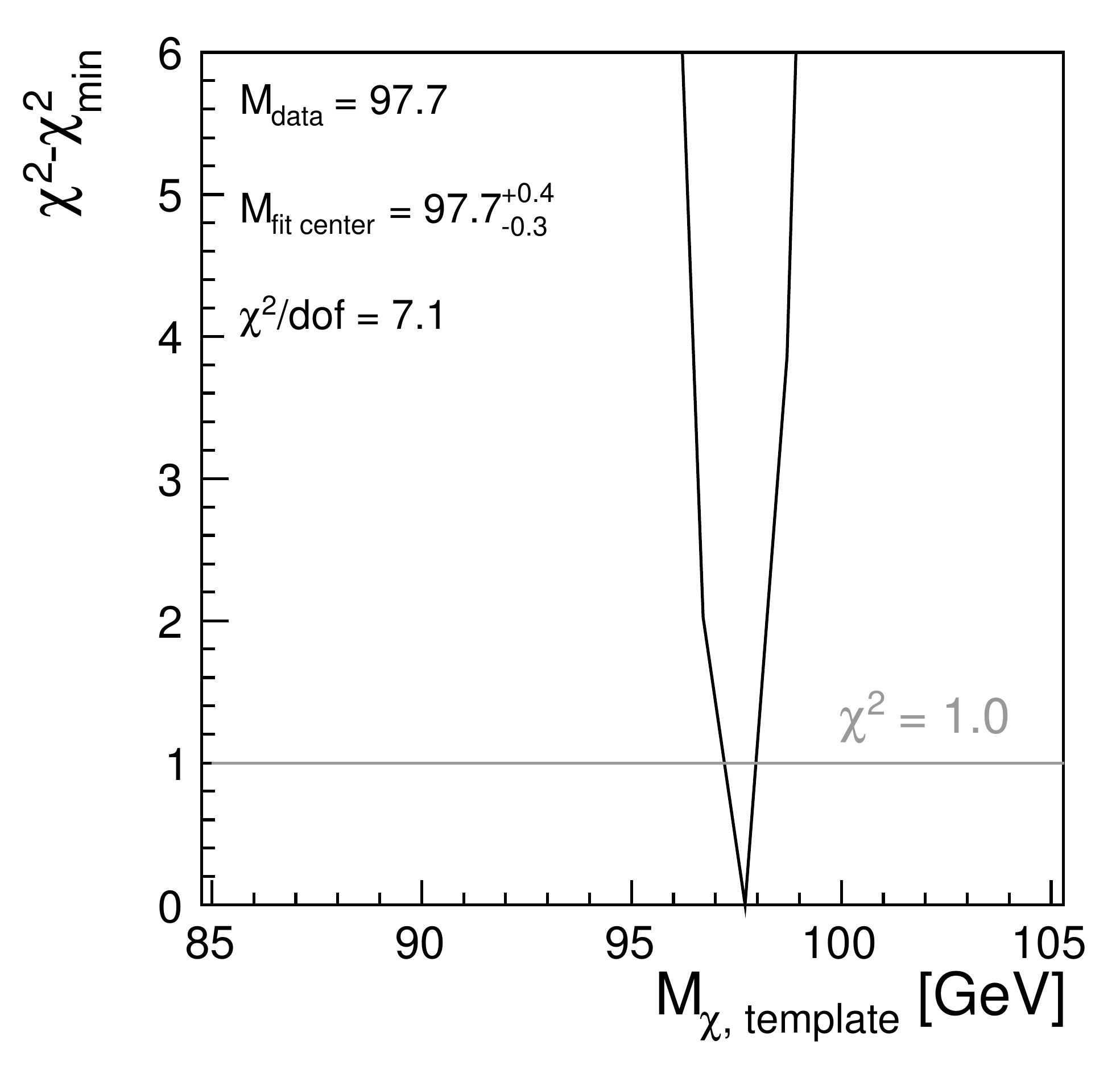}
 \end{center}
  \caption{\label{fig:WIMP} Determination of $m_{\tz_1}$ from a template fit to the photon 
energy spectrum in $\eeto \tz_1 \tz_1 \gamma$. From~\cite{bib:thesis_bartels}. } 
\end{figure}

As can be seen in Fig.~\ref{fig:WIMP}, the statistical precision, here from a template fit to the 
energy spectrum of the ISR photons, is much lower already with this rather modest luminosity assumed
in the study. A more detailed analysis of the impact of the beam energy spectrum should be performed
in the future.

\subsection{Analysis of direct slepton production}
\label{subsec:sleptons}

By studying slepton-pair production in the continuum (i.e.\ well above the threshold), one
takes advantage of the known initial state at the ILC: In general, the endpoints
of the energies of the standard model particles $X$ and $Y$ (both assumed to have negligible mass) in the process\footnote{Usually, $X$ and $Y$ are particle and anti-particle.} 
$\eeto \widetilde{X} \widetilde{Y} \rightarrow X Y \XN{1} \XN{1}$
can be found to be
\begin{align}
E_{i^{max}_{(min)}} =&   { M_{\widetilde{i}} \over 2 } \left ( 1 - \left( { \MXN{1} \over M_{\widetilde{i}} } \right )^2 \right )\left (  \gamma  \begin{smallmatrix}+ \\ (-)\end{smallmatrix}  \gamma \beta  \right ) =
   {  M_{\widetilde{i}} \over 2 }  \left ( 1 - \left( { \MXN{1} \over M_{\widetilde{i}} } \right )^2 \right ) \left ( { E^{lab}_{\widetilde{i}}  \over  M_{\widetilde{i}} } \begin{smallmatrix}+ \\ (-)\end{smallmatrix} { |\bar{p}^{lab}_{\widetilde{i}}| \over  M_{\widetilde{i}} } \right ) 
\label{eq:genspartendp}
\end{align}
where 
$i$ is either $X$ or $Y$. Note that due to momentum conservation, $\bar{p}^{lab}_{\widetilde{X}} = - \bar{p}^{lab}_{\widetilde{Y}}$. 
If the two sparticles have the same mass, so that $ M_{\widetilde{X}} = M_{\widetilde{Y}} = M_{\widetilde{i}}$, also  $E^{lab}_{\widetilde{X}} = E^{lab}_{\widetilde{Y}} = \Ecms / 2 $, and 
$ |\bar{p}^{lab}_{\widetilde{i}}| = \sqrt{ (  \Ecms / 2 )^2 -  M^2_{\widetilde{i}}}$.
Hence, in the case of slepton pair production,   $\eeto \sle \sle' \rightarrow \ell \ell \XN{1} \XN{1}$, one finds that
\begin{align}
E_{\ell^{max}_{(min)}} &=  
   { \Ecms \over 4}  \left ( 1 - \left( { \MXN{1} \over M_{\sle} } \right )^2 \right ) 
   \left ( 1  \begin{smallmatrix}+ \\ (-)\end{smallmatrix}  \sqrt{1 -  4 \left ( {  M_{\sle} \over \Ecms   } \right )^2} \right )
\label{eq:sleptpairendp}
\end{align}
i.e.\ by determining these two endpoints, and using the knowledge of $\Ecms$, both
$\MXN{1}$\ and $M_{\sle}$\ can be determined.
Even in the case where the slepton is a $\ttau$, the endpoint of the spectrum of the $\tau$
decay products can be used to determine $M_{\ttau}$, provided $\MXN{1}$ is known from e.g.\
the $\seR$ and $\sMuR$ spectra, measured from the same data.

The key characteristics of slepton production and decay, which
single them out from the background, are:
\begin{itemize}\itemsep1pt \parskip0pt\parsep0pt 

\item only two leptons in the final state
\item large missing energy and momentum
\item large acollinearity, with little correlation to the energy of the
lepton
\item central production
\item no forward-backward asymmetry
\end{itemize}

Different backgrounds dominate for the lighter and the heavier sleptons:
for the $\stau{1}$, the background from photon-photon processes is important,
while $WW \rightarrow \ell \nu \ell \nu$ is less so; the
opposite is true for the $\stau{2}$, $\seL$ and $\sMuL$. 
For $\widetilde{e}$ and $\tmu$, the lower endpoints of the spectra are high enough
- both for the right-handed and left-handed states - that the photon-photon background
does not pose a problem.
The mass measurement relies on the detection of edges or endpoint in the
spectrum of the SM decay products.
Therefore,
the SUSY background is less important, since it
is only important that it is small and/or flat
in a narrow region around the edges. 
In case for $\widetilde{e}$ and $\tmu$ the SUSY background is indeed
found to be both small and flat in these regions.
In the case of the $\ttau$,
the SUSY background is dominated by
$\XPM{1}\XMP{1}$ and $\XN{2}\XN{2}$ production
with cascade decays over $\ttau$ sleptons,
which all yield an upper kinematic limit of the produced $\tau$ leptons
well below those of both $\stau{1}$ and $\stau{2}$ pair production,
so they have little influence on the determination of the
upper endpoints of the spectra. 

These considerations lead to the following selection criteria, valid
for all slepton studies: The events contain less than  10 charged particles,
and two lepton candidates. The total charge should vanish, while each of the
lepton-candidates should have opposite charge. The mass of each of the candidates
should be less than $M_\tau$. The total visible energy in the event, $E_{\mathrm{vis}}$,
should not exceed 300\GeV, and the missing mass, $M_\mathrm{miss}$, should be
larger than 200\GeV. No particle in the event should have momentum above 180\GeV.
Lepton-candidates are found using the DELPHI $\tau$-finder \cite{Abdallah:2003xe}, which
is designed to identify both isolated electrons or muons, and decays
of $\tau$ leptons.
It is also robust against extra activity in the detector coming from
beam-beam effects and overlaid low-\pt $\gamma\gamma$ events.

\subsubsection{Selectrons and smuons}
\label{subsubsec:selsmu}

As can be seen in Fig.~\ref{fig:xsect-ilc}, the selectron pair
production cross section is huge in our scenario due to the $t$-channel
neutralino exchange, allowing a very precise determination
of the masses and polarised cross sections in a short time.
At larger integrated luminosities, very precise measurements 
of the selectron and LSP masses can be obtained from the edges of the spectrum.
In addition to the generic slepton selection criteria presented above,
it is also demanded that both lepton candidates are identified as
electrons. 
To separate $\seR$ and $\seL$, the beam-polarisation is helpful:
for $\mathcal{P}_{+80,-30}$,  $\seR$ production is enhanced, while $\seL$
production is suppressed. For $\mathcal{P}_{-80,+30}$, the situation is the opposite.
In addition, since $\seL$ is appreciably heavier than than $\seR$, the decay 
products of the former are less boosted, and hence the detected leptons tend to be
less back-to-back in  $\seL$ events. 
A cut on the transverse momentum of one electron with respect to the other one, $\pt$$_{1(2)}^{\mathrm{rel}}$
is therefore applied to further separate the two cases: $\pt$$_{1}^{\mathrm{rel}} + \pt$$_{2}^{\mathrm{rel}}$
should be less than 30\GeV for  $\seR$ candidates, and larger then 100\GeV for
$\seL$ candidates.
With these cuts, the selection efficiency is 51\,\% for $\seR$ and 47\,\% for $\seL$.

\begin{figure}[htb]
  \begin{center}
\subfigure[]{\includegraphics[width=0.49\linewidth]{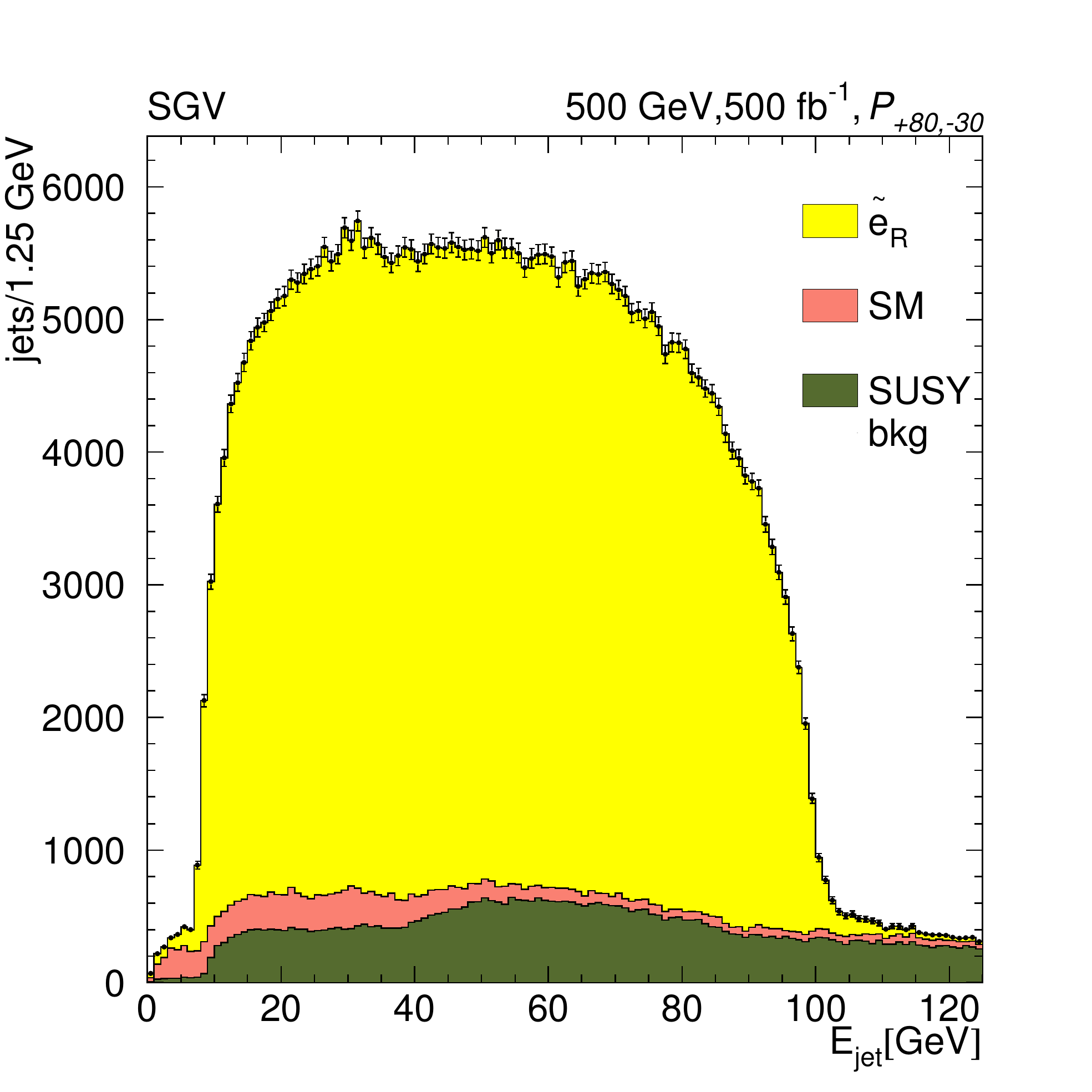}}
\subfigure[]{\includegraphics[width=0.49\linewidth]{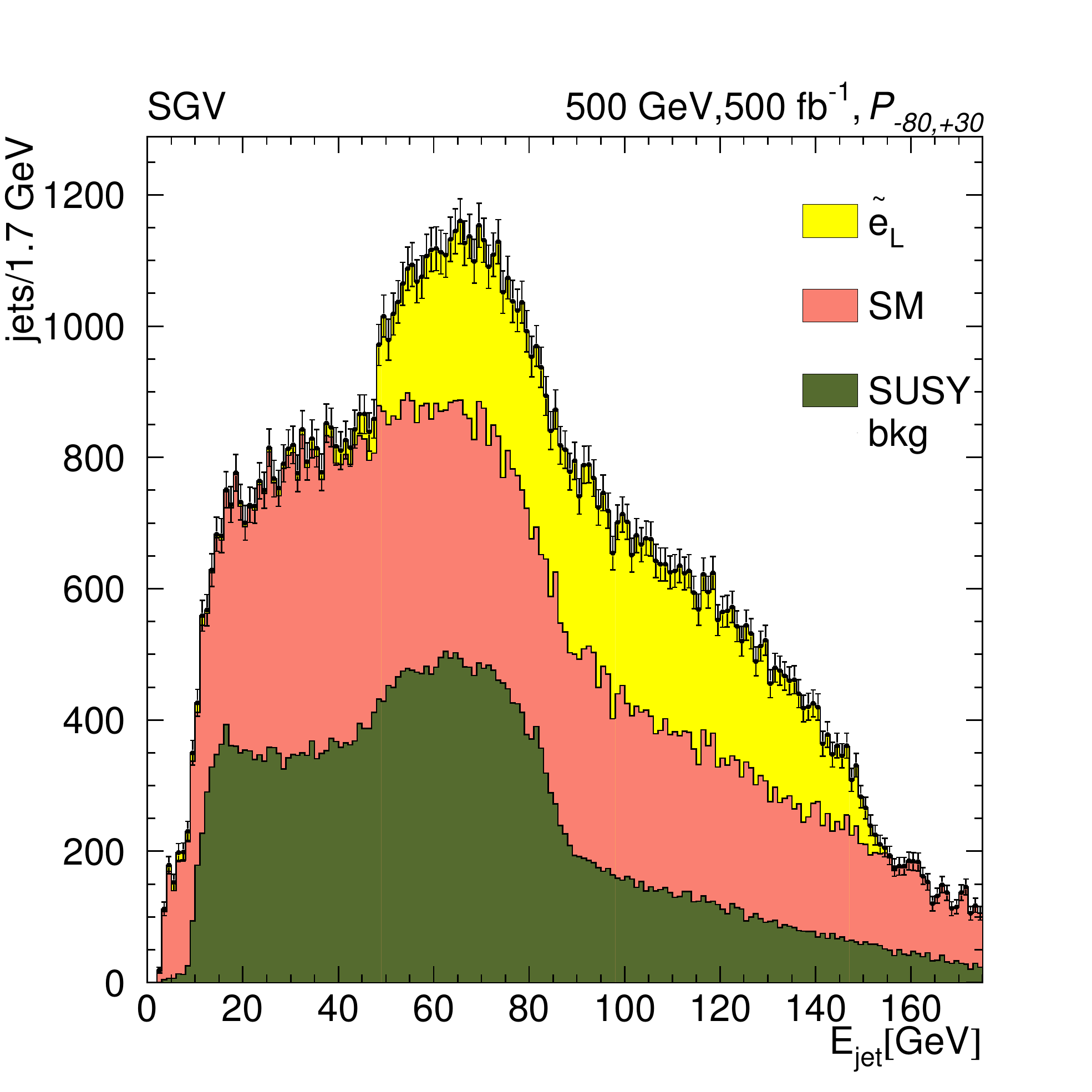}}
 \end{center}
  \caption{\label{fig:selspect} Electron spectra from selectron decays and background, with electrons from \seR decays (a) and
  electrons from \seL decays (b).}
\end{figure}
Figure~\ref{fig:selspect}
shows the spectra of electron energies in selected di-electron events
after collecting
500\fbinv of data for each of the beam-polarisations  $\mathcal{P}_{-80,+30}$  
and  $\mathcal{P}_{+80,-30}$.
The $\seR$ signal stands out above quite small SM and SUSY background for
beam-polarisation $\mathcal{P}_{+80,-30}$. The $\seL$ signal 
is less clean both due to smaller
signal cross section (cf.\ Fig.~\ref{fig:xsect-ilc}), larger background and lower
efficiency, but 
is nevertheless quite prominent. In particular, both edges are detectable.
This will still be the case even for substantially smaller branching ratios for the
direct decay to lepton and LSP, compatible with LHC limits.

The position of the edges is determined by sub-dividing the full data sample in
sub-sets, and finding the most and least energetic lepton in each sub-set,
after excluding a certain fraction of the extreme cases. The size of the sub-samples,
and the fraction to be excluded is optimised to yield the lowest possible
uncertainty on the endpoints. The resulting endpoints averaged over the sub-samples
show a bias. This bias has been corrected for by means of a toy Monte Carlo procedure;
the uncertainty on the SUSY masses is determined with this procedure as well.
In this way, we obtain
$\MXN{1}   = 95.47  \pm 0.16\GeV$ and $ M_{\sel{r}} = 126.20 \pm 0.21\GeV$ from the $\seR$ spectrum.
The true masses in STCx are  $\MXN{1} =95.59\GeV$ and $M_{\seR} = 126.24\GeV$.
   
The large SUSY background to $\seL$ is mostly
$\seR \seL$ production\footnote{Equation \ref{eq:genspartendp} shows that the lighter state receives a lower fraction of the
total initial energy than the heavier state in $\seR \seL$ production,
so that the upper edge from the $\seR$ decay is lower than what it is
in the symmetric case in $\seR \seR$ production seen in Fig.~\ref{fig:selspect}a.}. 
This channel gives important information
on the neutralino mixing, since e.g.\ in case of light higgsinos the $t$-channel would be strongly suppressed by
the small electron Yukawa coupling. In particular, if both beams are given 
left-handed polarisations, only the $\eeto\seR^-\seL^+$ process is possible. 
As this reaction proceeds exclusively via neutralino exchange in the $t$-channel, 
its size gives insight to the neutralino mixing~\cite{MoortgatPick:2005cw}.

\begin{figure}[htb]
  \begin{center}
\subfigure[]{\includegraphics[width=0.49\linewidth]{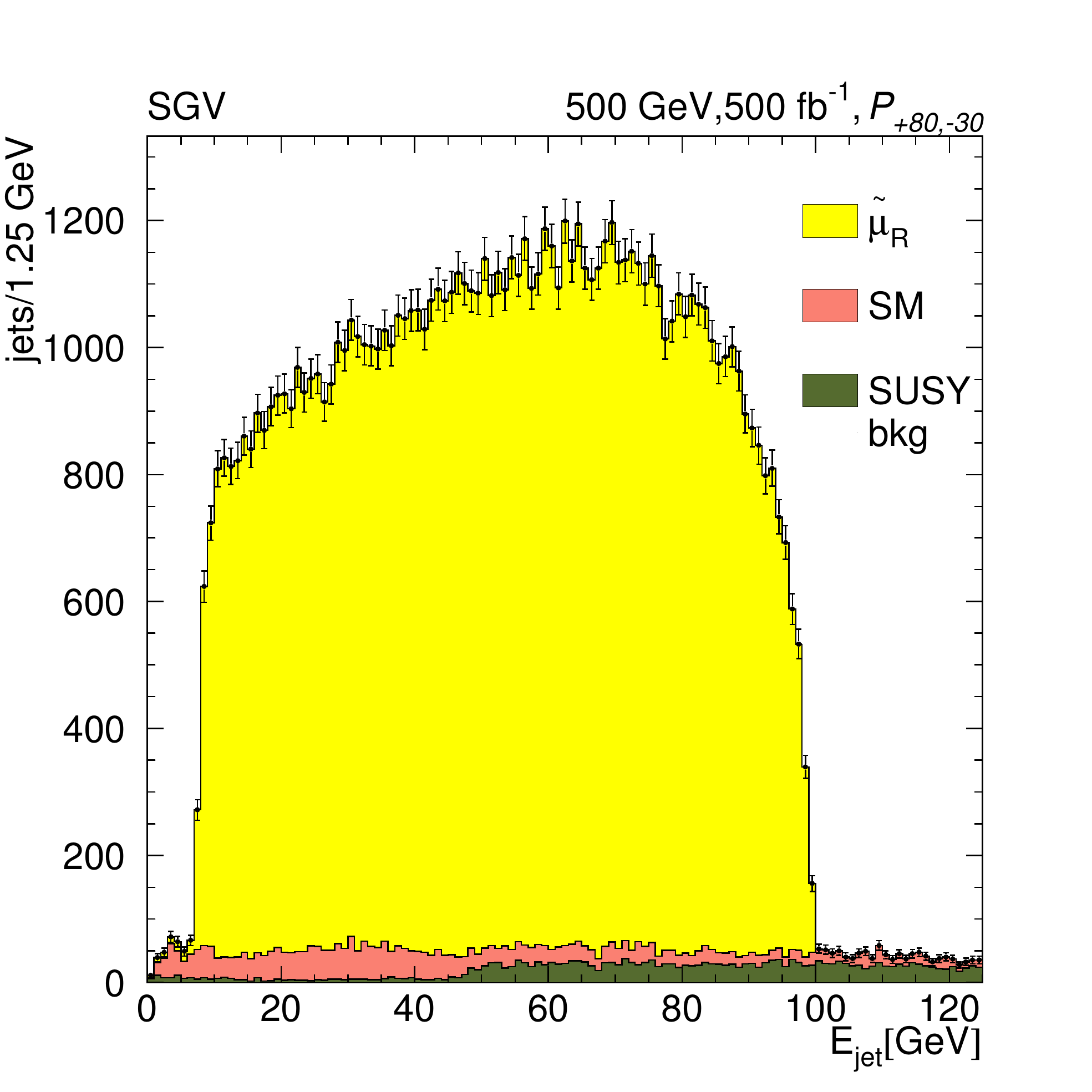}}
\subfigure[]{\includegraphics[width=0.49\linewidth]{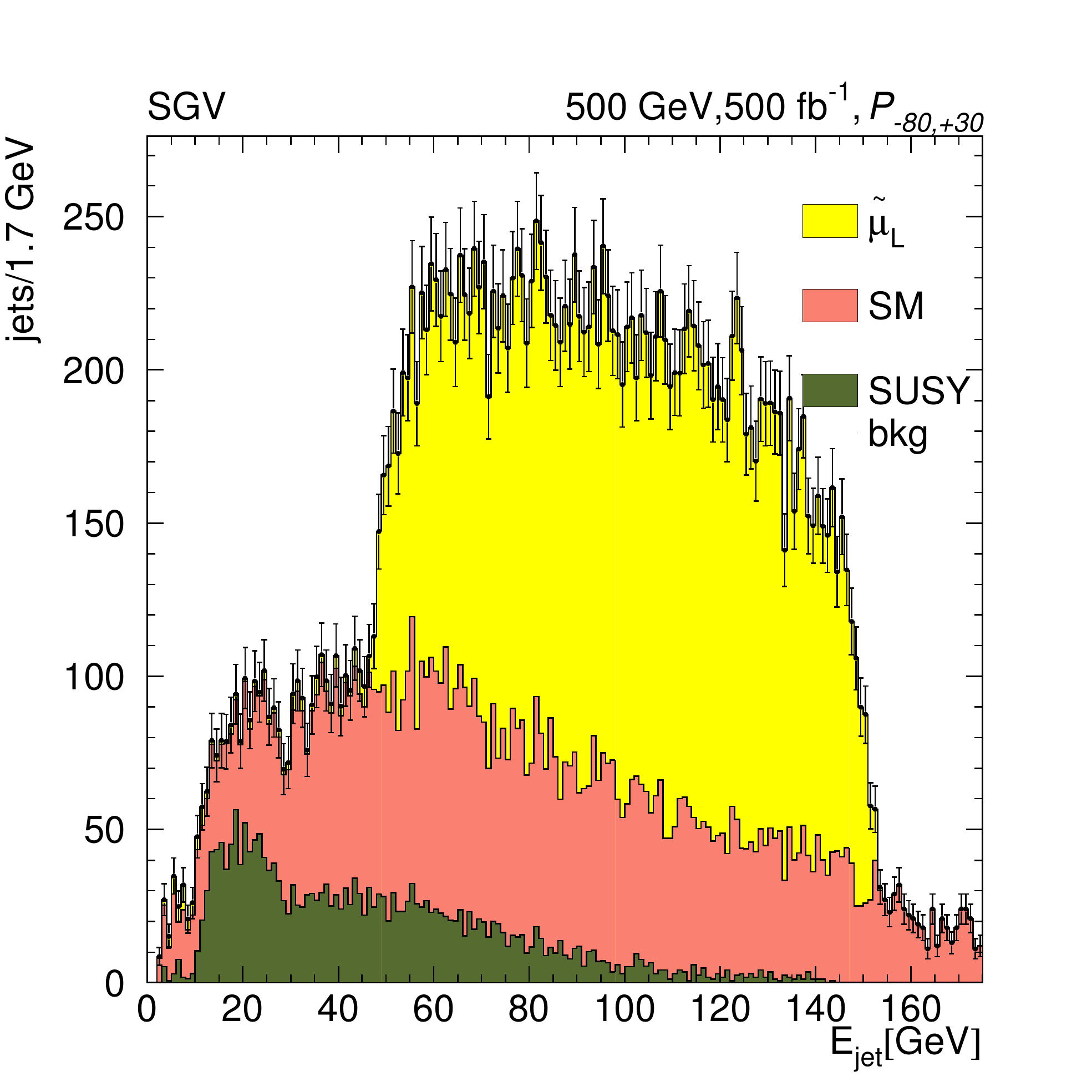}}
 \end{center}
  \caption{\label{fig:smuons} Muon spectra from smuon decays and background. Fig.~(a): muons from $\sMuR$ decays and 
  (b) muons from $\sMuL$ decays.}
\end{figure}
Figure~\ref{fig:smuons} shows the spectra of muon energies in di-muon events, under the same
conditions ($\int \mathcal{L} dt = 500\fbinv$ at  each of the beam-polarisations).
In this analysis, the generic selection is supplemented by demanding that
both lepton candidates are identified as muons.
The same criteria as for $\widetilde{e}$ are used to separate $\sMuR$ and $\sMuL$ candidates.
The selection efficiency
is larger for $\tmu$ than for $\widetilde{e}$: it is 65\,\% and 60\,\% for $\sMuR$~and $\sMuL$, respectively.
This is due to the t-channel contribution to  $\widetilde{e}$: even though the sleptons 
aren't highly boosted, their decay products nevertheless carry the imprint of the
initial slepton angular distribution. Therefore, the cuts designed to remove backgrounds
at low angles to the beam-axis or with low missing transverse momentum tend to
remove more $\widetilde{e}$ events than  $\tmu$ ones.
Using the same procedure to extract the edges,
the LSP and smuon masses can be determined to $\MXN{1} = 95.47 \pm 0.38\GeV$
and $M_{\sMuR} = 126.10 \pm 0.51\GeV$, once again in good agreement with the true masses
in the STCx model:  $\MXN{1} =95.59\GeV$ and $M_{\sMuR} = 126.16\GeV$.
It can be noted that, as expected, the SUSY background to $\sMuL \sMuL$ production is much lower
than for $\seL \seL$ production, and that a significant signal
would be expected even if the branching ratio to the direct decay would only be
a few percent.
Combining the measurement of the LSP mass from the right-handed selectron
and smuon analyses yields an uncertainty of $\sigma_{\MXN{1}} = 147\MeV$,
ie.\ slightly above 1 permille.




\begin{figure}[htb]
  \begin{center}
\subfigure[]{\includegraphics[width=0.49\linewidth]{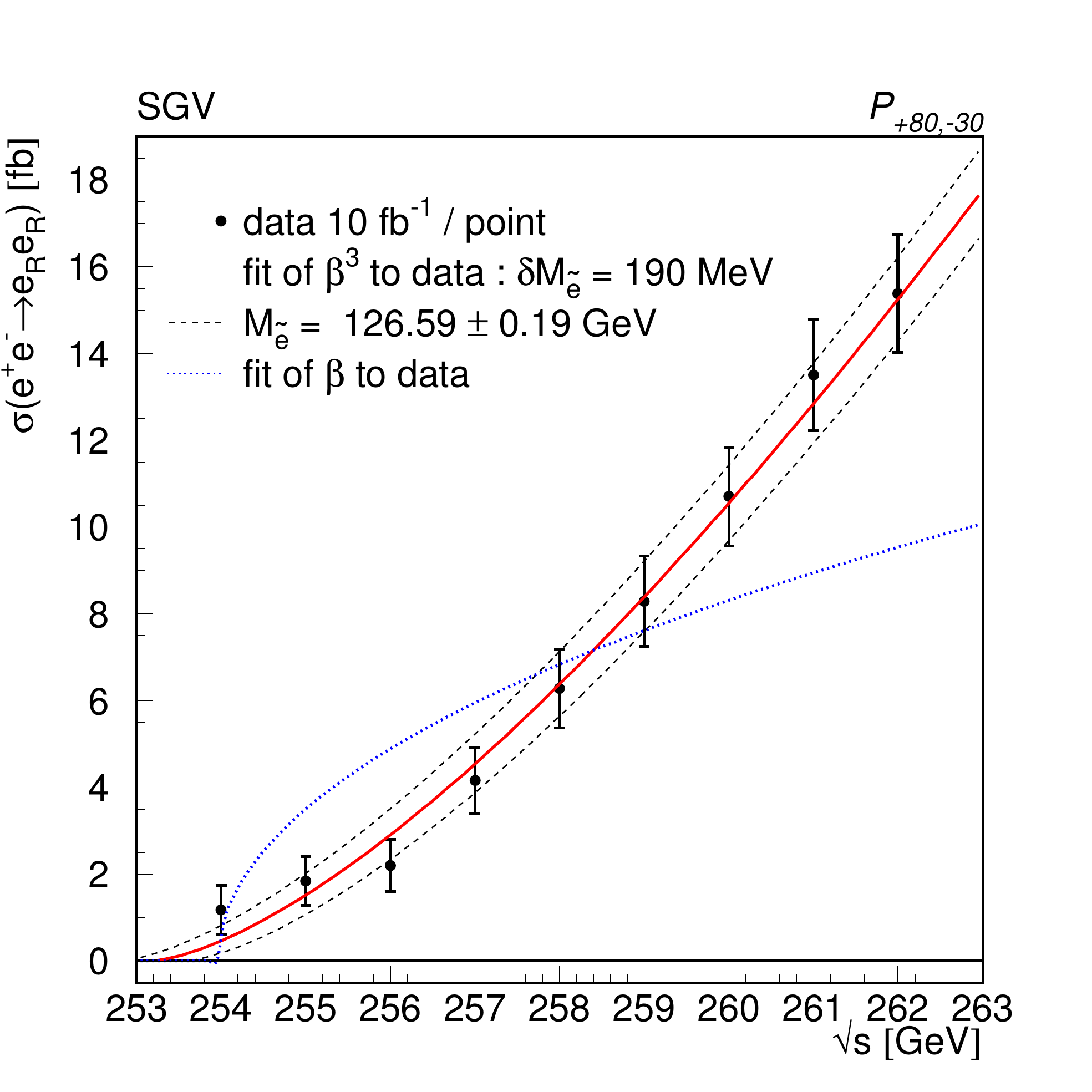}}
\subfigure[]{\includegraphics[width=0.49\linewidth]{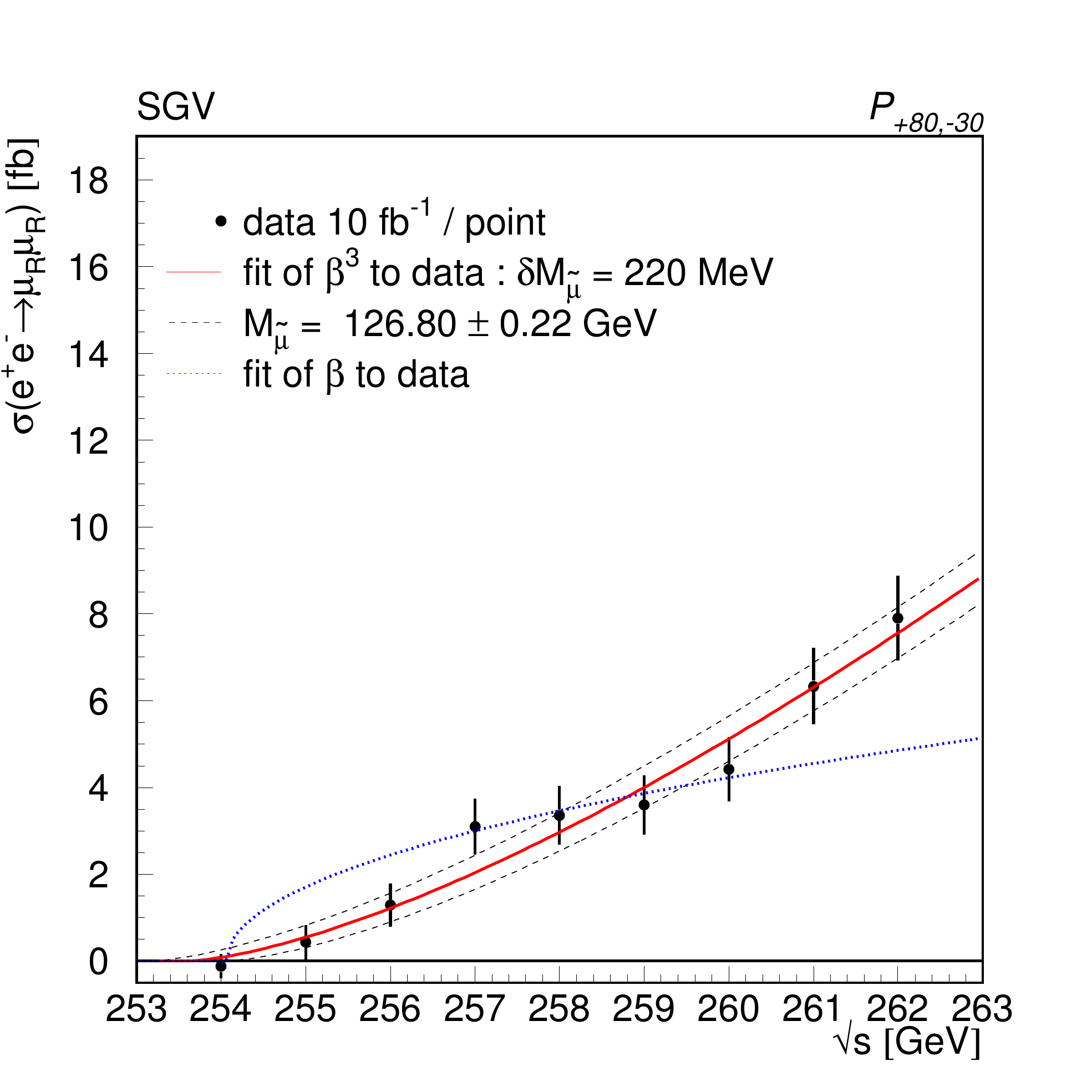}}
 \end{center}
  \caption{\label{fig:slept_thresh} Threshold scans at the $\eeto \seR \seR$ (a)  and
      $\eeto \sMuR \sMuR$ (b) thresholds.}
\end{figure}

In addition to the mass determination from the spectrum edges,
the mass of both \seR and \sMuR can be determined by scanning the production threshold near 
$250\GeV$, as illustrated by Fig.~\ref{fig:slept_thresh}.
Close to threshold, the cross section is obviously small, but on the other hand, the signal is very clean:
since the sleptons are produced almost at rest, and they undergo two-body decays, the decay products are
almost mono-energetic. In addition, in STCx the mass-difference between the LSP and \seR and \sMuR is
rather large, so that even in a decay at rest the produced leptons have momentum $\sim$ 25\GeV.
Hence, by selecting events with two opposite-sign, same-flavour leptons with large acoplanarity, and with
momentum in a narrow window, a large efficiency, low background sample can be obtained, and an significant excess 
of events can be obtained quite close to the threshold.
For STCx, it was demanded that the two leptons should have momentum between 20 and 37\GeV,
and both the acoplanarity and acollinearity angles should be below 3.1 radians.
With these cuts, the efficiency for the signal is between 85 and 95\,\%, and the signal-to-background
ratio is above 1 for almost all points, the exception being the lowest $\Ecms$ for the \sMuR.

Sleptons are scalars - if not the new physics the observations have revealed is not supersymmetry.
Therefore, it is certain that slepton-pairs are produced in a P-wave, and hence that the rise of
the cross section with increasing $\Ecms$ is proportional to 
$\beta^3 =  \left [ 1 -  4 \left ( {  M_{\sle} / \Ecms   } \right )^2 \right ]^{3/2} $.
Investing a few months of ILC beam-time\footnote{Note that this centre-of-mass energy range is also optimal for 
studying the properties of the 
higgs boson with the model-independent recoil-mass method.}, the mass of \seR can be 
determined to $\sim 190\MeV$, and at the
same time that of \sMuR to $\sim 220\MeV$, by fitting the observed background-subtracted 
cross section to $\beta^3( \Ecms)$.

In addition, one can test the hypothesis that the observed states are spin-1/2 particles,
rather than scalars.
If the particles indeed have spin-1/2, the pairs are produced in an S-wave, and the rise of the cross section  
with increasing $\Ecms$ would be  proportional to $\beta$ rather than $\beta^3$.
The dashed curve in Fig.~\ref{fig:slept_thresh} is the best fit of  $\beta( \Ecms)$ to
the data.
For the \seR case the fit-probability is $ < 10^{-9}$, while it is $7\times 10^{-5}$
in the \sMuR case. The spin-1/2 hypothesis would therefore be excluded by the data.

%
\subsubsection{The $\ttau$-sector}
\label{subsubsec:staus}

Especially in $\ttau$-coannihilation scenarios, a precise determination of the 
$\ttau$ sector is essential in order to be able to predict the expected  
relic density with sufficient precision to test whether the $\XI0{1}$ is indeed the dominant 
Dark Matter constituent. 
The $\stau{1}$-pair production  is different from \seR- or $\sMuR$-pair production
in several aspects: The mass difference to the LSP is much smaller,
meaning that the $\tau$ spectrum is softer than the spectrum of the leptons
in \seR or \sMuR decays. In addition, $\tau$ leptons decay, further
softening the spectrum of observed particles,
and making the particle identification requirements less effective in
background suppression. 
This leads to a signal that much more resembles that of $\gamma\gamma$ events,
but also more resembles di-boson events decaying to $\tau\nu\tau\nu$.
The generic slepton selection therefore needs to be supplemented
by several further criteria to reduce these sources of background:
The requirements on $E_{\mathrm{vis}}$ and $M_\mathrm{miss}$ are strengthened to
$<120$ and $> 250\GeV$, respectively, and the visible mass, $M_\mathrm{vis}$,
should be below $M_{Z}-5$, which reduces the  di-boson background.
The cosine of the direction of the missing momentum is required to be between -0.8 and 0.8,
$M_\mathrm{vis}$ should be above 20\GeV, and the total energy observed below
30 degrees to the beam-axis should not exceed 2\GeV. This selection
reduces the $\gamma\gamma$ background, which is then further decreased by
a cut on the likelihood that the event is a $\gamma\gamma$ event.
Finally, to reduce the SUSY background from  \seR- or \sMuR-pair production,
as well as from di-boson events, it is required that the event is not identified
as a di-electron or di-muon event. With these cuts, the selection efficiency
for $\stau{1}$-pair production is 17\,\%.

Only the upper endpoint can be measured in  $\stau{1}$ production:
due to the decay of the $\tau$, the lower endpoint is only visible
as a knee in the spectrum of the decay-products of the $\tau$.
Because of the small mass-difference between the $\stau{1}$ and $\XN{1}$,
this knee is in a region where the spectrum is strongly distorted by
the cuts removing the $\gamma\gamma$ background.
Contrary to the case for $\widetilde{e}$ or $\tmu$, the upper kinematic limit is
not an edge, but the endpoint of the spectrum of $\tau$ decay-products.
This endpoint is determined by fitting the background in the region
well above the endpoint and then fitting a signal contribution in the data
above the extrapolated background fit. 

\begin{figure}[htb]
  \begin{center}
\subfigure[]{\includegraphics[width=0.45\linewidth]{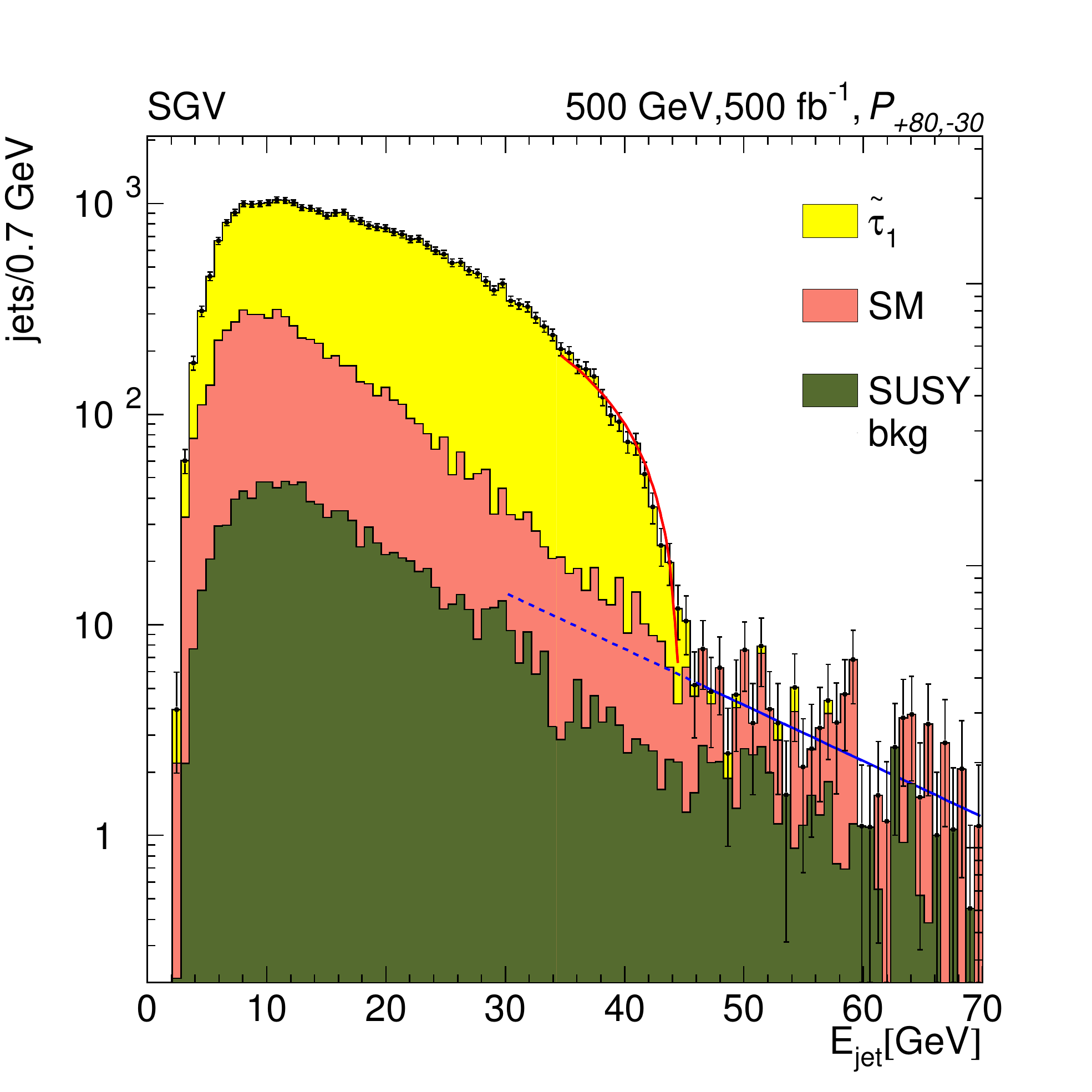}}
\subfigure[]{\includegraphics[width=0.45\linewidth]{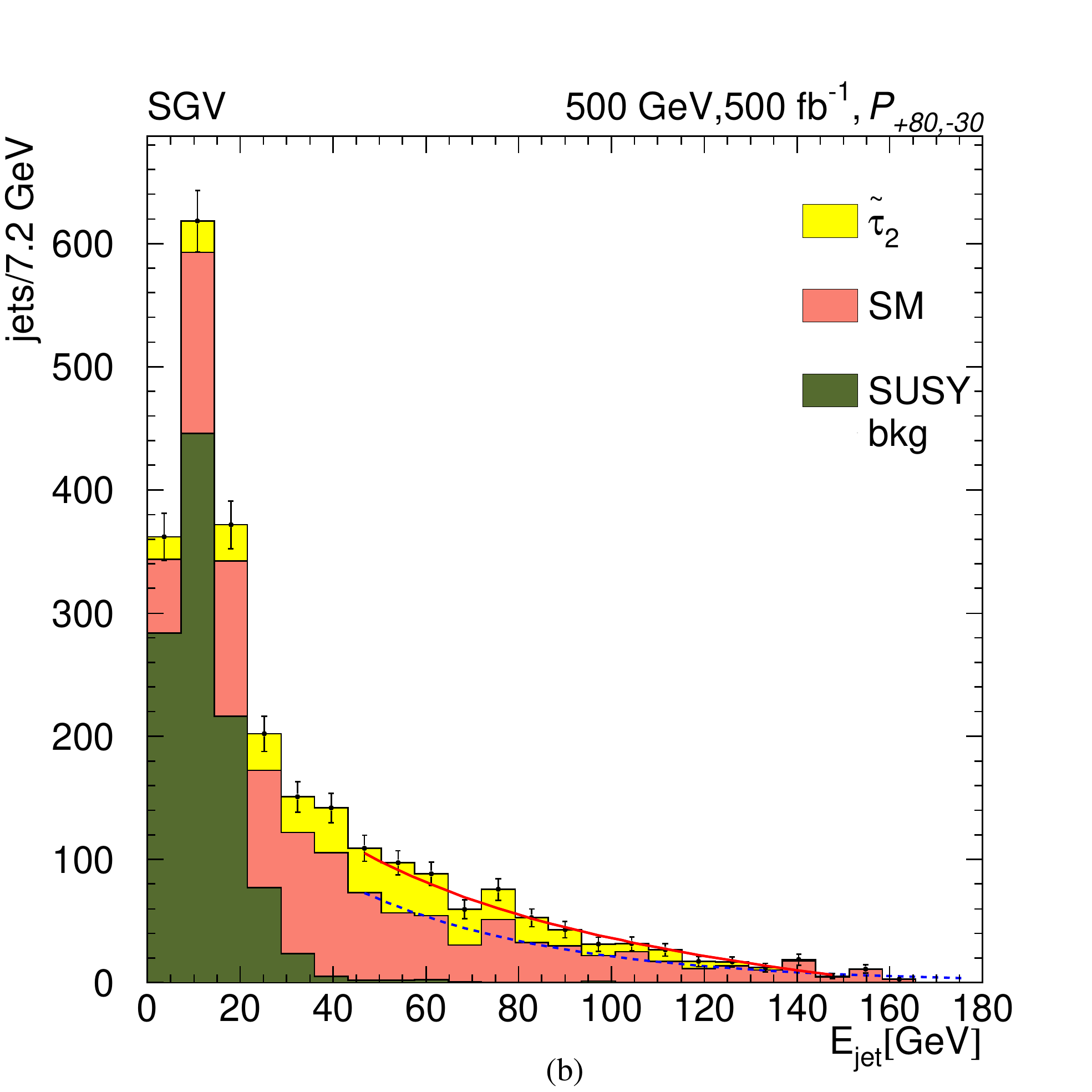}}
 \end{center}
  \caption{\label{fig:stc-ilc} The $\tau$ spectra in \stau{1} decays and
  SM as well as SUSY background, with an endpoint fit (a), and $\tau$ spectra in \stau{2} decays
 and background, with endpoint fit (b). }
\end{figure}
Figure~\ref{fig:stc-ilc} shows the energy-spectrum at $\Ecms = 500\GeV$ 
of selected $\tau$-jets
for an integrated luminosity of 500\fbinv, and polarisation $\mathcal{P}_{+80,-30}$.
In the case of $\stau{1}$ production, the endpoint
could be determined to be E$_\mathrm{endpoint}=44.49^{+0.11}_{-0.09}$\GeV,
corresponding to an uncertainty of $M_{\stau{1}}$ of 200\MeV, if an uncertainty on
the LSP mass of $\sim$ 100\MeV is assumed.
Similarly, the \stau{2} mass could be determined with an uncertainty of 5\GeV.

In~\cite{Bechtle:2009em}, where a model quite similar to STCx has been studied,
it is found that, in addition to $M_{\stau{1}}$, the production cross section for both these modes can be
determined at the level of 4\,\%, and the polarisation
of $\tau$-leptons from the \stau{1} decay, which gives access to the $\ttau$ and 
$\XI0{1}$ mixing\footnote{Interaction of sfermions and gauginos conserve chirality, while the Yukawa
 interaction of the higgsinos flips chirality.}, 
could be
measured with an accuracy better than 10\,\% from shape of the $\pi$ spectrum in the  $\tau\to\pi^+\nu_{\tau}$ mode or
to better than 5\,\% by a template fit of $R=E_{\pi}/E_{jet}$ in the $\tau^-\to\rho^-\nu_{\tau} \to \pi^- \pi^0\nu_{\tau}$ (and c.c.)
mode.

\subsubsection{The sneutrinos}
While the vast majority of sneutrino decays proceeds to a completely invisible final state, the $\sNu_{\tau}$ has a branching fraction of about $5\%$ to $\stau{1} W$. In this situation there are two possible strategies: a) search for
the completely invisible final state via its recoil against a hard photon from initial state radiation in analogy to Section~\ref{subsubsec:WIMP} and b) select $\sNu_{\tau}$ pair events where at least one visible decay occurs. Neither case has yet been
studied in detailed simulations, but we will sketch the strategies here.

In case of the ISR recoil, the situation is more difficult than for $\XI0{1}$ pair production since the cross section for the low-background polarisation $\mathcal{P}_{+80,-30}$ is about one order of magnitude smaller than for the $\XI0{1}$
case. At the same time, due to the larger mass of the sneutrinos, the energies of the photons are smaller, thus buried in the steep shoulder of the $\XI0{1}$ spectrum~\cite{Bartels:2012rg}. Therefore, the $\mathcal{P}_{-80,+30}$ combination
seems more promising. In this case, the sneutrino pair production cross section is an order of magnitude larger than the cross section of $\XI0{1}$ pair production, and of roughly the same size as the $\XI0{1}$ cross section in the other polarisation, c.f.\ Table~\ref{tab:xsect-ilc}. The price to pay is a background from SM neutrino pair production which
is about a factor of $5$ larger than in the $\mathcal{P}_{+80,-30}$ case~\cite{Bartels:2012ex}. This means that
the control of systematic uncertainties becomes even more important than in the classic $\XI0{1}$ case. However, due to
the lower photon energy endpoint, a large part of the photon energy spectrum will be signal free, which should
give an excellent possibility to constrain absolute normalisation uncertainties as well as shape uncertainties, arising e.g.\ 
due to the finite knowledge of the beam energy spectrum. Therefore, based on the experience from radiative neutralino production, we expect that
both cross-section and mass measurements are possible, but would need a quantitative study.

The visible decays of the $\sNu_{\tau}$ lead to a
rather unique final state with a $\tau$, a (hadronic) $W$ and large missing four-momentum. SM background from $WW \to 
\tau \nu_{\tau} q \bar{q}'$ as well as the main SUSY background from $\XIPM{1}\XIPM{1} \to
\stau{1} \nu_{\tau} \XI0{1} W$ can be reduced very effectively by choosing $\mathcal{P}_{+80,-30}$, since the 
polarisation dependence for the pure $s$-channel $\sNu_{\tau}$ is much weaker. Furthermore, both backgrounds
feature different kinematics from the signal, since the $\tau$ and the $W$ stem from different hemispheres, and not from
the same as in the signal case, which can be exploited efficiently once the masses of the $\stau{1}$ and $\XIPM{1}$ have been measured. The SM $WW$ background can be further suppressed by exploiting the forward-backward asymmetry~\cite{Bechtle:2009em}. Once a sufficiently clean signal has been selected, the $\sNu_{\tau}$ mass can
be determined from the endpoints of the $W$ energy spectrum (c.f.\ Section~\ref{subsubsec:chipm1}), based on the
$\stau{1}$ mass measured in direct $\stau{1}$ pair production as described above. Similar to the ISR recoil case, this analysis seems feasible, but awaits a detailed simulation study for quantitative conclusions.

Thus, the currently most obvious way to access the $\sNu$ mass are the \XIPM{1} cascade decays discussed in Section~\ref{subsubsec:chipm1}.

\subsection{Analysis of sleptons in cascade-decays}

\begin{figure}[htb]
  \begin{center}
\subfigure[]{\includegraphics[width=0.4\linewidth]{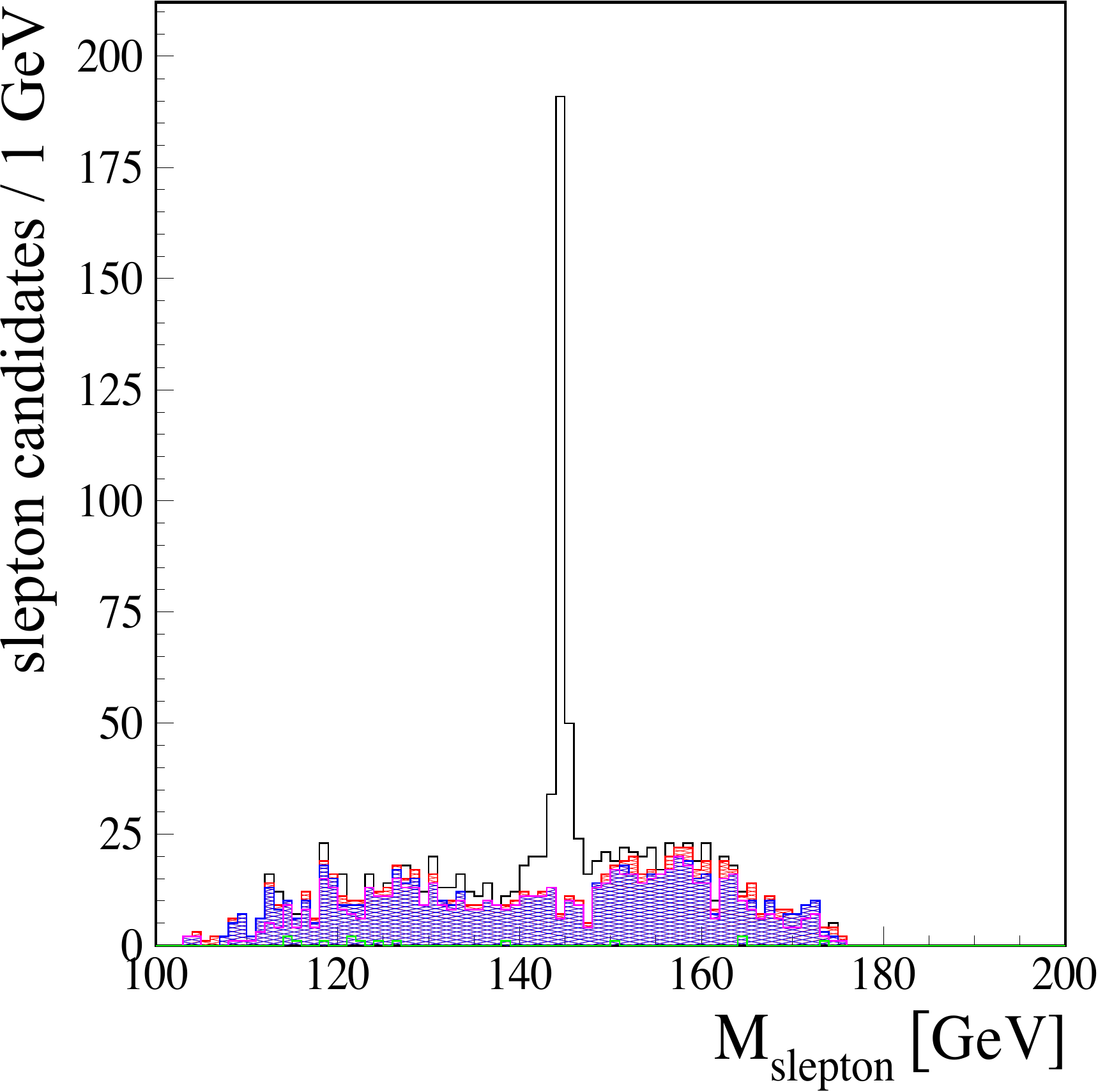}}
\hspace{1cm}
\subfigure[]{\includegraphics[width=0.4\linewidth]{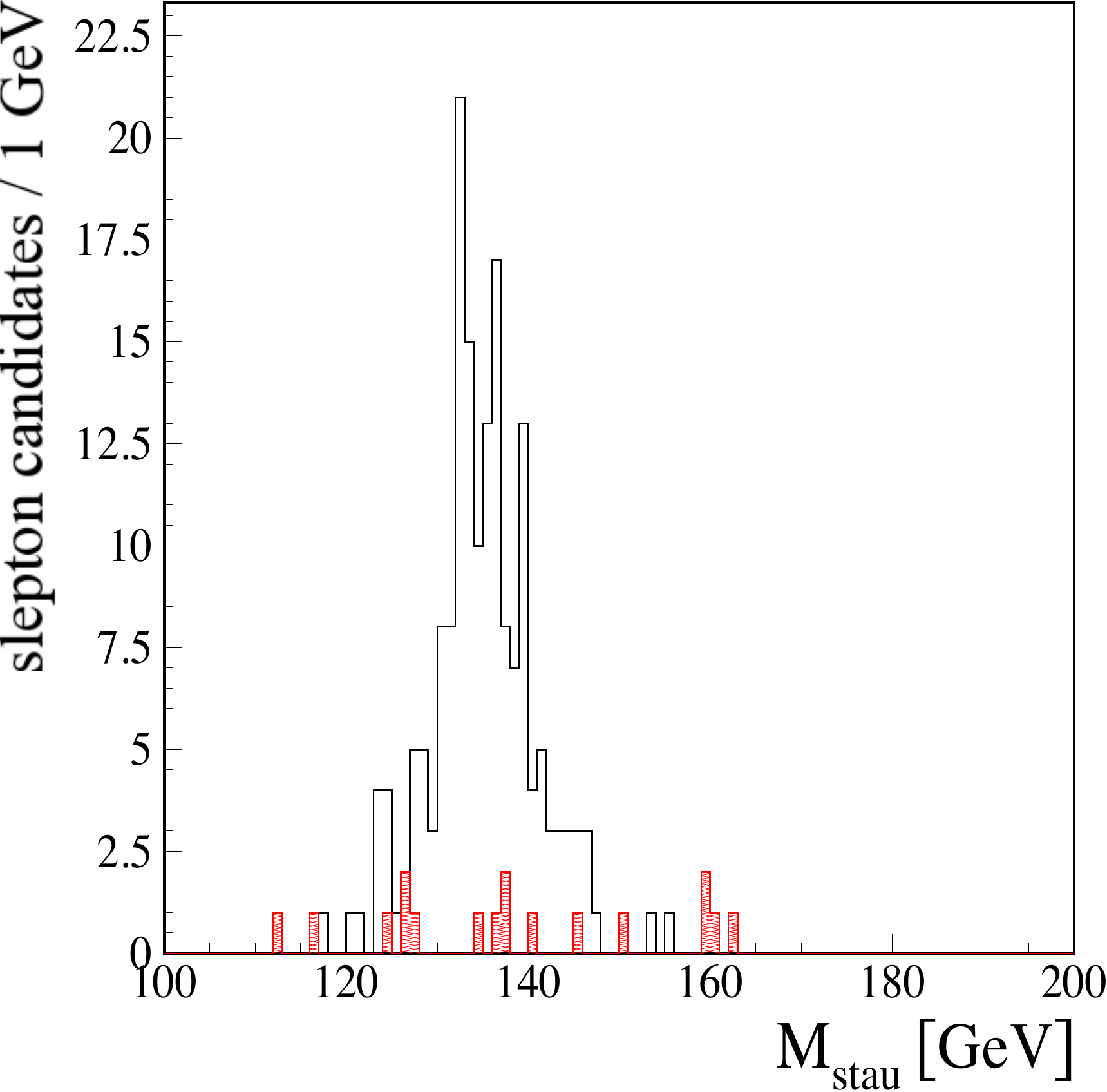}}
\end{center}
\caption{\label{fig:kinrec}Reconstruction of slepton masses from $\XI0{2}\XI0{2} \to \tilde{l} l   
\tilde{l} l$ in SPS1a, which has a very similar spectrum to our case. 
We show the reconstructed \sMuR mass (a) and the reconstructed \stau{1} mass (b). From~\cite{Berggren:2005gp}.}
\end{figure}

A particularly interesting channel for the determination of
slepton properties is $\eeto \XI0{2}\XI0{2}$ and the \XI0{2} decay to \sMuR$\mu$ (or 
equivalently to $\seR$e), even if the branching ratio is at the level of a few percent as in STCx.
These cascade decays can be fully kinematically constrained at the ILC,
and would promise to yield even lower uncertainties on the
\sMuR and \seR masses than the threshold scans,
of the order of 25\MeV. This is estimated on an earlier study in a scenario with about twice as 
large branching ratios for the considered decay mode, where a precision of 
$10\MeV$~\cite{Berggren:2005gp} was found. The corresponding distribution of the reconstructed 
\sMuR mass is shown in  Fig.~\ref{fig:kinrec}(a), including
all SM and SUSY backgrounds.

Also for the study of the $\stau{1}$, the channel \eeto\XI0{2}\XI0{2} and the \XI0{2} decay to \stau{1}$\tau$ 
is quite powerful. This decay channel has a much larger
branching-ratio than the electron and muon channels, but the presence of the neutrinos in the decay of the $\tau$
prevent the exact kinematic reconstruction of the events.
However, approximate reconstruction is still possible, 
as shown in Fig.~\ref{fig:kinrec}(b). Potentially, this method could yield comparable 
results to a threshold scan. 

%% file: Interpretation.tex
\section{LHC-ILC Interplay} \label{sec:interpretation}

In this section, we employ the simulation studies based on the STCx scenarios to illustrate 
how discoveries and measurements at the LHC and a future linear collider like the ILC could
work together to gain as precise knowledge as possible about the origin of the beyond-the-Standard-Model (BSM) observations.

\subsection{Discoveries at the LHC}
\label{subsec:LHCdisc}
By construction, several sectors of the STCx spectra offer
discovery opportunities at the LHC. For instance the left-handed selectron and smuon
could be seen early if their branching ratio for the direct decay $\slel \to l \XN{1}$ is large, c.f.\ Section~\ref{subsec:massdecay}.
The right-handed sleptons, on the other hand, are much harder to detect at the LHC, due to the -- for the same mass values -- lower production rates, and in particular in the STCx points due to their smaller
mass difference, which leads to softer leptons in the visible final state. The $\sTau_1$ and $\sTau_2$
will be very hard to observe.

\begin{figure}[htb!]
\centering
\subfigure[]{\includegraphics[width=0.49\linewidth]{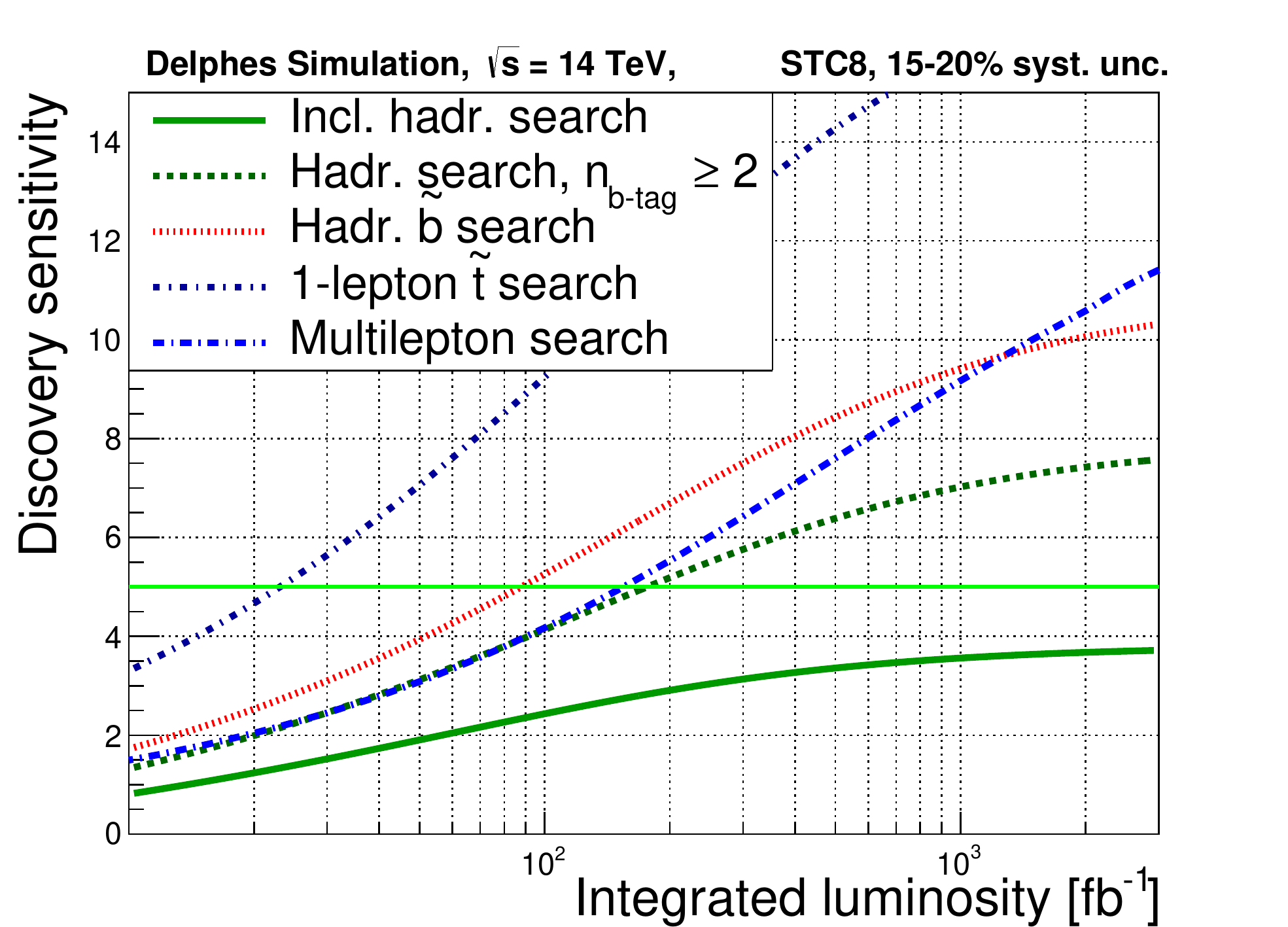}}
\subfigure[]{\includegraphics[width=0.49\linewidth]{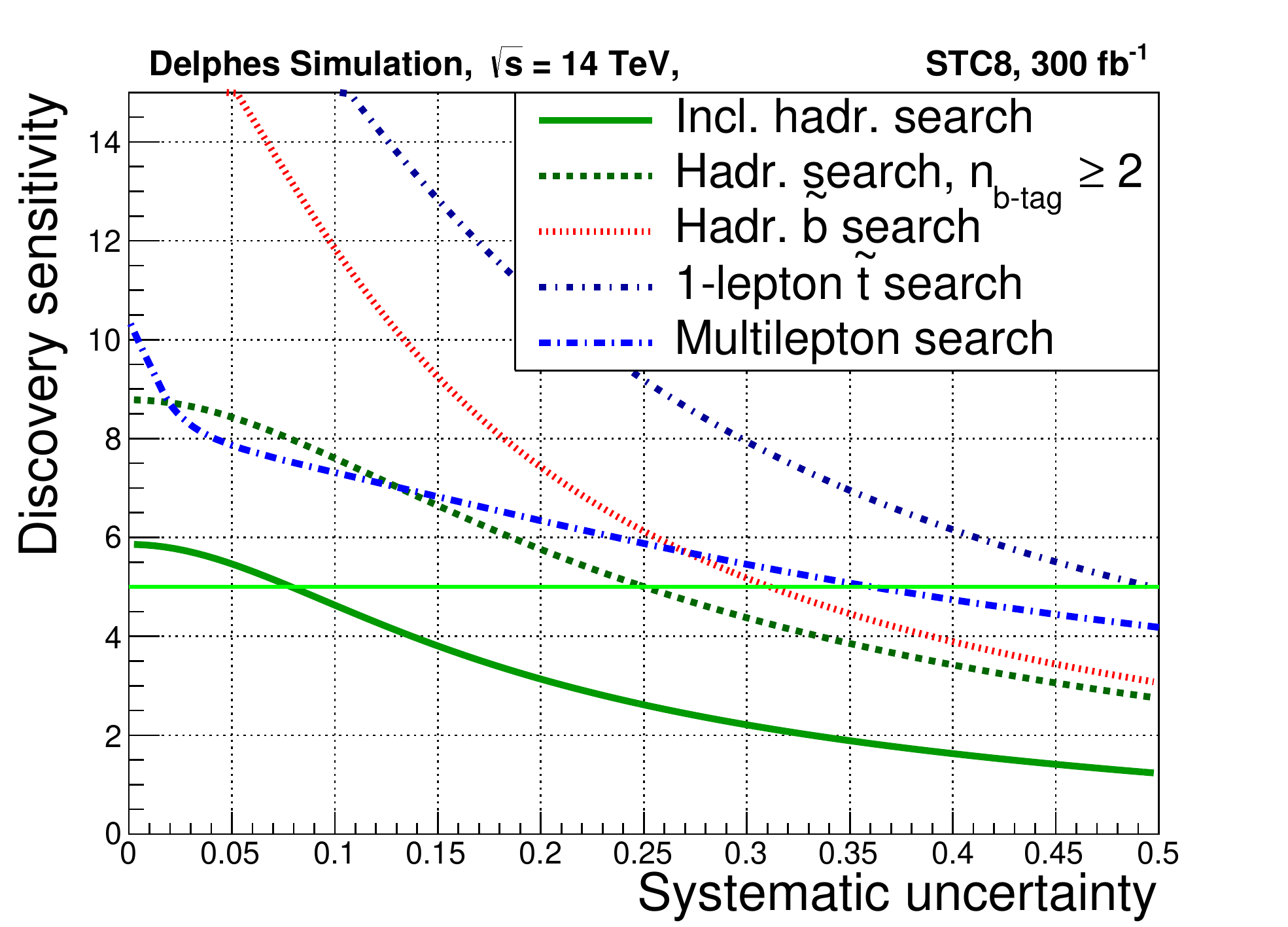}}\\
\subfigure[]{\includegraphics[width=0.49\linewidth]{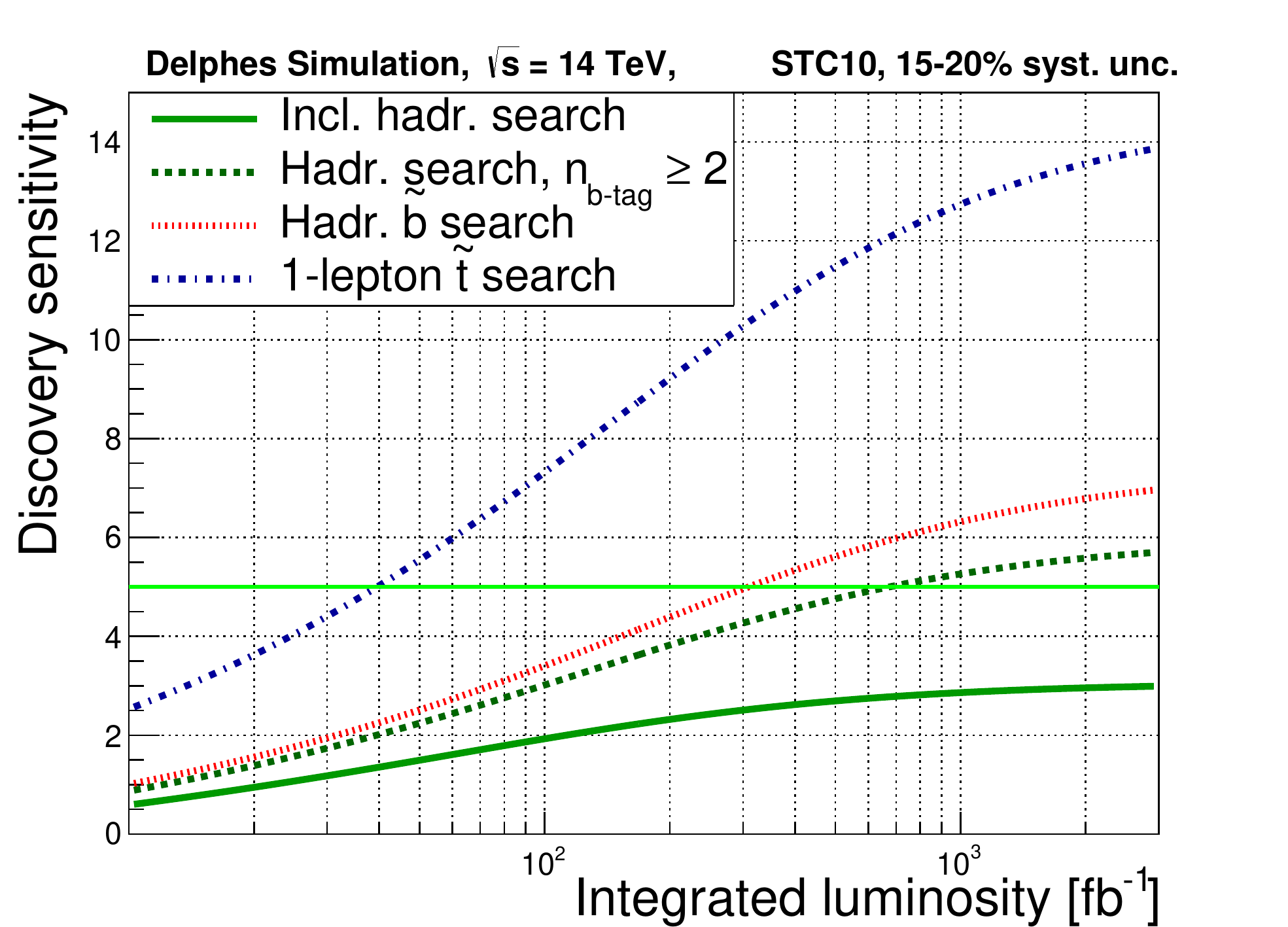}}
\subfigure[]{\includegraphics[width=0.49\linewidth]{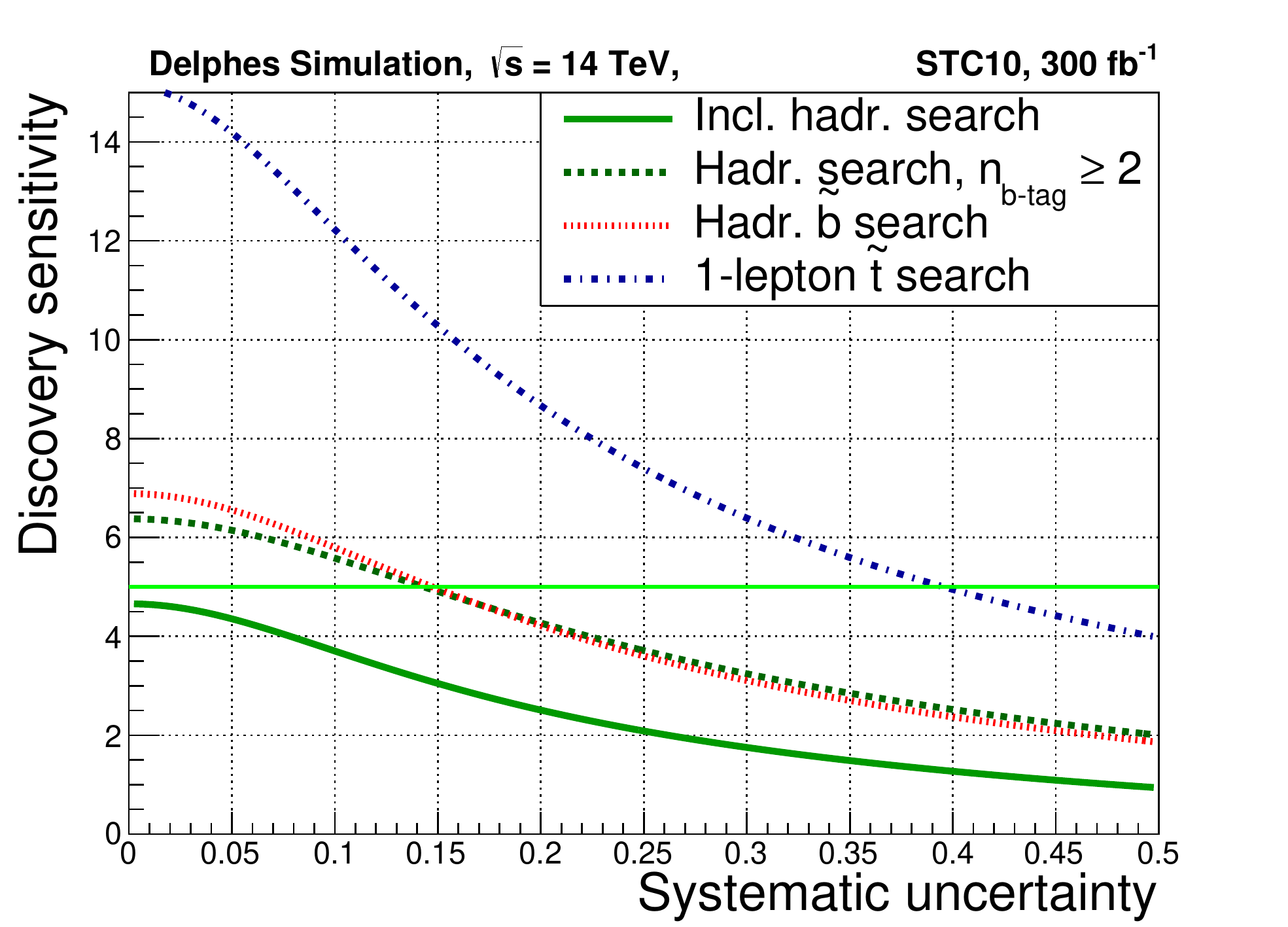}}
\caption{LHC searches: Expected significance as a function of luminosity and of the systematic uncertainty, respectively, for the two signal scenarios STC8 (a,b) and STC10 (c,d). For plot (a) and (c) we assume a
systematic uncertainty of 15\,\% for the bottom- and top-squark search, and 20\,\% for the other searches. As 
the spectrum of the electroweak sparticles is the same for STC8 and STC10, we show the multilepton analysis
which focuses on the electroweakinos only for STC8.}
\label{fig:LHCsignificances}
\end{figure}

Figure~\ref{fig:LHCsignificances} summarises the significances obtained by the analyses presented
in Section~\ref{sec:LHC} as a function of the integrated luminosity and as a function of the
assumed systematic uncertainty on the background estimate.
The first significant deviation of the coloured sector of the STCx scenarios 
from the SM would appear in the single-lepton channel at the LHC, caused by a mix of top- and bottom-squark pair 
production, as well as squark-gluino production. Assuming a systematic uncertainty of 15\,\% on the background 
prediction, a $5\,\sigma$ deviation could be visible with
less than $50$\,\fbinv. The amount of integrated luminosity required rises strongly
with the masses of the top and bottom squarks, as can be seen by comparing Fig.~\ref{fig:LHCsignificances}(a) 
with the lower top and bottom-squark masses in model STC8 to Fig.~\ref{fig:LHCsignificances}(c), where the same
is shown for the model STC10 with ~200\GeV higher top- and bottom-squark masses. From the existing LHC analyses,
one would not necessarily expect a specialised search for direct top-squark production to be one of the 
first discovery channels. However, the limits on direct top-squark production are usually determined 
for one or maybe two decay channels, assuming that direct top-squark pair production is the only production 
mechanism of SUSY particles, while in the full models there are several production and decay channels 
contributing to the discovered signal. In our case, we see also decays of top and bottom squarks to final 
states with higher-mass neutralinos and charginos, and contributions from gluino-pair production. If, however,
the gluino mass would be higher, the direct top- and bottom-squark production would be the main contribution,
leading to a lower significance and later discovery.

In case of STC8, the dedicated search for bottom squarks decaying directly into a b-quark 
jet and the LSP would report a $5\,\sigma$
deviation from the SM with about $90$\,\fbinv (assuming a systematic uncertainty
of $15$\,\%). With the higher bottom-squark mass of STC10,
however, a substantially larger amount of integrated luminosity, 300\fbinv, would
be required. This is due to the fact that the bottom-squark selection has a much higher purity, 
in particular it is less sensitive to gluino-squark production than the top-squark
search. Thus the difference in integrated luminosity required to observe a signal in these
channels does not necessarily correlate with the masses of the targeted squarks.

After an integrated luminosity of $150$\,\fbinv, the multi-lepton channel would report
a $5\,\sigma$ deviation from the SM in either STC8 or STC10. Since b-tagging does not play a role in 
this analysis, it will be highly likely that this deviation is not caused by third-generation 
partners, but by lighter sparticles, i.e.\ electroweakinos and/or sleptons. The variation
of the excess over the numerous signal regions gives a hint that several different 
sparticles with sizeable mass differences contribute to the overall signal. For the expected significance shown in Fig.~\ref{fig:LHCsignificances} we combine the results of all search 
regions~\cite{0954-3899-28-10-313,Junk:1999kv,CMS-NOTE-2011-005}. 
The prediction for the non-prompt background is the most controversial part of this analysis, 
since the fake rate has to be assumed. With the fake rate increased by a factor of $2$, the
discovery would require an integrated luminosity of
$200$\,\fbinv instead of $150$\,\fbinv. 

A further hint towards the existence of rather light third-generation partners 
would come a little later from the inclusive hadronic search: While this
channel would remain at the $2$ to $3\,\sigma$-level before the High-Luminosity LHC when no b-tagging
is applied, this could be enhanced in case of STC8 to the $5\,\sigma$-level after less than 200\,\fbinv by requiring two
jets to be identified as b-quark jets, as shown in Fig~\ref{fig:LHCsignificances}. In STC10, about 600\,\fbinv would
be needed for a discovery in this channel.
In addition, it will be extremely difficult to separate the contributions from direct top- and bottom-squark production 
and gluino decays via the third-generation squarks in this analysis. 

All search channels show strong dependency on systematical uncertainties at different pile-up and higher luminosity scenarios, as shown in Fig.~\ref{fig:LHCsignificances}(b) and (d). In particular
the single-lepton stop (Section~\ref{sec:LHCstop}) and the hadronic searches would profit significantly from an improved 
understanding of the background level beyond the $15-20$\,\% assumed as default here.
The hadronic search requiring two $b$-tagged jets (Section~\ref{subsec:LHCsbot}) and the hadronic bottom-squark search 
(Section~\ref{sec:LHChadronic}) are most susceptible to systematic uncertainties. For STC8-like masses, they would
miss the $5\,\sigma$-level with 300\fbinv if the systematic uncertainty on the background was larger than $25$\,\% 
and $30$\,\%, respectively, while in case of STC10 a $5\,\sigma$ discovery is only possible if the backgrounds are controlled 
to better than $15$\,\%. The single-lepton search is, not surprisingly, more robust and would still reach
$5\,\sigma$ with uncertainties of $50$\,\% in case of STC8 and $40$\,\% in case of STC10. The sensitivity 
of the multilepton search depends on the significance the 45 different search bins. With decreasing systematic 
uncertainty, the contribution of the low \MT signal regions gains significance, leading a steeper rise in the sensitivity can be seen for very small systematic uncertainties, as shown in Fig.~\ref{fig:LHCsignificances}(c). However, we expect 
systematic uncertainties of the order of 20\,\% for this analysis.

\subsection{Signal characterisation at the LHC}
\label{subsec:LHCchar}

Once a clear deviation from the SM has been discovered, the immediate question will
be the origin of this deviation. Given the fact that at this stage neither the masses 
of the produced sparticles, nor their decay chains and branching ratios are known, 
it seems highly unlikely that the contributing production modes could be identified 
at that stage.


The most promising candidates for isolation of a sufficiently pure sample of an individual
decay chain are the direct slepton decays $\slel \to l \XN{1}$ and the direct bottom-squark decay
$\sBot_1 \to b \ninoone$. These could then be used to obtain information on the masses of the produced
sparticles via kinematic edges.

For instance one can use the so-called boost-corrected contransverse mass (\mCT)~\cite{mCT1,mCT2},
which is defined as 
\begin{eqnarray*}
\mCT^{2}(j_1, j_2) &=& [\et(j_1)+\et(j_2)]^2-[{\ptvec}(j_1)-{\ptvec}(j_2)]^{2} \\ \nonumber
   & = & 2\pt(j_1)\pt(j_2) (1+\cos\Delta\phi(j_1,j_2))  \;\; ,\nonumber
\end{eqnarray*}
can be used for extracting information from events in which
two heavy particles decay into a visible part $j_i$ and missing energy. This mass is
invariant under equal and opposite boosts of the parent particles in the transverse
plane. For parent particles produced with small transverse boosts, \mCT is bound from 
above by an analytic combination of particle masses.

In the case of the direct bottom-squark decay, the distribution of the
contransverse mass of the two jets,  $\mCT(j_1,j_2)$, is expected to show an edge at 
\begin{equation}
\mCT^{\mathrm {edge}} = (m_{\sBot_1}^{2}-m_{\ninoone}^2)/m_{\sBot_1} \;\;, \label{eq:edge}
\end{equation}
which equals $780$\GeV for STC8, and $1000$\GeV for STC10.

\begin{figure}[!htb]
  \begin{center}
  \subfigure[]{\includegraphics[width=0.45\linewidth]{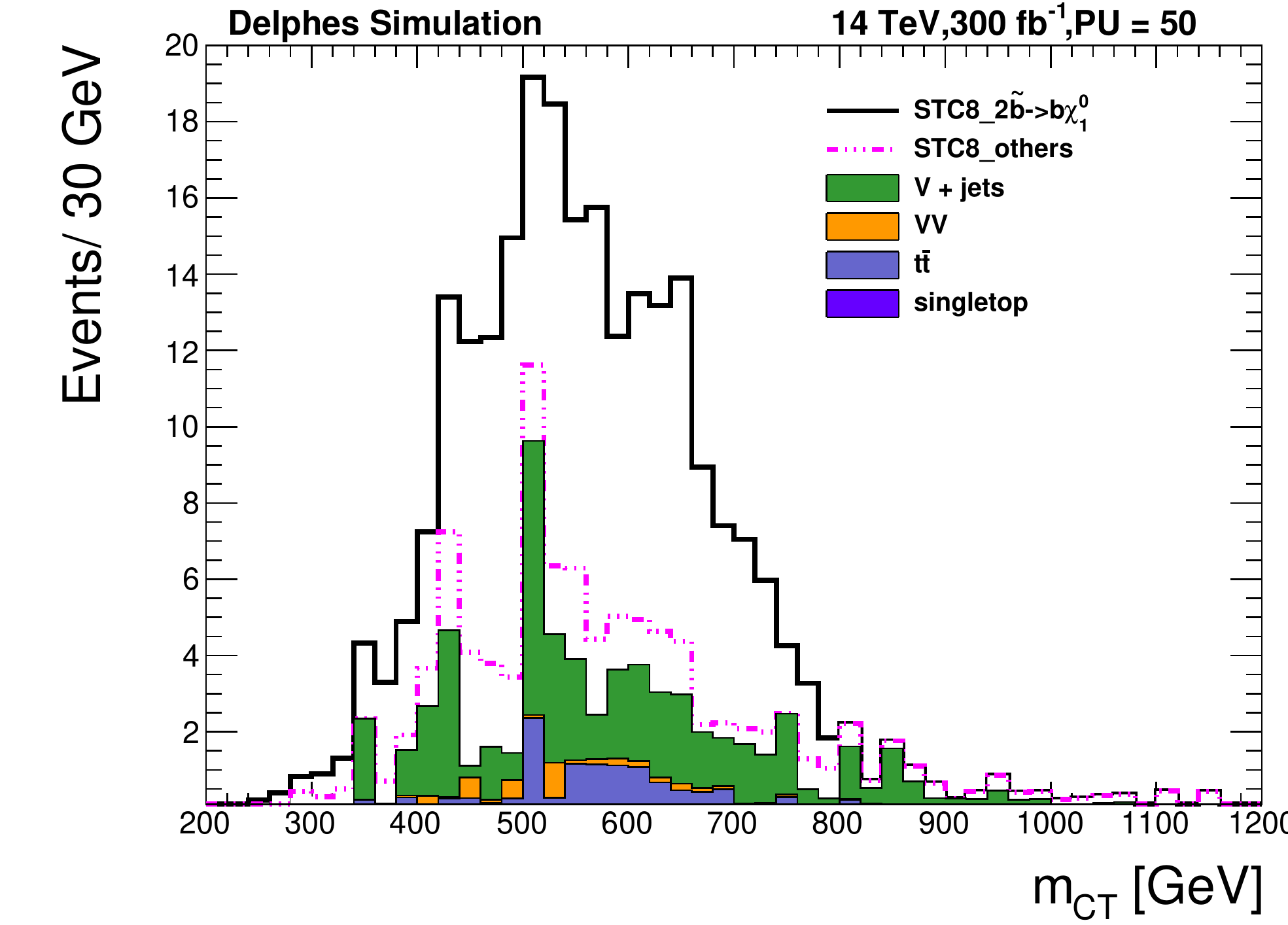}} 
  \subfigure[]{\includegraphics[width=0.45\linewidth]{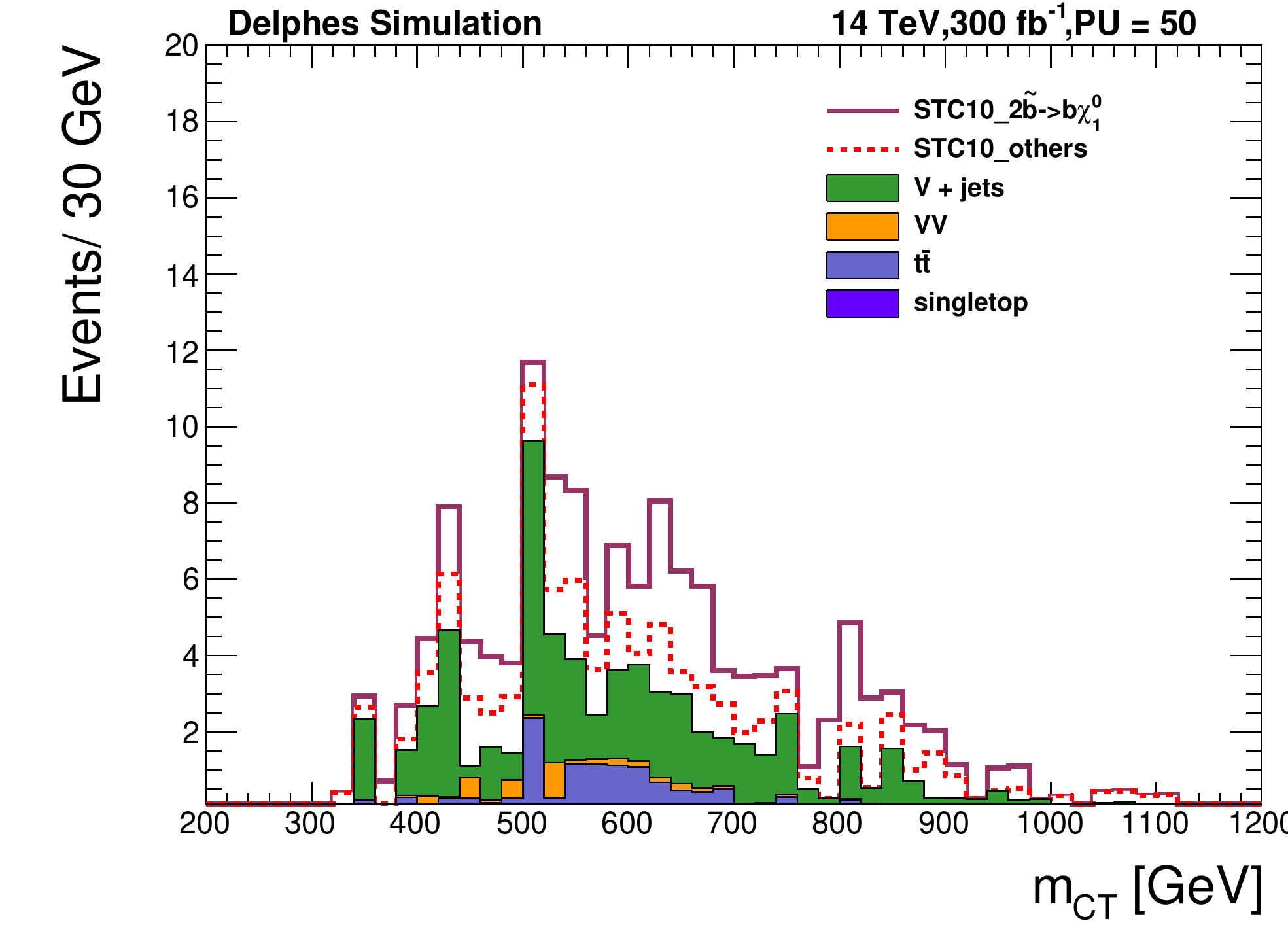}} 
  \end{center}
\caption{LHC bottom-squark search: Comparison of the \mCT distribution for signal and background
  from SM and other SUSY processes for STC8 (a) and STC10 (b). All SUSY processes from the STCx models
  are considered, with the signal being stacked on top of the stacked background.
}
\label{fig:MCT}
\end{figure}

The expected distributions are shown in Fig.~\ref{fig:MCT} for an integrated luminosity of $300$\,\fbinv 
after the selection presented in Section~\ref{subsec:LHCsbot}. With this amount of integrated luminosity, 
the edge position in model STC8 could be determined, while for STC10 this would be possible only with the 
luminosity that can be reached at the HL-LHC, $3000\fbinv$.

The edge is smeared out by the jet energy resolution, which has still to be determined for the $14\TeV$ data with 
the corresponding jet reconstruction at higher pileup. Assuming a resolution of 10\,\% for $\pt(j)=300-500\GeV$ jets 
(which is a typical transverse momentum for jets from bottom-squark decays, as shown in Fig.~\ref{fig:sbottom}),  
motivated by a corresponding 7\TeV measurement~\cite{JME-10-014}, we can deduce a resolution of
about $15$\,\% for the \mCT variable, corresponding to about $120\GeV$ in STC8 and $150\GeV$ in STC10.
The edge position gets further diluted or even biased by the SM background as well as by the SUSY background, 
in combination with fluctuations due to low statistics. While the situation could 
be partly improved by removing the SM background based on prediction from simulation, which is expected to be well understood with 300\fbinv of data, the SUSY background remains as a priori unknown and can  
distort the determined endpoint as shown in the following. 

In order to determine the mass edge, one could e.g.\ exploit a typical edge-finder like the so-called 
edge-to-bump method~\cite{Curtin:2011ng}. To test the capability of this method for the model STC8, 
we generate 1000 times pseudo-data from 
the distribution shown in Fig.~\ref{fig:MCT}(a), and determine the edge position with the edge-to-bump method.
The mean of the fitted value is at 832\GeV with an RMS of 114\GeV,
which is compatible with the expected value of 780\GeV. This RMS is expected from the resolution of the \mCT 
variable. With ten times more statistics from the HL-LHC, this uncertainty could be reduced roughly by a factor
of three.
Without further knowledge about the \ninoone, the determined endpoint can be interpreted as 
lower limit on the bottom-squark mass using Equation~\ref{eq:edge}, assuming a mass-less neutralino,
of $832 \pm 114\GeV$. Higher masses of the neutralino would lead to higher bottom-squark masses.

Analogously, an edge in the contransverse mass spectrum would be expected from the decays of
$\sell$ and $\smul$ near $\mCT^{\mathrm {edge}}(l_1,l_2) = 175\GeV$. Without any information from ILC, this would yield a lower limit on the slepton mass.

\subsection{Discoveries at the ILC}
\label{subsec:discILC}
The first channel to manifest itself at the ILC depends on the assumed running scenario. If the ILC
starts out as a higgs factory at $\Ecms=250\GeV$, then 
$\eeto \ttau_1 \ttau_1$ and $\tz_1 \tz_1 \gamma$ would
be the first observable channels, while $\te_R$- and $\tmu_R$-pair production is just beyond reach. 
The measurement of the $\ttau$ mass, however, would be challenging close to threshold, 
since both upper and lower
edge of the $\tau$-lepton energy spectrum would be in the region affected by background from 
multi-peripheral two-photon processes.

\begin{figure}[htb]
  \begin{center}
\includegraphics[width=0.49\linewidth]{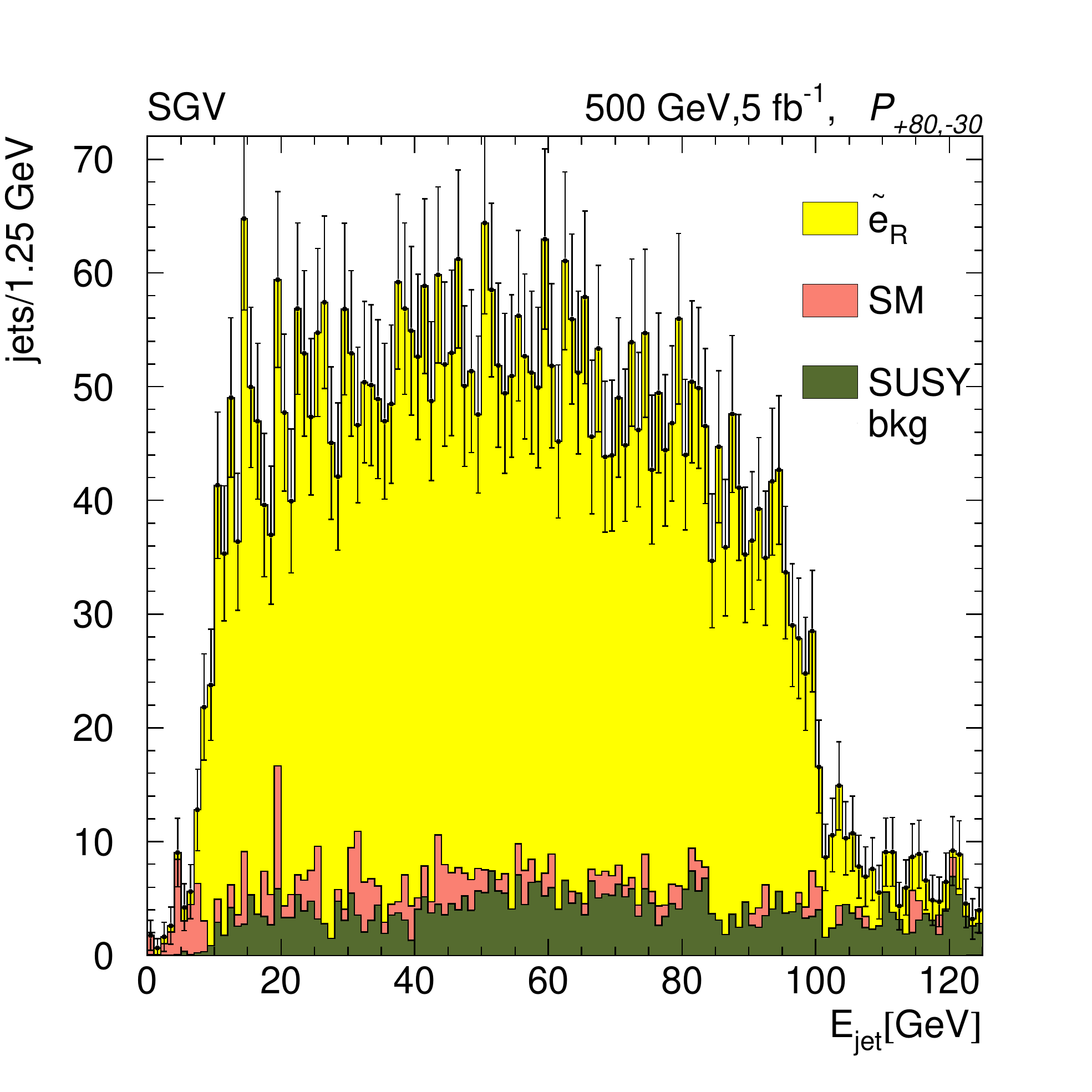}
 \end{center}
  \caption{\label{fig:selectrons} Momentum spectrum 
of events with $e^+e-$ and missing $4$-momentum. The assumed luminosity of
  $5\,$fb$^{-1}$ corresponds to one week data taking at design luminosity.} 
\end{figure}

On the other hand, the LSP mass and pair-production cross section could be measured at least
with a few percent precision from the energy (or recoil mass) spectrum of the accompanying 
initial state radiation photons as discussed in Section~\ref{subsubsec:WIMP}. Although that study 
has been performed at a higher centre-of-mass energy, the quality of the mass determination is 
expected to be comparable at $\Ecms=250\GeV$, since the photon energy spectrum less 
spread out while the size of the cross section is similar. 
Since the neutralino pair-production is dominated by $t$-channel selectron exchange,
the mass of the lighter selectron and its helicity can be determined from the measurement of the 
polarised cross sections. However, this requires a sizeable amount of integrated luminosity collected
with right-handed electrons and left-handed positrons in order to control the SM neutrino background.

In the recently published running scenario~\cite{paramgroup}, the ILC starts up in its TDR baseline configuration 
at $\Ecms=500\GeV$. In this case, the picture changes drastically with already a modest amount of data,
since very first evidence for BSM at the ILC would be observed after a few 
days of running from selectron pair production. Figure~\ref{fig:selectrons} 
shows the selectron signal expected with only $5$\fbinv, allowing already a first determination of
both the selectron and LSP masses according to the technique described in Section~\ref{sec:ILC}. 
Since the lighter set of sleptons must have
a $100$\,\% branching ratio to lepton and LSP, their production cross sections can be measured unambiguously. 
In particular for the selectron and smuon case, where the mixing
is typically small, the coupling can be extracted once the mass is known, thus enabling
the first verification that the couplings of the exotic states equal those of their
SM partners -- a fundamental property of supersymmetry!

\begin{figure}[!htb]
  \begin{center}
\subfigure[]{\includegraphics[width=0.45\textwidth]{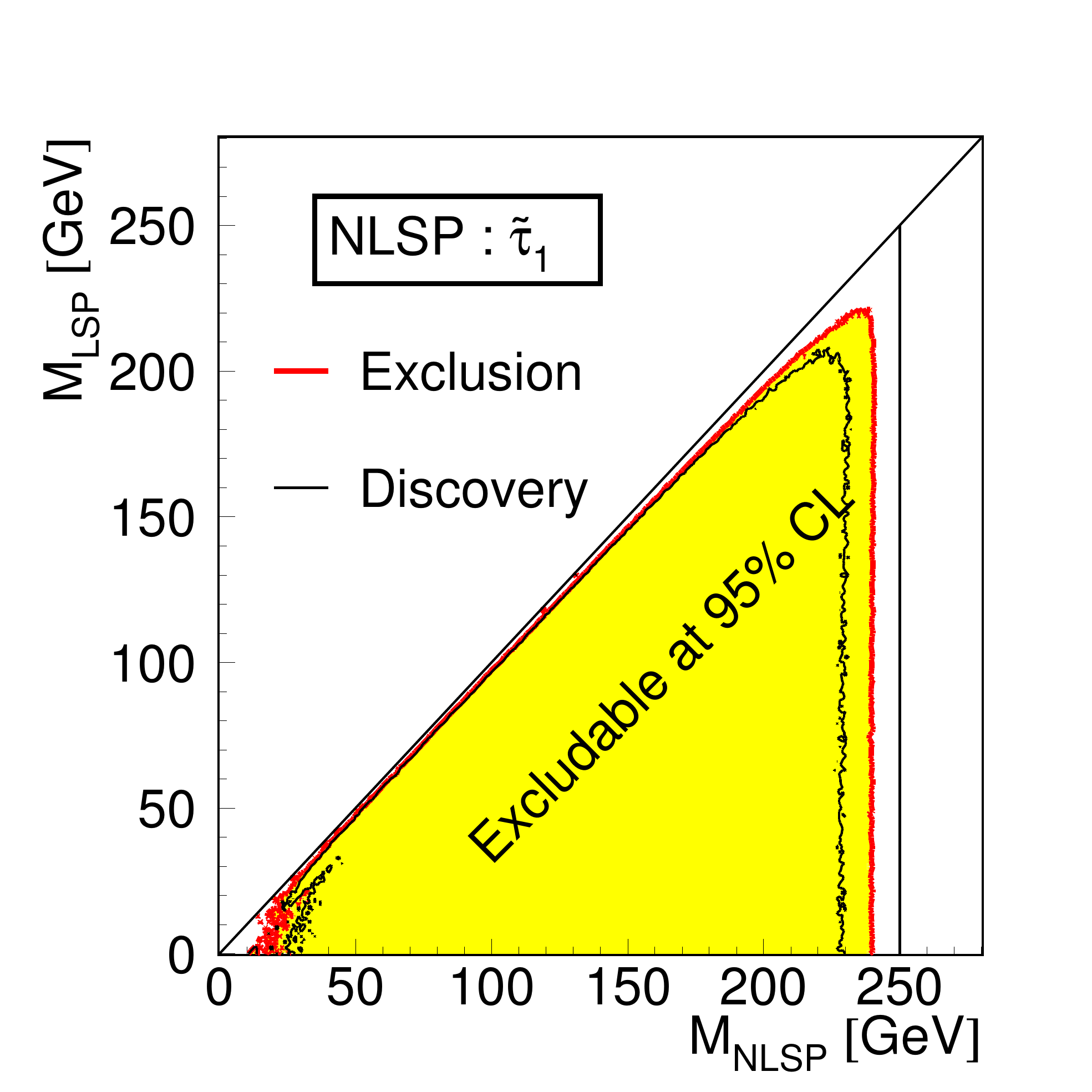}}
\subfigure[]{\includegraphics[width=0.45\textwidth]{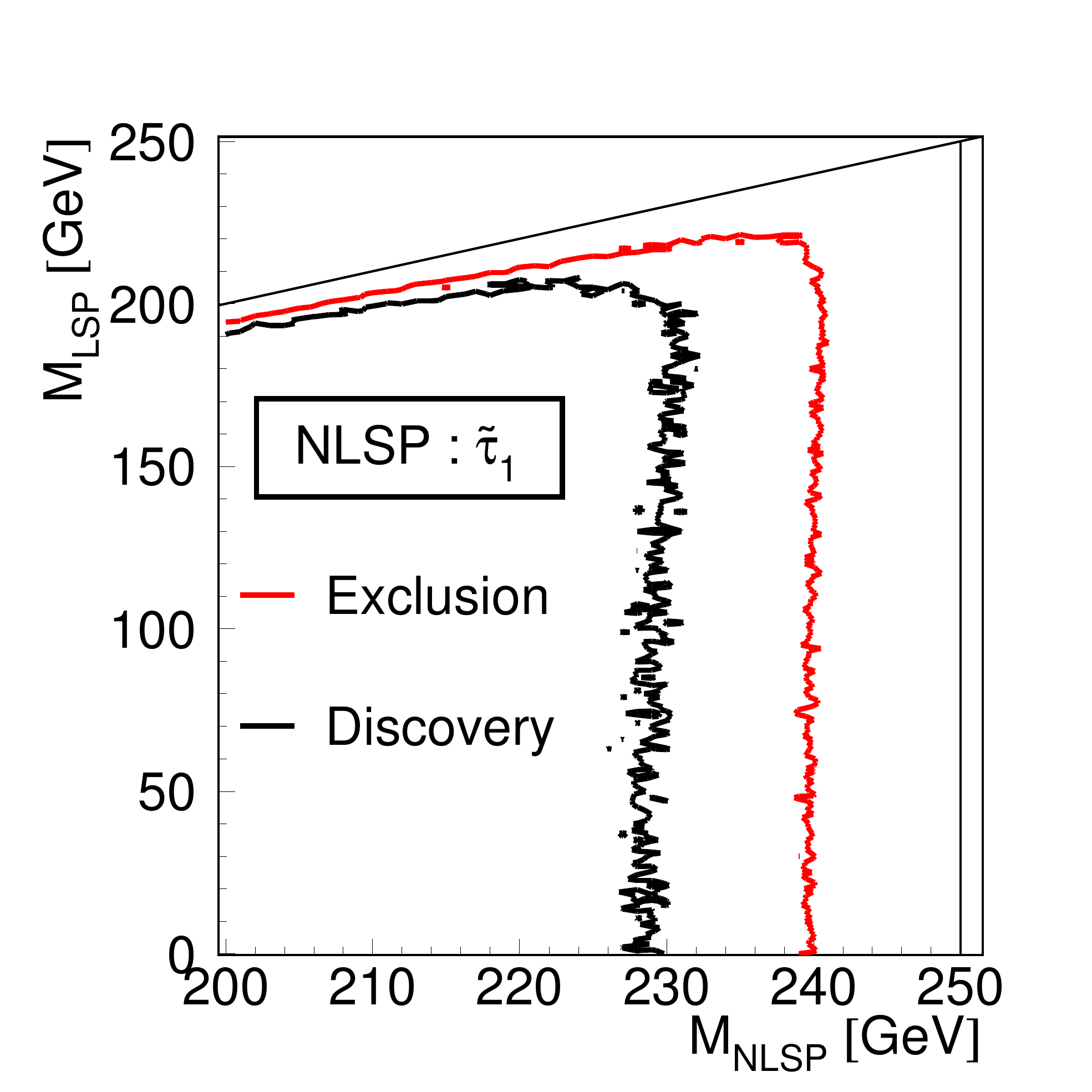}}
  \end{center}
  \caption{Discovery-reach for  a $\stau{1}$  NLSP after 
collecting 500 fb$^{-1}$
at $\Ecms$ = 500 GeV at the ILC. Left: full scale, Right: zoom to last few GeV before the
kinematic limit. From \cite{Berggren:2013vna}.
}
\label{fig:stauexcl}
\end{figure}

Beyond the specific case of STCx, any sparticle lighter than half the centre-of-mass energy of an
$e^+e^-$ collider will be observable.
If furthermore, like in the STCx models, $R$-parity is conserved so that the LSP is stable, the NLSP
will have a 100\% BR to the LSP and the corresponding SM particle.
This means that the search for NLSP pair production at the ILC can be seen as a 
loop-hole free search for SUSY \cite{Berggren:2013vna}.
The case of a general STC-like model, where the NLSP is the $\stau{1}$, 
is illustrated in Fig.~\ref{fig:stauexcl}, showing the exclusion and discovery reach for the 
NLSP at the ILC is shown
in the plane of the only free parameters, i.e.\ $\MXN{0}$ and $M_{\mathrm{NLSP}}$.
It should be pointed out that the case with the NLSP being the $\stau{1}$ is one of the most experimentally
difficult ones, due to the need to detect $\tau$-leptons in the final state.

\subsection{Signal characterisation at the ILC}
\label{subsec:addILC}
According to the currently favoured running scenario for the ILC~\cite{paramgroup}, $500$\,\fbinv would be collected
at $\Ecms=500\GeV$ within the first $4$ years of operation. For these data, a sharing
between the four possible configurations with different polarisation signs is foreseen:
40 \% of the data would be collected at each of the configurations $\mathcal{P}_{-80,+30}$ and
$\mathcal{P}_{+80,-30}$, while 10 \% would be collected at each of  the $\mathcal{P}_{-80,-30}$ and
$\mathcal{P}_{+80,+30}$ ones.
This would allow a first assessment of all SUSY particles with
masses below $250\GeV$, including their masses and mixings, with a precision typically a factor $1.6$ worse
than then values quoted in section~\ref{sec:ILC}.

After the initial run at $\Ecms=500\GeV$, it is foreseen the lower the centre-of-mass energy
to scan the top-pair production threshold and to run near the Zh threshold for a high-precision
determination of the higgs boson mass and its coupling to the Z boson. These runs would also be of
high interest for SUSY spectroscopy: 

Running near the Zh threshold could include a scan of the
threshold for pair production of the lighter sleptons as shown in Fig.~\ref{fig:slept_thresh},
which provides mass measurements of the $\sel{R}$ and $\smu{R}$ with precisions of 190\MeV and
220\MeV, respectively. The shape of the threshold as well as angular distributions identify
the produced partners of electrons and muon as scalar particles~\cite{Freitas:2004re}.

At or slightly above the \ttbar threshold, the cross
 section for $\sTau_1 \sTau_2$ mixed production could be measured with high precision due to the absence of
the $\sTau_2$-pair production background which limits the purity of this measurement at $\Ecms=500\GeV$. 
This provides an interesting possibility to determine the $\sTau$ mixing and $\tan{\beta}$. Together with the 
masses of the $\stau{1}$ and the $\XI0{1}$ and the $\XI0{1}$ mixing determined at $\Ecms=500\GeV$ (c.f.\ Section~\ref{subsubsec:staus} and~\ref{subsubsec:selsmu}), these are important inputs for the prediction of the 
dark matter relic density within the MSSM, since in the STCx scenarios LSP pair annihilation and 
$\stau{1}$-coannihilation contribute about equally to the cosmic dark matter annihilation. 
By comparing the predicted value for the relic density to the cosmologically
observed one, the $\XI0{1}$ can be either identified as the sole constituent of 
dark matter, or as being responsible for only a fraction of the observed relic density. This
has been explicitly demonstrated in a SUSY scenario with an electroweak sector very similar to the
STCx cases~\cite{Bechtle:2009ty}.

After a luminosity upgrade, the integrated luminosity at 500\GeV is then foreseen to be increased
to 4\abinv, with the same polarisation sharing as before, enabling a full BSM precision program, which 
reaches the permille-level for many observables. While the currently assumed running scenario is 
based on guaranteed measurements only, it will be adjusted to future discoveries by scheduling further 
threshold scans as well as runs at dedicated energies not far above important thresholds. Such special 
runs give e.g.\ important information for parameter determination in the neutralino sector, which in turn
enables predictions of the masses of the kinematically not accessible states~\cite{Bharucha:2012ya}.
The results of the initial run at $\Ecms=500\GeV$ will give decisive input to determining the detailed
running strategy.


\subsection{Combining ILC and LHC}
With ILC information on the lower part of the spectrum, the situation at the LHC
changes drastically. While the clear advantage of the LHC is the larger kinematic reach for 
the production of the more heavy sparticles, disentangling their signals is challenging due to the multitude of decay modes, most of them with small branching ratios. Thus the information from
the ILC might provide the key to access all the information contained in the LHC data, on one hand
by exploiting the knowledge of the lower parts of the decay chain in the data analysis, on the other by combining LHC and ILC results in global pMSSM fits.
This has been pointed out in several previous studies~\cite{Weiglein:2004hn, Blair:2005ui, Bechtle:2009ty}. Most of them, however, are based on scenarios with a light, sub-TeV, coloured sector,
which is by now excluded by LHC measurements. Thus we highlight here the specific possibilities
arising in the more challenging STCx benchmark points.

First of all, the determined edge position from bottom-squark
pair production could be turned into a measurement of the bottom-squark mass by adding the information 
from ILC about the \ninoone mass, which can be determined at the ILC with a negligible uncertainty.
Using Equation~\ref{eq:edge}, we find a bottom-squark mass of $843 \pm 115\GeV$, which is compatible 
with the true value of 795\GeV, as well as with the value obtained for a mass-less LSP. Even with HL-LHC
precision of $30-40$\GeV, the hypothesis of a mass-less LSP would yield a bottom-squark mass compatible
with the model value. However, already for an LSP mass of 200\GeV which could still be accessed at the ILC
with $\Ecms=500\GeV$ from radiative neutralino production (c.f.\ Section~\ref{subsubsec:WIMP}), the bias
due to the zero-mass hypothesis rises to $50$\GeV and thus surpasses the statistical uncertainty on $\mCT^{\mathrm {edge}}$.


In addition, we could use the measured bottom-squark mass to calculate the cross section for this process, 
and compare it to the cross
section times branching fraction determined above
from counting the events in the signal region. While the mass determination is independent of the spin of the newly discovered particle,
the measured event rate that can be related to the cross section (times branching fraction) depends on the
spin of the new particle and therefore help identify whether the newly discovered particle is actually a SUSY particle or not. 
Final conclusions could only be drawn if the branching fraction can be determined in addition. For the latter,
input from the ILC on the lower-mass sparticle spectrum might be exploited as well.

In the electroweakino sector, since now not only the masses, but also the mixings and decay modes of
the lighter electroweakinos are known, they enable a significant improvement of the analyses of LHC data. In case of the multi-lepton analysis, 
the signal contribution from the lighter
electroweakinos can be subtracted, giving access to the heavier electroweakinos not accessible at
the ILC with $\Ecms=500\GeV$. 

One example is given in the following. We want to study the decay of a \chipmtwo $\rightarrow$ \chipmone Z $\rightarrow \ell^{\pm}$ $\sNu$ $\ell^+\ell^-$. Therefore, we select events with three leptons and a large amount of missing transverse energy. The opposite-sign same flavour lepton pair $\ell^+\ell^-$ covers the full information about the Z-boson and can be identified by requiring the invariant mass as close as possible to the Z-boson mass. In the following we refer to the remaining lepton $\ell^{\pm}$ from the \chipmone decay when mentioning the lepton, and assume that the masses of the \chipmone and $\sNu$ will be known with sufficient
precision from ILC measurements, as discussed in section~\ref{subsubsec:chipm1}. The invisible decay products of the $\sNu$ lead to signatures with sizeable missing transverse energy. Unfortunately this is not the only source of missing transverse energy, since the decay of the other initial sparticles will also produce at least one LSP, hiding the information about the momentum of the $\sNu$. In order to avoid this, we boost into the \chipmone rest frame, where the lepton has a fixed energy E' which can be calculated by requiring energy conservation:
\begin{equation}
 E'_{lep}=\frac{(\MXC1)^2+(m_{lep})^2-(\msnu)^2}{2 \MXC2} \approx 8.6\GeV \; ,
\end{equation}
with the masses known from ILC: \MXC1=206.1\GeV, \msnu=197.3\GeV and $m_{lep}\approx 0\GeV$. The transformation from the lab frame into the \chipmone rest frame is done via the boost vector $\vec{\beta}$:
\begin{equation}
 E'_{lep}=\gamma E_{lep} - \gamma \vec{\beta} \cdot \vec{P_{lep}},
\end{equation}
with $\gamma=1/\sqrt{1-\vec{\beta}^2}$. This vector has three degrees of freedom, the magnitude and two angles. For the angles we assume that the boost is parallel to the lepton. This is a good approximation in the most cases, since the \chipmone decay does not add much to the boost of the lepton. With this assumption we can calculate $\gamma$:
\begin{equation}
 \gamma=\frac{1}{2} \left( \frac{E'_{lep}}{E_{lep}}+ \frac{E_{lep}}{E'_{lep}}\right ) \; ,
\end{equation}
and have the full information about the boost vector $\vec{\beta}$. As next step we boost the \chipmone back into the lab frame ($\vec{P}_{\chipmone}=+\gamma \vec{\beta}$ \MXC1). As last step we add the four-momenta of the Z boson and the \chipmone in order to calculate the invariant mass of \chipmtwo. 

\begin{figure}[!h]
\centering
\subfigure[]{\includegraphics[width=0.44\linewidth]{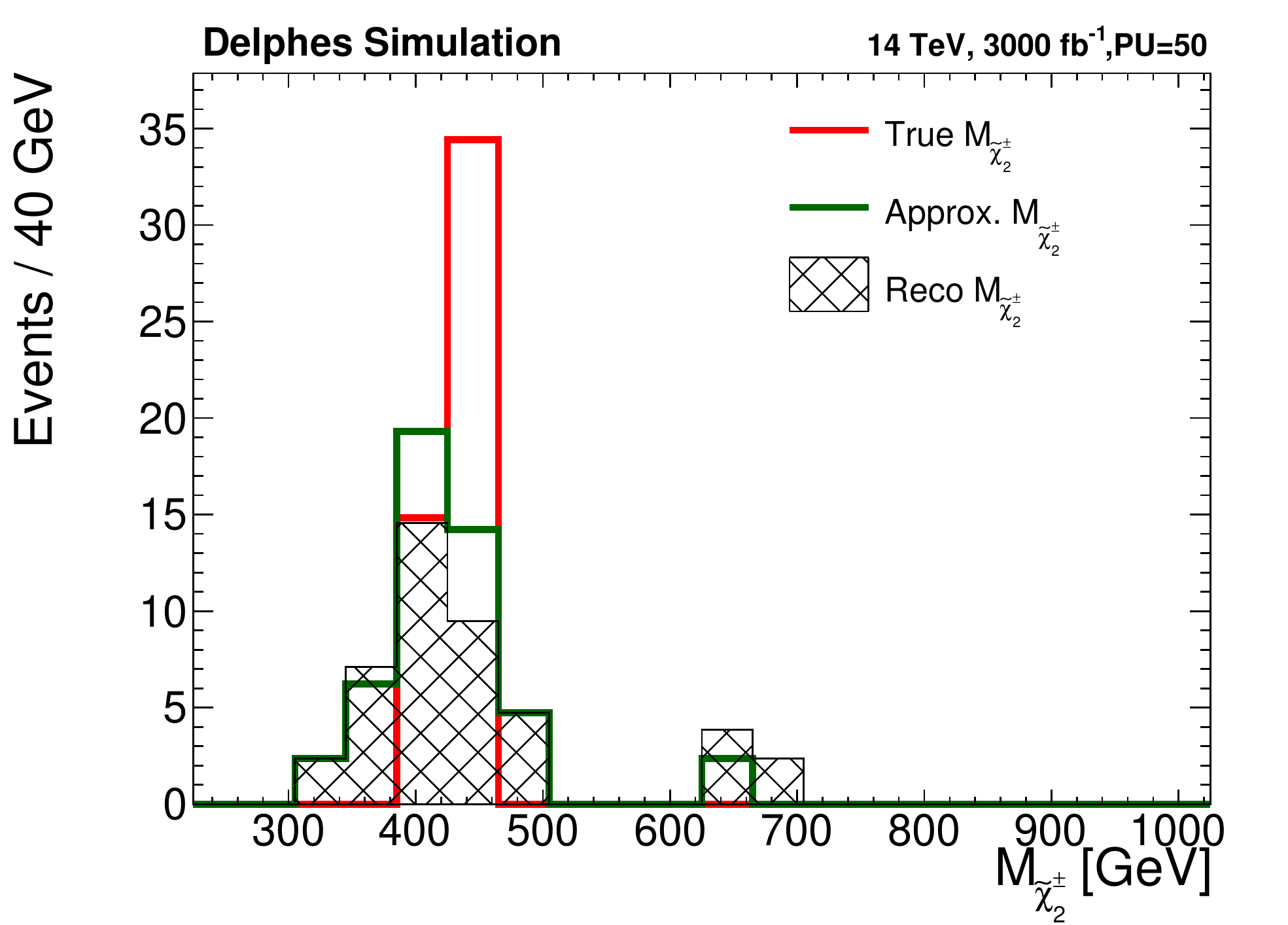} }
\subfigure[]{\includegraphics[width=0.44\linewidth]{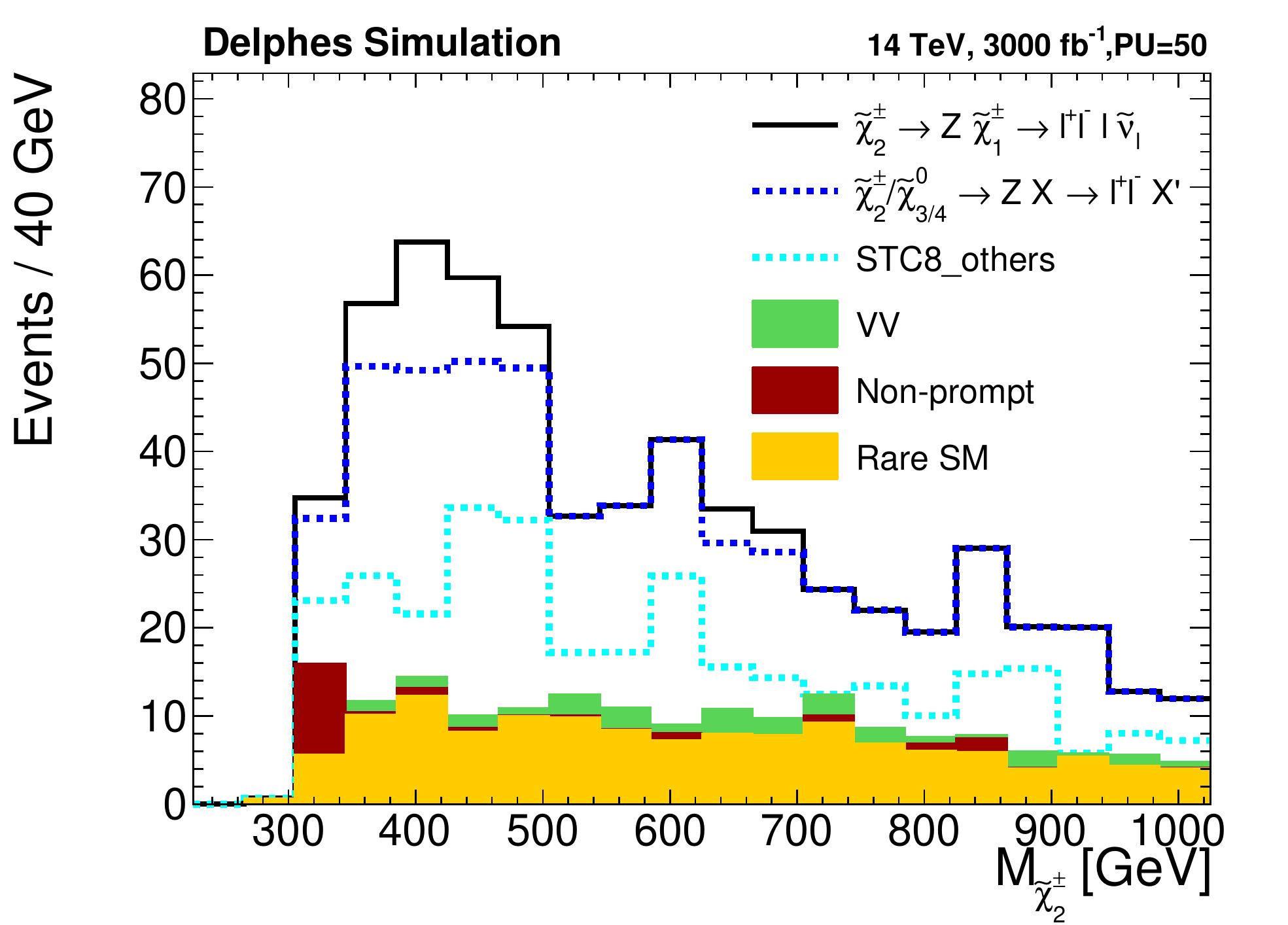} }\\
\caption{ \chipmtwo mass information at LHC. The left plot (a) shows the MC truth information of the invariant mass obtained with the procedure described in the text as well the reconstructed \MXC2 for those events. The right plot (b) shows the reconstructed \MXC2 for all events we would expect. There the SM and the SUSY events are stacked together.  
}
\label{fig:chargino2_mass}
\end{figure}

Figure~\ref{fig:chargino2_mass} shows the the \MXC2 by using the MC level information as well as reconstructed objects. 
The resolution gets smeared by two different effects: First, the $E'_{lep}$ is not fixed because of the natural width of 
the \MXC1 and \msnu. Second, the boost vector $\vec{\beta}$ is not perfect parallel to the lepton. Therefore, we expect to measure the \MXC2 to be $412 \pm 43\GeV$.
In order to suppress the SM background we require exactly three leptons
 ($\pt = 25/15/10\GeV$), one opposite sign same flavour lepton pair with an invariant mass between 84\GeV and 96\GeV, at 
least one b-tagged jet ($\pt> 100\GeV$ and $\eta < 2.4$), more than three jets ($\pt> 40\GeV$ and $\eta < 2.4$) where 
the leading jet has $\pt> 120\GeV$ and $\MET > 200\GeV$. With this selection our signal consists dominantly of \chipmtwo
 produced by stop decays, which has the best signal to background ratio. With a slightly softer selection we would also 
have sensitivity to direct \chipmtwo\Cnm{} production, which also can be accessible at HL-LHC. Per construction the \MXC2 
must be larger than $\sqrt{(\MXC1)^2+(M_{Z})^2} > 200\GeV$. As first observation we see much more SUSY events than we 
would expect from our targeted decay chain. Most of these events include heavy electroweakinos (\chipmtwo, \XN3, \XN4) 
decaying via Z bosons. The third lepton stems from the other decay products, which in most cases are $\tau$ leptons. 
This signature carry mass information about the heavy charginos and would be worth to be studied with the hadronic $\tau$ lepton 
final states. The other SUSY events passing this selection contains slepton decays, which has a kinematic edge signature 
close to the Z mass. This kind of background could be studied further if one requires the invariant mass of the opposite-sing 
same-flavour lepton pair exactly at the position of the edge. Overall the peak from our targeted decay chain around $420\GeV$ is visible.

The obtained knowledge of the \XIPM{2} and \XI0{{3,4}} masses in turn would add significantly
to the physics case of a $1$-TeV-upgrade of the ILC, or of an even higher energy $e^+e^-$ collider
like CLIC. Furthermore, the full knowledge of the
electroweakino masses, mixings and decay modes could provide the decisive information to isolate a signal from 
top-squark production, e.g.\ in it's largest decay mode $\sTop_1 \to b \XIPM{2}$,
enabling a determination of the $\sTop_1$ mass.

%% file: Summary.tex
\section{Conclusions} \label{sec:conclusion}

In this paper we discussed the complementarity and interplay of a proton-proton collider, the LHC,
and and electron-positron collider, the ILC, in discovering new particles and in determining 
their properties. As example we used an
$R$-parity and $CP$ conserving supersymmetric model where the \stau{1} is the next-to-lightest
supersymmetric particle and has a small mass difference of about $10\GeV$ to the lightest 
supersymmetric particle, the LSP, which is the lightest neutralino \ninoone. In such a scenario, 
$\stau{1}$-coannihilation allows for a sufficiently small dark matter relic density. All sleptons and 
electroweakinos have masses below $500\GeV$, while the lightest coloured sparticles, the lighter
top and bottom squarks, have masses around $800$ or $1000\GeV$. All other coloured sparticles
are much heavier, up to $2\TeV$.

We showed that such a scenario can be easily discovered at the LHC running at $13/14\TeV$.
In particular, this is true for the heavier selectrons and smuons, the heavier electroweakinos, 
as well as the lighter top and bottom squarks, if their masses are not much higher than a \TeV.
Depending on the search channel, on the exact masses and on the achieved control of systematic 
uncertainties, the integrated luminosities to discover deviations from the SM expectation due 
to production of these sparticles range from $50$ and $1000\fbinv$. The earliest discovery
would come from the single-lepton stop search, followed by hadronic searches including \botq-tags.
Inclusive hadronic searches would require systematic uncertainties to be controlled better than
$10\%$ in order to achieve a $5\,\sigma$ discovery.

However, in most cases the observed deviations from the SM expectation cannot be attributed to
a single process, but result from a mixture of e.g.\ diverse electroweakino processes or, 
in case of the coloured sector, of stop, sbottom and gluino production. The best chances are to
identify a single process with sufficiently high purity, in order to learn something about the properties
of the produced sparticles originating from direct decays of the produced sparticle to its standard-model
partner and the LSP. For the investigated scenarios, the direct decay has a sizeable branching 
fraction in case of the heavier sleptons and the lighter bottom squark. Here, a kinematic
edge can be isolated e.g.\ in the contransverse mass distribution. Without further knowledge of 
the mass of the LSP, the position of this edge can be converted into a lower limit on the sparticle mass. 

As example, we analysed the case of $\sbone \to \botq \ninoone$, and found that the edge in the 
contransverse mass distribution for the model with bottom squark mass could be determined with an 
uncertainty of about $115\GeV$ with an integrated luminosity of $300\fbinv$ at $14\TeV$. Assuming that 
systematic uncertainties play only a minor role, this could be improved to about $30\GeV$ at the High-Luminosity
LHC. For higher bottom squark masses of about 1\TeV, the discovery is still possible, however the uncertainties
on the mass determination will be significantly larger due to strongly reduced number of signal events.

The ILC on the other hand would complement these spectacular discoveries at the LHC with a systematic precision analysis of the lower part of the spectrum, actually discovering some of the lighter states.
In this paper, we especially present up-to-date detector simulation studies of the slepton sector
at the ILC, both in the continuum and in threshold scans. In particular a sufficiently light selectron
will lead to a striking signal within a few weeks of ILC operation. Already after the $500\fbinv$ collected in the first $4$ years of ILC operation, this would allow a permille-level determination of the LSP mass, and
the masses of the right-handed sleptons, improving further with the $4\abinv$ foreseen for the full ILC program. In conjunction with threshold scans and operation with different beam polarisations, all sleptons and the lighter half of the electroweakino spectrum can be characterised with at least percent-level precision,
including masses and mixing angles. This would demonstrate that the observed new particles are indeed
of supersymmetric nature, and at least some of the parameters of the underlying SUSY model can be determined.

This information from the ILC then creates new opportunities to analyse the LHC data.
Obviously, a precise determination of the LSP mass from the ILC can be employed to turn
edge determinations into mass measurements. Even better, the detailed information from the
ILC can be used in the analysis of the data themselves to disentangle the contributions
of different production modes and to reconstruct quantities which are sensitive to masses of the
heavier sparticles with more complex decay chains. As an example, we illustrated this in case
of the \chipmtwo. With the knowledge of the masses of the \chipmone and the \sNu from the ILC,
the signal can be isolated from the electroweakino mix and its mass can be reconstructed on an
event-by-event basis with a resolution of about $50\GeV$ with an integrated luminosity of 
$300\fbinv$ at $14\TeV$, with corresponding improvements at the High-Luminosity LHC. 

Although we studied the capabilities and the interplay of LHC and ILC based on a specific example, 
many aspects are transferable to other scenarios which comprise new particles in the kinematic reach
both colliders. In this context, it should be noted that the LHC in many cases the sensitivity to
the lightest new physics states is significantly smaller than to some of the more heavier states,
and that some of the lighter states might even await explicit discovery at a lepton collider.
We finally conclude that the combination of LHC and ILC data could reveal significantly more 
information about the properties and the origin of new particles than the results from either 
collider alone.

%% file: Appendix.tex
\section{Appendix}

Cutflow tables and numbers of events in the signal regions for the different LHC searches are given in this appendix.
The cutflow and event yields of the SM backgrounds and the signal samples for several signal regions in the hadronic search are shown in Table~\ref{tab:had_app}.  

\begin{table} [!h]
\caption{LHC all-hadronic inclusive search: Background and signal event yields corresponding to 300/fb. The vector boson denoted by "V" refers to W, Z and $\gamma$} 
\centering 
\begin{tabular}{l | r r r r | r | r r} 
\hline \hline
\textbf{Cutflow} & \textbf{\ttbar+jets} & \textbf{V+jets} & \textbf{VV+jets} & \textbf{Top+jets} & \textbf{Total SM} &\textbf{STC8} & \textbf{STC10} \\ 
\hline
Preselection & 3.93$\times10^7$ & 3.94$\times10^8$ & 4.5$\times10^6$ & 5.7$\times10^6$ & 4.43$\times10^8$ &27695 & 11510\\
HT $>$ 1000 GeV & 1.6$\times10^6$ & 1.04$\times10^7$ & 236392 & 120527 & 12.3$\times10^6$ &8261 & 3246\\ 
MHT $>$ 500 GeV & 8503 & 108477 & 4795 & 483 &122458&3502 & 1895\\
$|\Delta \phi (Jets, MHT)|$ & 1094 & 22256 & 934 & 43 & 24328 & 912 & 519\\ 
\hline \hline
\end{tabular}
\label{tab:had_app}
\end{table}

The cutflow with the number of expected signal and background events for the bottom squark search is given in
Table~\ref{tab:baselineEventYields}.

\begin{table}[!htb]
\caption{LHC bottom squark search: Event yields for the different SM processes and STC models after applying each cut, the numbers are normalised to total 300\fbinv.}
\begin{tabular}{ l | r r r r | r | r r }\hline \hline
\textbf{Cutflow} &  \textbf{VV+jets}  & \textbf{\ttbar+jets}  &  \textbf{V+jets}  &  \textbf{Top+jets} & \textbf{Total SM}    & \textbf{STC8} & \textbf{STC10} \\ \hline
Preselection & 86940500 & 173553000 & 11212700000 & 34843100 & 11508036600 & 1160340 & 1202280
\\ \hline
Lepton(e,$\mu$)veto & 63661400 & 86844100 & 9884430000 & 24245100 & 10059180600 & 936391 & 974139
\\ \hline
N-Jets $>=$2 & 19224300 & 77540200 & 3895480000 & 16818900 & 4009063400 & 191599 & 185023
\\ \hline
$1^{st}$Jet$\pt>$300\GeV & 688728 & 2606200 & 42888600 & 388099 & 46571627 & 27418 & 18690
\\ \hline
$2^{nd}$Jet$\pt>$200\GeV & 437937 & 1602890 & 31104200 & 237408 & 33382435 & 15593 & 9458
\\ \hline
$2^{nd}$Jet$\pt<$70\GeV & 124572 & 207703& 9029620 & 60479 & 9422374 & 2801 & 2130
\\ \hline
N-bjets =2 & 543 & 29836 & 32045 & 3290 & 65715 & 521 & 159
\\ \hline
\ETm $>$ 450\GeV& 5.9 & 386 & 72 & 14 & 478 & 214 &  82
\\ \hline
\MT $>$ 500\GeV & 4.1 & 12 & 51 & 0.27 & 67 & 184 & 72 \\ \hline
\HT $>$ 850\GeV & 1.4 & 2.8 & 24 & 0.05 & 28 & 74 & 44
\\ \hline \hline

\end{tabular}
\label{tab:baselineEventYields}
\end{table}

Table~\ref{tab:cutflow_stop} shows the expected number of SM
background and SUSY signal events after each step of the event
selection for the top squark search. 

\begin{table}[htb!]
  \caption{LHC top squark search: The event yields for the
 inclusive signal samples and several SM processes with 300\fbinv at 14\TeV with 50 pileup interactions.
}

\begin{center}
\begin{tabular}{l | r r r r | r | r r }
\hline
\hline
\textbf{Cutflow}  &  \textbf{VV+jets}  & \textbf{Cutflow}  &  \textbf{VV+jets} s &  \textbf{Top+jets} &  \textbf{Total SM}  & \textbf{STC8} & \textbf{STC10} \\ \hline
Exactly 1 lepton & 10141300 & 43887300 & 548959000 & 7665270 & 610652870 & 114405 & 112180 \\ \hline
Njets $>=$ 5 & 57851 & 3298680 & 827110 & 29067 & 4212708 & 4149 & 1911 \\ \hline
1 or 2 b jets & 19249 & 2405190 & 232049 & 21492 & 2677980 & 2642 & 1040 \\ \hline
$\MET$ $>$ 400 GeV & 219 & 5339 & 1601 & 62 & 7222 & 642 & 410 \\ \hline
$\Delta \phi >$ 0.8 & 194 & 3990 & 1397 & 45 & 5629 & 536 & 338 \\ \hline
Centrality $>$ 0.6 & 117 & 2275 & 822 & 24 & 3240 & 418 & 282 \\ \hline
$\mT>260\GeV$ & 4 & 94 & 0 & 0 & 98 & 194 & 134 \\ \hline
$\MTtW>260\GeV$ & 2 & 26 & 0 & 0 & 28 & 155 & 113 \\ \hline
\hline
\end{tabular}
\label{tab:cutflow_stop}
\end{center}
\end{table}

The number of expected events for each search region in the multilepton channel is given in Table~\ref{tab:3l}.

\begin{table*}[!h]
\caption{LHC Multilepton search: Expected events for an integrated luminosity of \Lum = 300\fbinv. The definition of the search channels (ch) is given in the text and Fig.~\ref{fig:multilepton_plots}. EWK refers here to electroweakly produced SUSY particles. \label{tab:3l}}
\begin{center}
\begin{tabular}{c |c | c |c|c c c c}
\hline \hline
 \textbf{Ch} &      \textbf{Total SM $\pm$ unc.} &      \textbf{STC8} &  $Z_{\mathrm {Bi}}$ &    \ninoone\chipmone &  \XPM2\Cnm &    \textbf{Other EWK}  &     \textbf{No EWK} \\\hline
\multicolumn{7}{l}{m$_{\ellell}<$ 75\GeV}& \\\hline
1 &     10900 $\pm$  3300 &     88 &    $<$0.5 &        76 &    7 &     3 &     2        \\
2 &     5900 $\pm$  960 &       130 &   $<$0.5 &        110 &   10 &    10 &    0        \\
3 &     1390 $\pm$  340 &       140 &   $<$0.5 &        110 &   20 &    10 &    10       \\
4 &     1290 $\pm$  210 &       26 &    $<$0.5 &        16 &    7 &     2 &     2        \\
5 &     348 $\pm$  121 &        19 &    $<$0.5 &        10 &    6 &     2 &     1        \\
6 &     45.1 $\pm$  32.4 &      8.5 &   $<$0.5 &        2.3 &   4.5 &   0.5 &   1.2      \\
7 &     469 $\pm$  125 &        26 &    $<$0.5 &        9 &     12 &    4 &     2        \\
8 &     29.6 $\pm$  6.8 &       9.6 &   0.9 &   2.5 &   5.7 &   0.6 &   0.8      \\
9 &     1.26 $\pm$  0.41 &      1.0 &   $<$0.5 &        0.2 &   0.6 &   0.0 &   0.3      \\
10 &    21.4 $\pm$  3.2 &       6.6 &   1.0 &   1.2 &   4.6 &   0.6 &   0.1      \\
11 &    4.48 $\pm$  1.72 &      2.7 &   0.6 &   0.5 &   1.9 &   0.3 &   0.0      \\
12 &    0.0262 $\pm$  0.0095 &  0.3 &   $<$0.5 &        0.0 &   0.15 &  0.15 &  0.0      \\
13 &    1.06 $\pm$  0.19 &      1.1 &   0.6 &   0.0 &   0.4 &   0.6 &   0.0      \\
14 &    0.89 $\pm$  0.263 &     0.3 &   $<$0.5 &        0.0 &   0.3 &   0.0 &   0.0      \\
15 &    0.0137 $\pm$  0.0048 &  0 &     $<$0.5 &        0 &     0 &     0 &     0        \\
\hline
\multicolumn{7}{l}{75\GeV $<$ m$_{\ellell}<$ 105\GeV}& \\\hline
1 &     111000 $\pm$  16000 &   97 &    $<$0.5 &        79 &    11 &    5 &     3        \\
2 &     45900 $\pm$  7700 &     170 &   $<$0.5 &        140 &   20 &    10 &    10       \\
3 &     7490 $\pm$  1390 &      210 &   $<$0.5 &        140 &   50 &    10 &    10       \\
4 &     4640 $\pm$  490 &       26 &    $<$0.5 &        12 &    10 &    2 &     1        \\
5 &     994 $\pm$  278 &        31 &    $<$0.5 &        13 &    13 &    4 &     1        \\
6 &     55.4 $\pm$  40.3 &      16 &    $<$0.5 &        2 &     10 &    2 &     2        \\
7 &     444 $\pm$  75 & 30 &    $<$0.5 &        9 &     17 &    2 &     1        \\
8 &     26.2 $\pm$  5.0 &       26 &    2.9 &   6 &     16 &    4 &     2        \\
9 &     1.91 $\pm$  0.47 &      2.0 &   1.0 &   0.2 &   1.5 &   0.0 &   0.4      \\
10 &    16.4 $\pm$  1.4 &       6.7 &   1.4 &   0.9 &   4.7 &   0.8 &   0.4      \\
11 &    5.01 $\pm$  0.93 &      3.0 &   1.0 &   0.6 &   1.6 &   0.0 &   0.8      \\
12 &    0.058 $\pm$  0.0176 &   0.12 &  $<$0.5 &        0.0 &   0.0 &   0.0 &   0.12     \\
13 &    1.77 $\pm$  0.2 &       0.15 &  $<$0.5 &        0.0 &   0.15 &  0.0 &   0.0      \\
14 &    2.32 $\pm$  0.35 &      0.31 &  $<$0.5 &        0.0 &   0.31 &  0.0 &   0.0      \\
15 &    0.113 $\pm$  0.034 &    0 &     $<$0.5 &        0 &     0 &     0 &     0        \\
\hline
\multicolumn{7}{l}{m$_{\ellell}>$ 105\GeV}& \\\hline
1 &     2380 $\pm$  320 &       22 &    $<$0.5 &        11 &    7 &     3 &     1        \\
2 &     1720 $\pm$  240 &       34 &    $<$0.5 &        18 &    10 &    4 &     2        \\
3 &     614 $\pm$  157 &        61 &    $<$0.5 &        21 &    24 &    10 &    6        \\
4 &     217 $\pm$  47 & 10 &    $<$0.5 &        3 &     6 &     1 &     1        \\
5 &     57.6 $\pm$  11.5 &      11 &    0.7 &   2 &     6 &     3 &     1        \\
6 &     13.7 $\pm$  3.5 &       8.0 &   1.2 &   1.4 &   3.9 &   1.8 &   0.8      \\
7 &     70.2 $\pm$  6.0 &       15 &    1.4 &   1 &     13 &    2 &     0        \\
8 &     12.6 $\pm$  2.7 &       15 &    2.6 &   2 &     10 &    2 &     1        \\
9 &     0.812 $\pm$  0.174 &    1.4 &   1.0 &   0.2 &   0.7 &   0.2 &   0.4      \\
10 &    11.7 $\pm$  1.4 &       3.5 &   0.8 &   0.2 &   2.3 &   1.1 &   0.0      \\
11 &    1.94 $\pm$  0.41 &      2.8 &   1.4 &   0.3 &   1.5 &   0.3 &   0.7      \\
12 &    0.0391 $\pm$  0.0114 &  0 &     $<$0.5 &        0 &     0 &     0 &     0        \\
13 &    1.36 $\pm$  0.14 &      0 &     $<$0.5 &        0 &     0 &     0 &     0        \\
14 &    0.674 $\pm$  0.176 &    0.78 &  $<$0.5 &        0.0 &   0.15 &  0.15 &  0.47     \\
15 &    0.0137 $\pm$  0.0055 &  0.24 &  $<$0.5 &        0.0 &   0.0 &   0.0 &   0.24     \\
\hline \hline
\end{tabular}
\end{center}
\end{table*}